\documentclass[12pt]{article}
\usepackage{
graphicx} 
\usepackage{amsmath}
\usepackage{amsfonts}
\usepackage{amssymb}
\usepackage{mathrsfs}
\usepackage{setspace}
\usepackage{color}
\usepackage{graphicx}
\usepackage{epstopdf}
\usepackage{float}
\usepackage{multirow}
\usepackage{mathtools}
\usepackage[lofdepth,lotdepth]{subfig}
\usepackage{graphicx,dblfloatfix}
\usepackage{lipsum} 
\usepackage{amsthm}
\usepackage[table]{xcolor}
\usepackage{booktabs}
\usepackage{threeparttablex}
\usepackage{hhline}
\usepackage{subcaption}
\usepackage[export]{adjustbox} 
\usepackage{rotating}

\usepackage[round,authoryear,sort]{natbib} 

\usepackage[colorlinks,
            linkcolor=blue,
            anchorcolor=blue,
            citecolor=blue
            ]{hyperref}

\newcommand{\keywords}[1]{%
  \begin{center}
    \textbf{Keywords:} #1
  \end{center}
}
\newtheorem{theorem}{Theorem}[section]

\newtheorem{lemma}{Lemma}[section]

\newtheorem{proposition}{Proposition}[section]

\numberwithin{equation}{section}
\newcommand{\QED}{\hfill $\Box$}
\newcommand{\COM}[1]{}
\newcommand{\abs}[1]{\left\lvert #1 \right\rvert}
\newcommand{\norm}[1]{\left\lVert #1 \right\rVert}
\newcommand{\pk}[1]{\mathbb{P} \left\{ #1 \right\} }
\newcommand{\expon}[1]{\exp\left(#1\right)}
\newcommand{\equaldis}{\stackrel{d}{=}}

\def\vk#1{\mathbf#1}

\def\I#1{\operatorname{\mathbb{I}}\{#1\}}

\def\E#1{\mathbb{E}\left \{#1 \right\}}
\def\fracl#1#2{\biggr(\frac{#1}{#2} \biggl) }
\def\d{\,\mathrm{d}} 
\def\e{\mathrm{e}} 
\def\I#1{\mathbb{I}\left (#1 \right)}

\def\R{\mathbb{R}}

\def\todis{\stackrel{d}\rightarrow}
\def\topb{\stackrel{p}\rightarrow}

\newcommand{\BQN}{\begin{eqnarray}}
\newcommand{\EQN}{\end{eqnarray}}
\newcommand{\BQNY}{\begin{eqnarray*}}
\newcommand{\EQNY}{\end{eqnarray*}}

\definecolor{c20}{RGB}{255,80,0}
\definecolor{c30}{rgb}{0.,0.,1.}
\definecolor{c40}{rgb}{1,0.1,0.7}
\definecolor{c50}{rgb}{1,0,0}
\definecolor{c60}{rgb}{1,0.9,0.1}

\RequirePackage{colortbl} 
\RequirePackage{soul}

\usepackage{authblk}
\usepackage{siunitx}
\usepackage{textcomp}
\usepackage[left]{lineno}
\title{Competing Accelerated Failure Time Models for Multiple Concurrent Failure Mechanisms}
\author[1,2]{Kai Wang}
\author[1,2]{Yuqin Mu}
\author[1]{Shenyi Zhang}
\author[*,3,4]{Zhengjun Zhang}
\author[*,1]{Chengxiu Ling}

\affil[1]{\small Academy of Pharmacy, Xi'an Jiaotong-Liverpool University, Suzhou, 215123, China}
\affil[2]{\small Department of Mathematical Sciences, University of Liverpool, Liverpool, L693BX, UK}
\affil[3]{\small School of Data Science and Artificial Intelligence, Beijing University of Chinese Medicine, Beijing 100029, China}

\affil[4]{\small Department of Statistics, School of Computer, Data \& Information Sciences, University of Wisconsin, Madison, WI 53706, USA}

\affil[*]{\small Corresponding authors: Chengxiu.Ling@xjtlu.edu.cn; zjz@stat.wisc.edu}

\date{}

\begin{document}

\maketitle
\begin{abstract}
The rising prevalence of complex diseases characterised by multiple coexisting and interacting etiological processes poses critical challenges for survival analysis and precision medicine, particularly as population ageing renders mutually exclusive models increasingly untenable. We propose a competing accelerated failure time (cAFT) framework to understand the individual-specific temporal dynamics of disease competition and interaction based on a "first-to-fail" principle. Specifically, we introduce an individualised, time-varying "winning probability" to quantify the relative contributions of latent causes and provide an interpretable basis for patient stratification within distinct subtypes. Consistency and asymptotic normality are established for the maximum likelihood estimation of the parameters, with practical implementation via an expectation-maximisation (EM) algorithm. We illustrate the model’s effectiveness and efficiency through numerical simulations and real-world applications, including biomarker discovery for 28-day survival in sepsis and overall survival in lung adenocarcinoma. Compared with standard AFT and Cox proportional hazards models, the cAFT model consistently improves predictive accuracy (C-index and iAUC gains of 5–10\%) and reveals subtype-dependent gene effects within distinct biological pathways across heterogeneous patient subgroups. Conclusively, the cAFT model provides deeper insights into patient prognosis and potential personalised therapeutic strategies.
\end{abstract}

\keywords{multiple causes, EM algorithm, regularisation, accelerated failure time model}

\section{Introduction}\label{sec: intro}
\doublespacing
Multiple causes are increasingly recognised as a defining feature of modern survival data, largely driven by the global rise in multimorbidity, particularly in ageing populations with coexistence of multiple chronic diseases \citep{bishop2023quantifying,dobson2023new}. Approximately 37.2\% of adults worldwide live with two or more chronic conditions, with prevalence exceeding 50\% among those aged over 60 years \citep{Chowdhury2023global}, and continuing to increase roughly 1\% annually \citep{breger2020estimating, LIN2024100260}. 
Consequently, mortality in contemporary populations is seldom attributable to a single isolated disease, but instead reflects the joint action of coexisting conditions and interacting biological pathways. Such complexity challenges conventional competing-risk frameworks that attribute each failure event to a single underlying cause, leading to underestimation of disease burden \citep{bishop2023quantifying} and misclassification of the underlying cause \citep{Johansson2000, mcgivern2017death}. 
Similar challenges arise in high-dimensional biomedical studies of complex diseases, where a single disease may involve multiple etiologies and, in fact, represent different routes or mechanisms that interact toward similar conditions in a patient \citep{ma2014modeling,Johansson2023precision,Jorgensen2020age}. 
Identifying a precise trigger in such cases often requires complex diagnostics, extensive biological assays, or prohibitively costly experiments, and even then, the underlying mechanisms may remain uncertain. From a modelling perspective, these survival outcomes arise from multiple concurrent mechanisms whose individual contributions are difficult to disentangle, motivating a framework that accommodates simultaneous causes and quantifies their relative importance.

{Recent methodological advances have shifted attention from single underlying causes to multiple contributing causes of failure. 
\citet{moreno2017survival} proposed a disease mixture model where a patient's hazard is expressed as a weighted combination of two pure states, one corresponding to a pre-specified disease of interest and the other aggregating all remaining causes, which has since been widely applied in epidemiological studies \citep{breger2020estimating, xie2020causes, anderson2023body, wijnen2022observed, dobson2023new}. A key component is the pre-assignment of weights, which combines rule-based attribution from death certificates with heuristic choices \citep{moreno2017survival} or external population-level information \citep{dobson2023new}. However, such predetermined, time-invariant weights can be restrictive in practice, as they do not account for subject-specific risk over time and may therefore misrepresent the contributions of different causes \citep{moreno2017survival}. Moreover, the formulation centred on a single disease of interest does not address a conceptually meaningful question in multiple-cause settings: for a given patient, which mechanism among several candidates predominantly drives the event. 
Identifying such dominance provides a natural way to characterise heterogeneity across patients. In precision medicine, similar forms of heterogeneity that manifest in treatment responses are studied using finite-mixture models with mixed densities, where distinct risk profiles arise across disease subtypes \citep{Wu2016subgroup, You2018subtype}.}

{Extreme linear (max/min-linear) models provide a natural foundation for addressing these challenges by introducing a competing factor structure: predictors enter linearly within groups, while a max/min operator across groups induces a structured nonlinearity \citep{cui2018max, Cui2021Max, Zhang2022genomic}. This formulation preserves interpretability while capturing heterogeneous effects through factor competition, enabling the identification of dominant mechanisms at the individual level. It bridges classical linear models and more complex nonlinear relationships, offering a transparent and flexible alternative to additive or black-box methods, with demonstrated utility in biomedical applications \citep{ zhang2021five,Zhang2022genomic}.}

{In this paper, we propose a competing accelerated failure time (cAFT) model for multiple causes, generalizing the extreme linear framework to survival analysis via the latent survival time formulation \citep{prentice1978analysis}. Each cause is represented as a competing factor, with covariate effects entering linearly into the corresponding latent survival times, and the observed event time defined as their minimum under a "first-to-fail" principle. This formulation accommodates concurrent mechanisms and introduces a data-driven, individualised (by covariates) and time-varying winning probability to quantify cause-specific contributions, addressing a central objective in multiple-cause mortality \citep{redelings2006comparison}. Meawhile, it enables interpretable quantification of cause-specific risk effects and direct estimation of survival times, supporting broad biomedical and epidemiological applications \citep{hosmer2008applied, butikofer2015two, newcombe2017weibull}.}

{The main contributions of this work are threefold. First, the competing structure provides a natural representation of multiple causes without imposing a single, rigidly prespecified cause of failure. It captures competitive interactions and cross-pathway biomarker effects that may be obscured by conventional AFT and Cox proportional hazards (PH) models (see Section \ref{sec: App}). Second, leveraging the extreme linear structure, the model accommodates a broad class of survival distributions through flexible error specifications and supports regularisation for high-dimensional biomarker data (Section \ref{sec: penal}), which are increasingly common in modern prognostic modelling \citep{You2018subtype}. Third, our approach yields interpretable measures of cause-specific contributions, interpretable as cause-specific burdens \citep{bishop2023quantifying}. These measures enable investigation of heterogeneous disease progression across mechanisms, and their variation across patients further supports stratification by dominant cause, providing mechanistic insights and facilitating personalised therapeutic strategies \citep{zhang2021five, Zhang2022genomic, liu2023five}, as illustrated in real applications (Section \ref{sec: App}).}

The paper is organised as follows. In Section \ref{sec: Model}, we introduce the competing AFT model. Section \ref{sec: EandP} describes the expectation-maximisation (EM) algorithm for maximum likelihood estimation, outlines sparsity penalisation, and presents the asymptotic properties. Sections \ref{sec: simu} and \ref{sec: App} present simulation studies and real-data applications to survival outcomes in sepsis and LUAD patients, respectively. Finally, Section \ref{sec: Dis} provides concluding remarks.
In the Appendices, Appendix A presents technical proofs and the unbiased prediction survival time estimator with approximation methods. Appendix B provides further applications to Alzheimer’s disease, hepatocellular carcinoma (HCC), and breast cancer. Appendix C reports model performance compared with the AFT, Cox PH, and adapted disease mixture models in simulations, along with additional simulations under violations of model assumptions.

\section{Model Specification}\label{sec: Model}
The AFT model specifies that the effect of a fixed covariate $\vk Z=(Z_1, \ldots, Z_p)^\top\in \mathbb{R}^p$ acts additively on the log-transformed survival time ($\log\widetilde T$). A linear model gives a location-scale form \citep[][Chapter 7]{kalbfleisch2011statistical}:
\begin{eqnarray*}
    \log \widetilde T \equaldis  \alpha+\vk Z^\top\boldsymbol\beta+ \sigma \epsilon,
\end{eqnarray*}
where $\alpha \in \mathbb{R}$ is the intercept, $\boldsymbol\beta \in \mathbb{R}^p$ is the vector of log-multiplicative effect parameters associated with the covariates $\vk Z$ on the survival time, $\sigma>0$ is a scale parameter, $\epsilon$ is a random error from a specified parametric distribution, and $\equaldis$ denotes equality in distribution.

Motivated by the extreme linear competing structure \citep{cui2018max, Cui2021Max, zhang2021five}, we propose a competing parametric regression model for survival time $\widetilde T$, whose log-scale is expressed as the minimum over $L$ competing factors (CFs), each denoting a potential contributing cause or disease mechanism.
\begin{eqnarray}\label{eq: Gumbel-location-scale}
   \log \widetilde T \equaldis \min (\alpha
_1+\vk X_1^\top \boldsymbol{\beta}_1 + \sigma_1 \epsilon_1,\ldots, \alpha_L+\vk X_L^\top \boldsymbol{\beta}_L + \sigma_L \epsilon_L),
\end{eqnarray}
where $\epsilon_l, \ l=1, 2, \ldots, L$ are independent and identically distributed errors with common distribution, and $\vk X_l = (X_j,\, j\in {A_l})$ is a $p_l$-dimensional sub-vector of candidate predictors $\vk X = ( X_1, \ldots,  X_p)^\top\in \mathbb{R}^p$. Here, $A_l \subset \{1, 2, \ldots, p\}$ with $|A_l| = p_l\in[1, p]$, and $|\cdot|$ denotes the cardinality of a set. The predictor vectors ($\vk X_l$'s) may overlap in their components but cannot be entirely identical, 
otherwise unidentifiability will be introduced. The intercepts $\alpha_1,\ldots,\alpha_L$ control the relative importance of each CF, where a sufficiently large intercept implies an approximately eliminated risk under the min-competing structure. 

The competing framework extends beyond the classical AFT model while preserving it as a special case of a single CF. For common choices of error ($\epsilon_l$) distributions, such as the standard Gumbel (for minima) or standard normal, the survival time $\widetilde T$ falls outside the AFT family, with the survival probability given by
\begin{eqnarray}\label{eq: T-survival}
    \pk{\widetilde T > t} =  \prod_{1\le l \le L} \pk{\epsilon_l > \frac{ \log t - \alpha
_l-\vk X_l^\top \boldsymbol{\beta}_l}{\sigma_l}}
     =   \prod_{1\le l \le L} \pk{\widetilde T_l > t},
\end{eqnarray}
where $\widetilde T_l,\ l=1,2,\ldots, L$, are potential survival times for contributing causes, following Weibull or lognormal distributions and conditionally independent given $\mathbf{X}_l$'s. Therefore, the hazard and survival functions of $\widetilde T$, denoted ${h}(t)$ and ${S}(t)$, satisfy the following equations
\begin{eqnarray}
\label{eq: marginal-hazard}
  \log {S}(t) = \sum_{l=1}^L \log {S}_l(t), \quad
    {h}(t) = -\frac{\d}{\d t}\sum_{l=1}^L \log {S}_l(t) = \sum_{l=1}^L {h}_l(t),
\end{eqnarray}
where ${h}_l(t)$ and ${S}_l(t)$ correspond to the hazard and survival functions of $\widetilde T_l$, respectively.
{The assumption of mutually independent errors ($\epsilon_l$'s) is not directly testable, since the joint distribution of latent survival times is not identifiable from the observed data \citep{prentice1978analysis}. However, the likelihood-based inference (Section \ref{sec: EM}) depends only on the joint survival function along the diagonal, and does not require strict mutual independence. The modelling assumption can thus be viewed as a working structural restriction on this diagonal survival that induces an additive hazard decomposition, at a cost of losing clear marginal interpretations \citep{andersen2012competing}.
Adequacy of the additive structure and parametric specification can be assessed via residual-based diagnostics \citep[][Chapter 11]{klein2003survival}. In this paper, Cox-Snell residual plots show good agreement under correct specification, whereas under dependent errors, the regression coefficient estimates remain relatively robust, but the diagnostics exhibit clear departures (Appendix C).
}

In contrast to classical competing-risk models that assume a single pre-specified cause drives each event, our framework allows multiple causes to act simultaneously without prior assignment of which will prevail. It leverages the data to determine the dominant mechanism via winning probabilities, defined as the ratio of the marginal hazard to the total hazard (cf. Eq.\eqref{eq: winning probability eta}), and naturally analogous to a softmax function (see Section \ref{sec: EandP}). 
This extends survival analysis beyond time-to-event modelling to identify the most likely driving mechanism and quantify the contributions of multiple latent processes. The resulting probabilities further enable attribution of survival times and patient stratification, providing both interpretability and a systematic approach to dissect heterogeneity in clinical outcomes (see Section \ref{sec: App}).

\section{Estimation and Prediction}\label{sec: EandP}
\subsection{Estimation via EM algorithm} \label{sec: EM}
We consider the setting of random censorship with data $(T_i, \delta_i, \vk X_i),\, i=1,\ldots, n,$ where $T_i = \min(\widetilde T_i, C_i)$, $\delta_i = \mathbb{I}(\widetilde{T}_i< C_i)$ is the failure indicator of the competing survival time $\widetilde{T}_i$ ({cf. Eq.\eqref{eq: Gumbel-location-scale}}),
and $C_i$ is an independent random censoring time. The vector $\vk X_{il} = (X_{il1}, \ldots,X_{ilp_l})^\top \in \mathbb{R}^{p_l}$ denotes covariates of the $i$-th sample in the $l$-th CF. We denote all parameters (location, coefficients, scale)  involved in each CF  for the survival time $\widetilde T$ as
\begin{eqnarray*}
    \boldsymbol\theta = (\boldsymbol\theta_1^\top,\ldots,\boldsymbol\theta_L^\top)^\top := (\alpha_1, \boldsymbol{\beta}_1^\top, \sigma_1, \ldots, \alpha_L, \boldsymbol{\beta}_L^\top, \sigma_L)^\top \in \Theta,
\end{eqnarray*}
where $\Theta = \prod_{l=1}^L (\R^{p_l+1}\times\R^+)$ stands for the parameter space. The likelihood function for the competing parametric survival model is given by (recall Eq.\eqref{eq: marginal-hazard})
\begin{eqnarray}\label{eq: lik}
    L(\boldsymbol\theta) &=& \prod_{i=1}^n f^{\delta_i}(T_i; \boldsymbol\theta, \vk X_i) S^{1-\delta_i}(T_i; \boldsymbol\theta, \vk X_i) = \prod_{i=1}^n h^{\delta_i}(T_i; \boldsymbol\theta, \vk X_i)S(T_i; \boldsymbol\theta, \vk X_i) \notag \\
    &=& \prod_{i=1}^n \left(\sum_{l=1}^L h_l(T_i; \boldsymbol\theta, \vk X_i) \right)^{\delta_i} \expon{ \sum_{l=1}^L\log S_l(T_i; \boldsymbol\theta, \vk X_i)} 
\end{eqnarray}
leading thus to the log-likelihood
\begin{eqnarray*}
    \ell(\boldsymbol\theta) = \log L (\boldsymbol\theta) = \sum\limits_{i=1}^n \delta_i \log \left(\sum\limits_{l=1}^L h_l(T_i;  \alpha_l, \boldsymbol{\beta}_l, \sigma_l,\vk X_{il}) \right)+ \sum\limits_{l=1}^L\log S_l(T_i; \alpha_l, \boldsymbol{\beta}_l, \sigma_l,\vk X_{il}) ,
\end{eqnarray*}
which is differentiable with respect to $\boldsymbol\theta$. We will show Theorem \ref{thm: 2.1} for the asymptotic properties of the maximum likelihood estimator (MLE) $\hat{\boldsymbol\theta}$ as the maximiser of $L(\boldsymbol\theta)$ in Eq.\eqref{eq: lik}.

Next, we develop an EM algorithm to optimise the group-wise likelihoods for $\boldsymbol\theta_l, \, l=1,2, \ldots, L$. Recall that when maximising the observed likelihood directly with parameter $\boldsymbol{\theta}$ and data $\boldsymbol t$ is complicated, the EM algorithm \citep{dempster1977maximum} provides an effective method by introducing an augmented variable, with the log-likelihood expressed as
\begin{eqnarray*}
\log L(\boldsymbol\theta|\boldsymbol t) = \mathbb{E}_{\boldsymbol\theta_0}\{\log L_C(\boldsymbol\theta|\boldsymbol t, \boldsymbol k)\}- \mathbb{E}_{\boldsymbol{\theta}_0}\{\log p( \boldsymbol k|\boldsymbol\theta,\boldsymbol t)\},
\end{eqnarray*}
{where $p( \boldsymbol
k|\boldsymbol\theta,\boldsymbol t) = p(\boldsymbol t , \boldsymbol k |\boldsymbol \theta)/p(\boldsymbol t|\boldsymbol\theta)$, and the expectation is taken with respect to $p( \boldsymbol k|\boldsymbol\theta_0,\boldsymbol t)$ with true parameter $\boldsymbol\theta_0$. To maximise the log-likelihood, we only need to iteratively increase the expectation of completed log-likelihood $ \mathbb{E}_{\boldsymbol\theta_0}\{\log L_C(\boldsymbol\theta|\boldsymbol t, \boldsymbol k)\}$, 
as $\mathbb{E}_{\boldsymbol\theta_0}\{\log p( \boldsymbol k|\boldsymbol\theta,\boldsymbol t)\}$ naturally decreases by Jensen's inequality when $p( \boldsymbol k|\boldsymbol\theta,\boldsymbol t)\ne p( \boldsymbol k|\boldsymbol\theta_0,\boldsymbol t)$.}

In this study, we introduce an augmented variable $K$, denoting the dominating group for the uncensored samples ($\delta =1$). Conditionally on $K_i=l$, the failure time holds for $T_i= \widetilde T_{il}$ (recall Eq.\eqref{eq: T-survival}), yielding the complete likelihood for $(\boldsymbol T, \boldsymbol K)$ under censoring:
\begin{eqnarray} \label{eq: Lc}
   L_C(\boldsymbol\theta)  = \prod_{i=1}^n \bigg\{\prod_{l=1}^L f^{\I{K_i=l}}(T_i, K_i=l; \boldsymbol\theta, \vk X_i)\bigg\}^{\delta_i}  S^{1-\delta_i}(T_i; \boldsymbol\theta, \vk X_i).
\end{eqnarray}
Noting the explicit form of survival function $S(t)$ specified in Eq.\eqref{eq: T-survival},
it follows that $K=l$ for uncensored samples is equivalent to {that} $\log T = \alpha_l+\vk X_{l}^\top \boldsymbol{\beta}_l+\sigma_l\epsilon_l
\le\alpha_k+\vk X_{k}^\top \boldsymbol{\beta}_k+\sigma_k\epsilon_k,\, \forall k\ne l$.  We have the joint survival distribution of $(T, K)$ given as
\begin{eqnarray*}
  \lefteqn{ \pk{T > t, K=l} = \pk{ \log T > \log t, K=l}}  \\
    &=& \pk{ \alpha_k+\vk X_{k}^\top \boldsymbol{\beta}_k+\sigma_k\epsilon_k >  \alpha_l+\vk X_{l}^\top \boldsymbol{\beta}_l+\sigma_l\epsilon_l > \log t, \, \forall k \ne l}\\
    & = & \int_{z>t}\prod_{k\neq l} \pk{\widetilde{T}_k > \log z} \d \pk{\widetilde{T}_l \le \log z} \quad \mbox{{by the total law of probability}} \\
        & = & \int_t^\infty h_l(z; \alpha_l,\boldsymbol{\beta}_l,\sigma_l,\vk X_{l})\exp\bigg( \sum\limits_{k=1}^L\log S_k(z; \alpha_k,\boldsymbol{\beta}_k,\sigma_k,\vk X_{k} )\bigg) \d z,\quad t>0,
\end{eqnarray*}
implying together with Eq.\eqref{eq: marginal-hazard} that  $f(t, K=l; \boldsymbol\theta, \vk X)  = h_{l}(t; \boldsymbol\theta, \vk X) S(t; \boldsymbol\theta, \vk X)$.
Combining Eqs.\eqref{eq: marginal-hazard} and \eqref{eq: Lc}, it is ready to give the complete log-likelihood function
\begin{eqnarray*}
 \ell_C(\boldsymbol\theta) &=& \sum\limits_{i=1}^n \delta_i\sum\limits_{l=1}^L \I{K_i=l} \log f(T_i, K_i = l; \boldsymbol\theta, \vk X_i)+\sum\limits_{i=1}^n (1-\delta_i) \log S(T_i; \boldsymbol\theta, \vk X_i) \\
&=& \sum\limits_{i=1}^n \delta_i \sum\limits_{l=1}^L \I{K_i=l}  \left\{\log h_l(T_i; \alpha_l, \boldsymbol{\beta}_l,\sigma_l,\vk X_{il})+ \sum\limits_{k=1}^L  \log S_k(T_i; \alpha_k, \boldsymbol{\beta}_k,\sigma_k,\vk X_{ik})\right\}\\
&&+\sum\limits_{i=1}^n (1-\delta_i) \sum\limits_{l=1}^L \log S_l(T_i; \alpha_l, \boldsymbol{\beta}_l,\sigma_l,\vk X_{il}).
\end{eqnarray*}
The conditional probability of {$K=l$} given $t$, interpreted as the winning probability of the $l$-th CF, is then expressed as
\begin{eqnarray}
    \label{eq: winning probability eta}\eta_l(t;\boldsymbol\theta, \vk X) = \pk{K=l|t,\boldsymbol\theta, \vk X}=  \frac{f(t, K=l;\boldsymbol\theta, \vk X)}{f(t;\boldsymbol\theta, \vk X)}= \frac{h_l(t;\alpha_l, \boldsymbol{\beta}_l,\sigma_l,\vk X_{l})}{\sum\limits_{k=1}^Lh_k(t;\alpha_k, \boldsymbol{\beta}_k,\sigma_k,\vk X_{k})}.
\end{eqnarray}
This leads to the following analytical form of the conditional expectation of the complete log-likelihood function $Q(\boldsymbol{\theta}|\boldsymbol\theta^{(m)})$ in the E-step of the EM algorithm, where $\boldsymbol\theta^{(m)}$ is the estimated $\boldsymbol\theta$ at step $m$ (the same for $\eta_{il}^{(m)}$):
\begin{eqnarray}
\label{eq: Q}
Q(\boldsymbol{\theta}|\boldsymbol\theta^{(m)})&=&\sum\limits_{i=1}^n  \sum\limits_{l=1}^L  \delta_i \eta_{il}^{(m)} \left\{\log h_l(T_i; \alpha_l, \boldsymbol{\beta}_l,\sigma_l,\vk X_{il})+ \sum\limits_{k=1}^L  \log S_k(T_i; \alpha_k, \boldsymbol{\beta}_k,\sigma_k,\vk X_{ik})\right\}  \notag\\
&&+\sum\limits_{i=1}^n  \sum\limits_{l=1}^L (1-\delta_i) \log S_l(T_i; \alpha_l, \boldsymbol{\beta}_l,\sigma_l,\vk X_{il})   \notag\\
&=& \sum\limits_{i=1}^n  \sum\limits_{l=1}^L  \delta_i \eta_{il}^{(m)} \log h_l(T_i; \alpha_l, \boldsymbol{\beta}_l,\sigma_l,\vk X_{il})+ \log S_l(T_i; \alpha_l, \boldsymbol{\beta}_l,\sigma_l,\vk X_{il}).
\end{eqnarray}
Clearly, we can rewrite the $Q$ function above as the sum of group-wise $Q_l$ functions below
\begin{eqnarray*}
    Q_l(\boldsymbol{\theta}|\boldsymbol\theta^{(m)}) := \sum\limits_{i=1}^n   \delta_i \eta_{il}^{(m)} \log h_l(T_i; \alpha_l, \boldsymbol{\beta}_l,\sigma_l,\vk X_{il})+ \log S_l(T_i; \alpha_l, \boldsymbol{\beta}_l,\sigma_l,\vk X_{il}).
\end{eqnarray*}
Since the parameters $\boldsymbol\theta_l$ appear only in $Q_l(\boldsymbol\theta| \boldsymbol\theta^{(m)})$, the M-step optimisation can be separated into $L$ independent group-wise problems. The EM algorithm thus iterates between updating the winning probabilities $\eta_l$ of each CF in the E-step and maximising all the group-wise conditional expectations $Q_l$ in the M-step.
Note that the objective function $Q_l(\boldsymbol{\theta}|\boldsymbol{\theta}^{(m)})$ may not be concave with respect to (w.r.t.) parameters $\boldsymbol\theta_l$, while the EM algorithm can still be applied by numerical optimisation methods (e.g., gradient descent), which only guarantee convergence to a local maximum. In Section \ref{sec: theo}, for nonconcave objective functions, we further present Theorem \ref{thm: local} for the existence of a sequence of local maximisers that converges to the true parameters (global maximiser).

Here, we present the competing versions of the two most commonly used parametric survival models: log-normal and Weibull. For the competing log-normal model, the objective function is concave w.r.t. $\mu_{il}: = \alpha_l+\vk X_{il}^\top \boldsymbol{\beta}_l$ and $\sigma_l$ \citep{Cui2021Max}, and is given by
\begin{eqnarray}\label{eq: Q-derivative-G}
    & Q^{\mathrm{LN}}_l(\boldsymbol{\theta}|\boldsymbol{\theta}^{(m)}) = \sum\limits_{i=1}^n  \delta_i \eta_{il}^{(m)} \log\left\{\frac{1}{\sigma_lT_i}\phi\fracl{ \log T_i - \mu_{il}}{\sigma_l} \right\}+(1-\delta_i \eta_{il}^{(m)}) \log \left\{ \bar{\Phi}\fracl{ \log T_i - \mu_{il}}{\sigma_l} \right\},
\end{eqnarray}
where $\phi(\cdot)$ and $\bar{\Phi}(\cdot)$ denote the density and survival function of the standard normal distribution. For the competing Weibull model, the objective function becomes non-concave:
\begin{eqnarray}\label{eq: Q-derivative-W}
    Q^{\mathrm{W}}_l(\boldsymbol{\theta}|\boldsymbol\theta^{(m)})
    =\sum\limits_{i=1}^n  \delta_i\left\{\eta_{il}^{(m)} \left(\frac{1}{\sigma_l}-1\right)\log T_i- \eta_{il}^{(m)}\log\sigma_l-\eta_{il}^{(m)}\frac{\mu_{il}}{\sigma_l}\right\}- \fracl{T_i}{\expon{\mu_{il}}}^{1/\sigma_l}.
\end{eqnarray}
In practice, these functions are augmented with penalisation terms stated below (Section \ref{sec: penal}) to support analyses in high-dimensional settings.

\subsection{Penalisation}\label{sec: penal}
In this section, we introduce sparsity by incorporating the lasso-type penalisation with tuning parameters $\lambda_1>0$ and $\lambda_2>0$,
\begin{eqnarray*}
\min_{\boldsymbol\theta} -Q(\boldsymbol\theta|\boldsymbol\theta^{(m)}) + \sum_{l=1}^L \big\{\lambda_1 \e^{-\alpha_l} +\lambda_2\norm{\boldsymbol{\beta}_l}_1\big\},
\end{eqnarray*}
where {$Q$ is given in Eq.\eqref{eq: Q}, and} $\norm{\cdot}_p$ denotes the $L_p$ norm for a vector.
The penalty on the regression coefficients in each CF is of standard lasso type, hence introducing within-group sparsity.
The penalty on the intercept is of exponential form $\e^{-x}$, which encourages $\alpha_l$ to increase, corresponding to diminishing the importance of the associated CF in the min-competing structure (Eq.\eqref{eq: Gumbel-location-scale}). Under this parametrisation, the intercept penalty would not penalise any intercept to exactly infinity and hence would not introduce group sparsity directly. However, group sparsity can still be observed, with a lower winning probability and a larger intercept, resulting in lower effectiveness for the corresponding CF.

With the lasso-type penalisation, the group-wise partial derivatives w.r.t. $\alpha_l$ and $\beta_{lj},\, j=1, \ldots, p_l$, are taken for the penalised objective function
\begin{eqnarray*}
    Q^*_l(\boldsymbol{\theta}|\boldsymbol\theta^{(m)}):=  -Q_l(\boldsymbol{\theta}|\boldsymbol\theta^{(m)}) +\lambda_1 \e^{-\alpha_l} +\lambda_2\norm{\boldsymbol{\beta}_l}_1.
\end{eqnarray*}
For the competing log-normal case in Eq.\eqref{eq: Q-derivative-G}), the partial derivatives are given by
\begin{align*}
\frac{\partial Q^{*\mathrm{LN}}_l(\boldsymbol{\theta}|\boldsymbol\theta^{(m)})}{\partial \alpha_l}
&= - \sum_{i=1}^n
\left\{ \frac{\delta_i{\eta_{il}^{(m)}} (\log T_i - \mu_{il})}{\sigma_l^2}
+ \frac{(1-\delta_i \eta_{il}^{(m)})\phi\big(\frac{ \log T_i - \mu_{il}}{\sigma_l}\big)}
{\sigma_l\bar{\Phi}\big(\frac{ \log T_i - \mu_{il}}{\sigma_l}\big)}\right\}
- \lambda_1 \e^{-\alpha_l},\\
\frac{\partial Q^{*\mathrm{LN}}_l(\boldsymbol{\theta}|\boldsymbol\theta^{(m)})}{\partial \beta_{lj}}
&= - \sum_{i=1}^n X_{ilj}
\left\{ \frac{\delta_i{\eta_{il}^{(m)}} (\log T_i - \mu_{il})}{\sigma_l^2}
+ \frac{(1-\delta_i \eta_{il}^{(m)})\phi\big(\frac{ \log T_i - \mu_{il}}{\sigma_l}\big)}
{\sigma_l\bar{\Phi}\big(\frac{ \log T_i - \mu_{il}}{\sigma_l}\big)}\right\}
+ \lambda_2 \operatorname{sign}(\beta_{lj}),\\
\frac{\partial Q^{*\mathrm{LN}}_l(\boldsymbol{\theta}|\boldsymbol\theta^{(m)})}{\partial \sigma_l}
&= - \frac{1}{\sigma_l}\sum_{i=1}^n \delta_i{\eta_{il}^{(m)}}
\left\{ \left(\frac{ \log T_i - \mu_{il}}{\sigma_l}\right)^2 -1\right\} + \frac{(1-\delta_i \eta_{il}^{(m)})(\log T_i - \mu_{il})\phi\big(\frac{ \log T_i - \mu_{il}}{\sigma_l}\big)}
{\sigma_l\bar{\Phi}\big(\frac{ \log T_i - \mu_{il}}{\sigma_l}\big)},
\end{align*}
where $\text{sign}(x)$ is the sign function, being $1,-1$ and zero for positive, negative and zero $x$.  
Analogously, for the competing Weibull case (Eq.\eqref{eq: Q-derivative-W}), the partial derivatives are given as
\begin{align*}
    \frac{\partial Q^{*\mathrm{W}}_l(\boldsymbol{\theta}|\boldsymbol\theta^{(m)})}{\partial \alpha_l} &=  \frac{1}{\sigma_l}\sum\limits_{i=1}^n \left\{\delta_i{\eta_{il}^{(m)}}-\fracl{T_i}{\expon{\mu_{il}}}^{1/\sigma_l} \right\} - \lambda_1 \e^{-\alpha_l},\\
    \frac{\partial Q^{*\mathrm{W}}_l(\boldsymbol{\theta}|\boldsymbol\theta^{(m)})}{\partial \beta_{lj}} &=  \frac{1}{\sigma_l}\sum\limits_{i=1}^n X_{ilj}\left\{\delta_i{\eta_{il}^{(m)}}-\fracl{T_i}{\expon{\mu_{il}}}^{1/\sigma_l}\right\} + \lambda_2 \text{sign}(\beta_{lj}),\\
    \frac{\partial Q^{*\mathrm{W}}_l(\boldsymbol{\theta}|\boldsymbol\theta^{(m)})}{\partial \sigma_l} &= -\frac{1}{\sigma_l^2}\sum\limits_{i=1}^n \Bigg\{ \delta_i \eta_{il}^{(m)}\left(\mu_{il}-\sigma_l-\log T_i \right) + (\log T_i-\mu_{il})\fracl{T_i}{\expon{\mu_{il}}}^{1/\sigma_l} \Bigg\}.
\end{align*}
{In practice, $\alpha_l$ and ${\beta}_{lj}$ are updated via gradient-based optimisation, implemented using the BFGS method in the \texttt{optim()} function from the \textsf{R} \texttt{stats} package. The penalty on coefficients induces shrinkage, driving small coefficients ($|\beta_{lj}|<\lambda_2$) toward zero. The scale parameter $\sigma_l$ is updated using bounded optimisation via the L-BFGS-B method in the \texttt{optim()} function.}

\subsection{Theoretical Results}\label{sec: theo}
{We assume some regularity conditions (Appendix A.1) to establish the main theoretical results. In what follows, all limits are taken as the sample size $n$ goes to infinity, with $\stackrel{p}{\to}$ and $\stackrel{d}{\to}$ denoting convergence in probability and distribution, respectively. Proofs for the competing Weibull model are developed and deferred to Appendix A.3. Extensions to other parametric families follow by verifying analogous likelihood conditions; an example of the competing log-normal model is presented in Appendix A.4.}

\begin{theorem}\label{thm: 2.1} Denote $\boldsymbol\theta_0$ as the true parameter and $\widehat{\boldsymbol\theta}_n$ as the maximum likelihood estimation of $L(\boldsymbol{\theta})$ given in Eq.(3.1).
Under assumptions A1$\sim$A4, we have
    $$
    \widehat{\boldsymbol\theta}_n \topb \boldsymbol\theta_0.
    $$
Under additional assumptions A5$\sim$A6, we have 
    $$
\sqrt{n}(\widehat{\boldsymbol\theta}_n - \boldsymbol\theta_0) \todis \mathcal{N}(0,{\mathcal{I}}^{-1}(\boldsymbol\theta_0)),
$$
where $\mathcal{I}(\boldsymbol\theta_0)$ is the Fisher information matrix.
\end{theorem}

\begin{theorem}\label{thm: EM}
Define the observed-data likelihood
\begin{eqnarray*}
L_0(\boldsymbol\theta) = \sum_{l=1}^L h_l(t_i; \alpha_l+ \vk X_{il}^\top  \boldsymbol\beta_l, \sigma_l)^{\delta_i} \expon{ \sum_{k=1}^L\log S_k(t_i; \alpha_k+ \vk X_{ik}^\top  \boldsymbol\beta_k, \sigma_k)}.
\end{eqnarray*}
Under assumptions A1 and A7, all the limit points of $\{\boldsymbol\theta^{(m)}\}$ of an EM algorithm are stationary points, and $L_0(\boldsymbol\theta^{(m)})$ converges monotonically to $L_0^*=L_0(\boldsymbol\theta^*)$ for some stationary point $\boldsymbol\theta^*\in \Theta$.

With the additional assumption A8, $\{\boldsymbol\theta^{(m)}\}$ converges to some $\boldsymbol\theta^*$ in  $\{\boldsymbol\theta \in \Theta:L_0(\boldsymbol\theta)=L_0^*\}$.
\end{theorem}

Under assumptions \textit{A1} and \textit{A7}, the compactness regularity condition \citep[Assumption 6]{wu1983} is achieved and $L_0(\boldsymbol\theta^{(m)})$ is bounded above, then $\{\boldsymbol\theta^{(m)}\}$ converges to stationary points by \citet[Theorem 2]{wu1983}. Assumption \textit{A8} specifies the stop rule for the algorithm, satisfying $\norm{\boldsymbol\theta^{(m+1)}-\boldsymbol\theta^{(m)}} < \varepsilon$. We set $\varepsilon=10^{-6}$ for simulation studies, and $\varepsilon=10^{-3}$ for real applications.

For competing parametric families in which the M-step involves a non-concave group-wise objective function $Q_l(\boldsymbol{\theta}|\boldsymbol\theta^{(m)})$ (e.g., the competing Weibull model), we show the convergence of a sequence of local maximisers to the global maximiser in Theorem \ref{thm: local}.

\begin{theorem}\label{thm: local} Let ${\boldsymbol{\theta}}_0^{(m+1)}$ be the maximiser of the objective function $Q(\boldsymbol{\theta}|\boldsymbol{\theta}^{(m)})$ defined in Eq.(3.4), within the $(m+1)$-th maximisation step of the EM algorithm.
Under assumptions A1$\sim$A5 and A9, for a sequence $\Delta_n \to 0$, $n\Delta_n \to \infty$ as $n\to \infty$, there exists a corresponding sequence of local maximisers $\widehat{\boldsymbol{\theta}}_n^{(m+1)}$, such that $\norm{\widehat{\boldsymbol{\theta}}_n^{(m+1)}-{\boldsymbol{\theta}}_0^{(m+1)}} \le \Delta_n$, satisfying
\begin{eqnarray*}
\norm{\widehat{\boldsymbol{\theta}}_n^{(m+1)}-{\boldsymbol{\theta}}_0^{(m+1)}} \topb 0.
\end{eqnarray*}
\end{theorem}

\section{Simulation Studies}\label{sec: simu}
{We evaluate the estimation procedure in Section \ref{sec: EandP} using simulations from the competing Weibull model. Model performance is assessed using expected survival times estimated via importance sampling (Appendix A.2), with metrics of concordance index \citep[C-index;][]{uno2011c}, time-dependent ROC curve, and integrated AUC \citep[iAUC;][]{uno2007evaluating}. Comparisons are made using the Weibull model, the Cox PH model, and an adapted disease mixture model \citep{moreno2017survival} that treats one CF as the "disease of interest" while aggregating the remaining CFs (see Appendix C.1). Additional simulations under dependent errors for the competing log-normal model are provided in Appendix C.2.
}

Specifically, we consider three scenarios with varying sample sizes, censoring rates, CF structures, and parameter settings (Table \ref{tab: param}), covering non-overlapping variables (Example 1), partially overlapping variables (Example 2) and overlapping variables with coefficients set to zero (Example 3). For all examples, the covariate vector for each sample, $\vk{X}_i = (X_{i1}, \ldots, X_{i6})$, is independently drawn from a multivariate standard normal distribution $\mathcal{N}(0, \mathrm{I}_6)$. 

\begin{table}[H]
    \centering
    \caption{Settings of true parameters, sample sizes $n$, and censored rates of competing Weibull survival model. The notation "-" in $\beta_j$'s means exclusion from the model.
    }
\resizebox{0.9\textwidth}{!}{
    \begin{threeparttable}
    \begin{tabular}{lccrrrrrrrr}\toprule
        & Censored rate& &$\sigma$ & $\alpha$  & $\beta_1$ & $\beta_2$ & $\beta_3$ & $\beta_4$  & $\beta_5$  & $\beta_6$  \\
        \midrule
     \multirow{3}{*}{\begin{tabular}{c}
        Example 1 \\$(n=1000)$
    \end{tabular}}&  \multirow{3}{*}{0\%, 10\%}   & CF1 & 1.0 & 1.6 & 1.2&  -&  -& -&  -& -\\
   & & CF2 & 1.0 & 1.2& -&  2.0& - & -&  -& -\\
 & & CF3 & 1.1 & 2.1& -& - & 1.0 & -&  -& -\\
    \midrule
     \multirow{3}{*}{\begin{tabular}{c}
        Example 2 \\
        $(n = 1500)$
    \end{tabular}}&\multirow{3}{*}{\begin{tabular}{c}
       0\%, 10\% \\ 20\%, 30\%
    \end{tabular}} &CF1 & 1.0& 1.0& $-$3.0& 2.0 & - & 1.0&  -& -\\
    & &CF2 & 1.0 & 1.5& 2.0& 2.0 & - & -&  -& -\\
    & &CF3 &1.1 & 1.0  & $-$2.0& 3.0& 2.0 &- &  -& - \\
   \midrule
     \multirow{3}{*}{\begin{tabular}{c}
        Example 3 \\
        $(n = 1500)$
    \end{tabular}}&\multirow{3}{*}{10\%, 30\%} &CF1 & 1.0& 1.0& $-$3.0& 2.0 & - & -&-& -\\
    & &CF2 & 1.0 & 1.5&-& 0.0 & 2.0 & 2.0&  -& -\\
    & &CF3 &1.1 & 1.0  & -& -& - &0.0 &  $-$2.0& 3.0 \\
\bottomrule
    \end{tabular}
    \end{threeparttable}
    }
      \label{tab: param}
\end{table}

In the estimation, the tuning parameters were selected via a grid search with $\lambda_1 \in [0,2]$ and $\lambda_2 \in [0,1]$, with results of $\lambda_1 = 0.5$ and $\lambda_2 = 0.2$ for Example 1, and $\lambda_1 = 2$ and $\lambda_2 = 1$ for Examples 2 and 3. The results are summarised in Table \ref{tab: est}, which remains stable across alternative selections of tuning parameters within the ranges. Across all examples, the estimators are virtually unbiased, the standard errors (SE) are relatively small, and a slight reduction in accuracy is observed as censorship increases. In Example 3, the two zero coefficients in the second and third CFs are successfully identified under 10\% censoring, while their estimates remain very close to zero under 30\%. For model comparison (Table 6 and Figure 5), the competing Weibull model consistently outperforms across all metrics and with stable accuracy over time. {Diagnostic results based on Cox-Snell residuals show good alignment, supporting the adequacy of the competing Weibull model (Figure 5).}

\begin{table}[H]
    \centering
    \caption{Estimated parameters with {mean (SE)} of the competing Weibull survival model for three examples in simulations with varying levels of random censorship (0\%, 10\%, 20\%, and 30\%). {Here, $n$ represents the total sample size and $c$ represents number of censored data.}
    }
\resizebox{1\textwidth}{!}{
    \begin{threeparttable}
    \begin{tabular}{lccllllllll}\toprule
        & $c$ & & $\widehat{\sigma}$  & $\widehat{\alpha}$  & $\widehat{\beta}_1$  & $\widehat{\beta}_2$  & $\widehat{\beta}_3$  & $\widehat{\beta}_4$  & $\widehat{\beta}_5$   & ${\widehat{\beta}_6}$  \\
        \midrule
    \multirow{6}{*}{\begin{tabular}{c}
        Example 1 \\
        $(n = 1000)$
    \end{tabular}}&  \multirow{3}{*}{$c=0$} & CF1 & 0.918 (0.051) & 1.626 (0.092) & 1.170 (0.082)&  -&  -& -&  -& -\\
    && CF2 & 0.994 (0.044) & 1.219 (0.089)& -&  2.028 (0.077)& - & -&  -& -\\
   & & CF3 & 1.078 (0.068) & 2.110 (0.139)& -& - & 1.042 (0.116) & -&  -& -\\
    \cmidrule{2-11}
   &  \multirow{3}{*}{$c=100$} & CF1 & 0.922 (0.057)& 1.623 (0.100) & 1.077 (0.091)&  -&  -& -&  -& -\\
   && CF2 & 0.970 (0.044)& 1.153 (0.077)& -&  1.948 (0.077)& - & -&  -& -\\
   & & CF3 & 1.127 (0.075) & 2.157 (0.151)& -& - & 0.838 (0.137) & -&  -& -\\
    \midrule
   \multirow{12}{*}{\begin{tabular}{c}
        Example 2 \\
        $(n = 1500)$
    \end{tabular}} &  \multirow{3}{*}{$c=0$}& CF1& 0.911 (0.038) & 0.951 (0.085)  &  $-$2.906 (0.071)& 2.002 (0.053)& - & 0.933 (0.052)&  -& -\\
   & & CF2 & 1.072 (0.038) & 1.395 (0.101)& 2.011 (0.082)& 2.007 (0.053) & - & -&  -& -\\
   & & CF3 & 1.101 (0.048)& 1.043 (0.129) & $-$2.000 (0.101)& 3.098 (0.070)& 2.035 (0.088) &-  &  -& -\\
  \cmidrule{2-11}
  & \multirow{3}{*}{$c=161$}& CF1& 1.025 (0.041)& 1.010 (0.101)  &  $-$2.977 (0.086)& 1.968 (0.062)& - & 0.977 (0.061)&  -& -\\
   & & CF2 & 1.011 (0.038)& 1.601 (0.115)& 2.011 (0.089)& 2.038 (0.063) & - & -&  -& -\\
   & & CF3 & 1.006 (0.043)& 0.941 (0.117) & $-$1.978 (0.094)& 2.946 (0.064)& 1.973 (0.081) &- &  -& - \\
   \cmidrule{2-11}
   & \multirow{3}{*}{$c=289$}& CF1& 1.025 (0.041) & 0.959 (0.109)  &  $-$2.958 (0.090)& 1.947 (0.068)& - & 0.954 (0.064)&  -& -\\
   & & CF2 & 0.981 (0.039)& 1.493 (0.123)& 1.956 (0.092)& 1.978 (0.070) & - & -&  -& -\\
  & & CF3 & 0.993 (0.044) & 0.927 (0.125) & $-$1.961 (0.095)& 2.947 (0.068)& 1.964 (0.083) &- &  -& - \\
  \cmidrule{2-11}
  & \multirow{3}{*}{$c=443$}& CF1& 1.020 (0.044)& 0.924 (0.120)  &  $-$2.927 (0.096)& 1.942 (0.075)& - & 0.964 (0.069)&  -& -\\
  &  & CF2 & 0.989 (0.043) & 1.569 (0.146)& 1.977 (0.103)& 2.048 (0.080) & - & -&  -& -\\
  & & CF3 & 0.990 (0.044)& 0.875 (0.132) & $-$1.953 (0.096)& 2.916 (0.073)& 1.947 (0.086) &-&  -& -  \\
    \midrule
    \multirow{6}{*}{\begin{tabular}{c}
      Example 3 \\
      $(n=1500)$
    \end{tabular}} & \multirow{3}{*}{$c=150$}& CF1& 0.997 (0.037) & 1.050 (0.087)  &  $-$3.000 (0.065)& 2.113 (0.062)& - & -&  -& -\\
    & & CF2 & 1.002 (0.046) & 1.474 (0.116)&- & 0.000 (0.069)& 1.919 (0.076) & 2.023 (0.082) & -&  -\\
   & & CF3 & 1.131 (0.041)& 0.999 (0.099) &-  &  -& - & 0.000 (0.069)& $-$2.008 (0.067)& 2.965 (0.073) \\
     \cmidrule{2-11}
  & \multirow{3}{*}{$c=446$}& CF1& 0.960 (0.039) & 0.936 (0.108)  &  $-$2.911 (0.075)& 2.100 (0.067)& - & -&  -& -\\
  &  & CF2 & 1.032 (0.053) & 1.453 (0.174)&- & 0.011 (0.084)& 1.881 (0.104) & 2.022 (0.107) & -&  -\\
   & & CF3 & 1.144 (0.045)& 1.005 (0.124) &-  &  -& - & 0.017 (0.077)& $-$1.998 (0.077)& 2.964 (0.086) \\
\bottomrule
    \end{tabular}
    \end{threeparttable}
}    \label{tab: est}
\end{table}

\section{Real Data Applications}\label{sec: App}
Sepsis, cancer, and Alzheimer's disease pose persistent challenges in clinical management due to complex aetiology and the coexistence of multiple interacting conditions that shape disease prognosis. Sepsis patients often experience concurrent pathophysiological processes such as hyperinflammation, immune suppression, and coagulation dysfunction. Cancer progression involves the interplay of multiple genetic and molecular pathways, further compounded by various comorbidities. Alzheimer's disease similarly presents with heterogeneous neuropathological features alongside age-related comorbidities. These complexities motivate analytical approaches that can disentangle distinct sources of risk. In this section, we apply the competing Weibull model to: (1) examining gene effects on 28-day survival among sepsis patients; and (2) analysing differential gene expression in relation to overall survival in LUAD patients. {Applications to Alzheimer’s disease, HCC, and breast cancer, along with the disease mixture model adaption and diagnostic plots, are provided in Appendix B.} Results from the competing log-normal model are consistent and omitted for brevity. 

{In all applications, the number of CFs and feature allocation are determined using a hybrid strategy that combines domain knowledge with data-driven evaluation. 
Features showing strong associations in a preliminary AFT or Cox model are grouped into the first CF to capture the primary risk structure. Features with prior evidence of heterogeneity across subpopulations, or with weak associations in the preliminary model but established biological relevance, are then assigned to a second CF or duplicated across the first and second CFs to reflect subgroup-dependent or pathway-specific mechanisms. Additional competing effects relative to this second CF are organised into a third CF to capture further heterogeneity. The remaining features are allocated across existing CFs to balance model fit and parsimony. This strategy yields a parsimonious and interpretable structure with good empirical fit. Based on these CFs, we further quantify individualised, time-varying contributions of each cause and enable patient stratification via attributed survival times.}

\subsection{Sepsis patients with GSE65682 \label{sec: sep}}
Sepsis is a life-threatening condition caused by a dysregulated host response to infection, often leading to multiple organ dysfunction or failure and posing a major global health burden \citep{scicluna2015molecular}. We analysed transcriptomic data from the GSE65682 dataset, comprising 761 adult patients with severe pneumonia or sepsis and 41 healthy controls. Whole blood samples were collected in intensive care units, and RNA was extracted using the PaxGene Blood RNA kit (Qiagen, Netherlands) and profiled on the GPL13667 [HG-U219] Affymetrix Human Genome U219 Array platform. The competing Weibull survival model was applied to assess the overall survival of 479 patients with recorded times during the 28-day follow-up, of whom 114 experienced mortality. Patient age and gender were included as covariates, and the analysis focused on 13 genes previously reported to be associated with sepsis prognosis \citep{zhang2025ai,martinez2021distinguishing}: CKAP4, FCAR, RNF4, NONO, PLEKHO1, BMP6, RNASE2, IGHG1, IL1R2, LCN2, LTF, MMP8, and OLFM4.

\begin{figure}[H]
    \centering
    \subfloat[]{\includegraphics[width=0.31\linewidth]{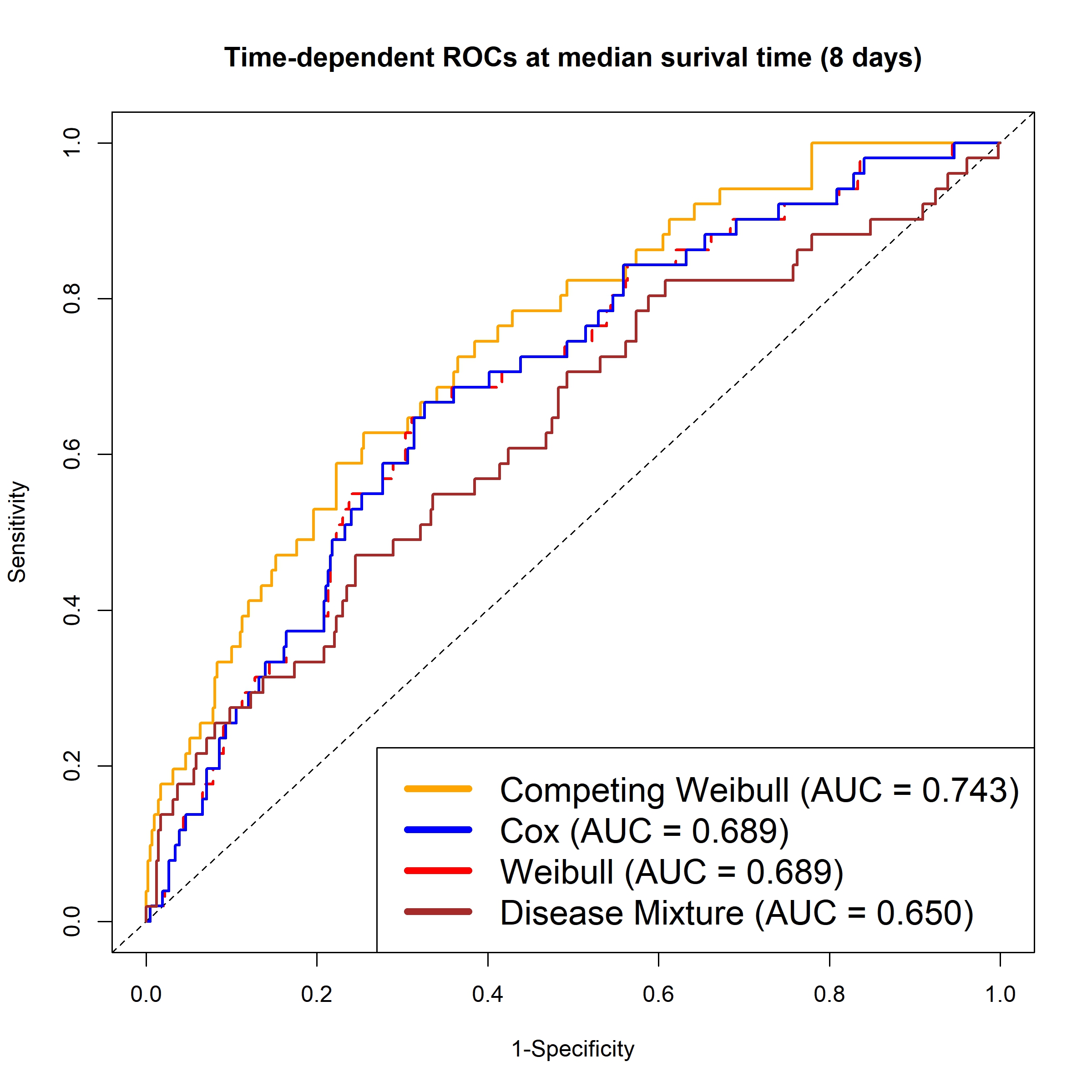}}
    \subfloat[]{\includegraphics[width=0.42\linewidth]{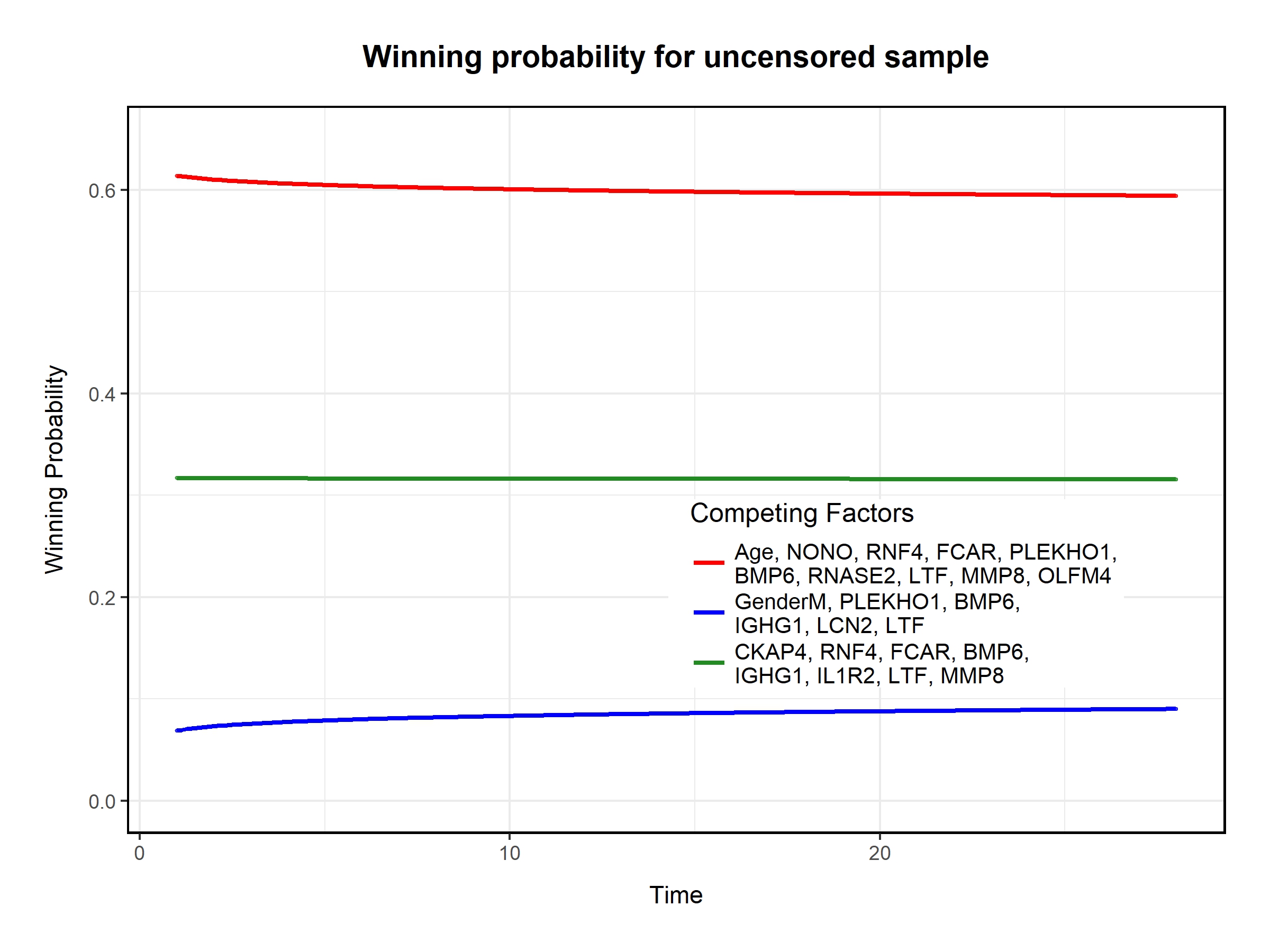}}
    \subfloat[]{\includegraphics[width=0.25\linewidth]{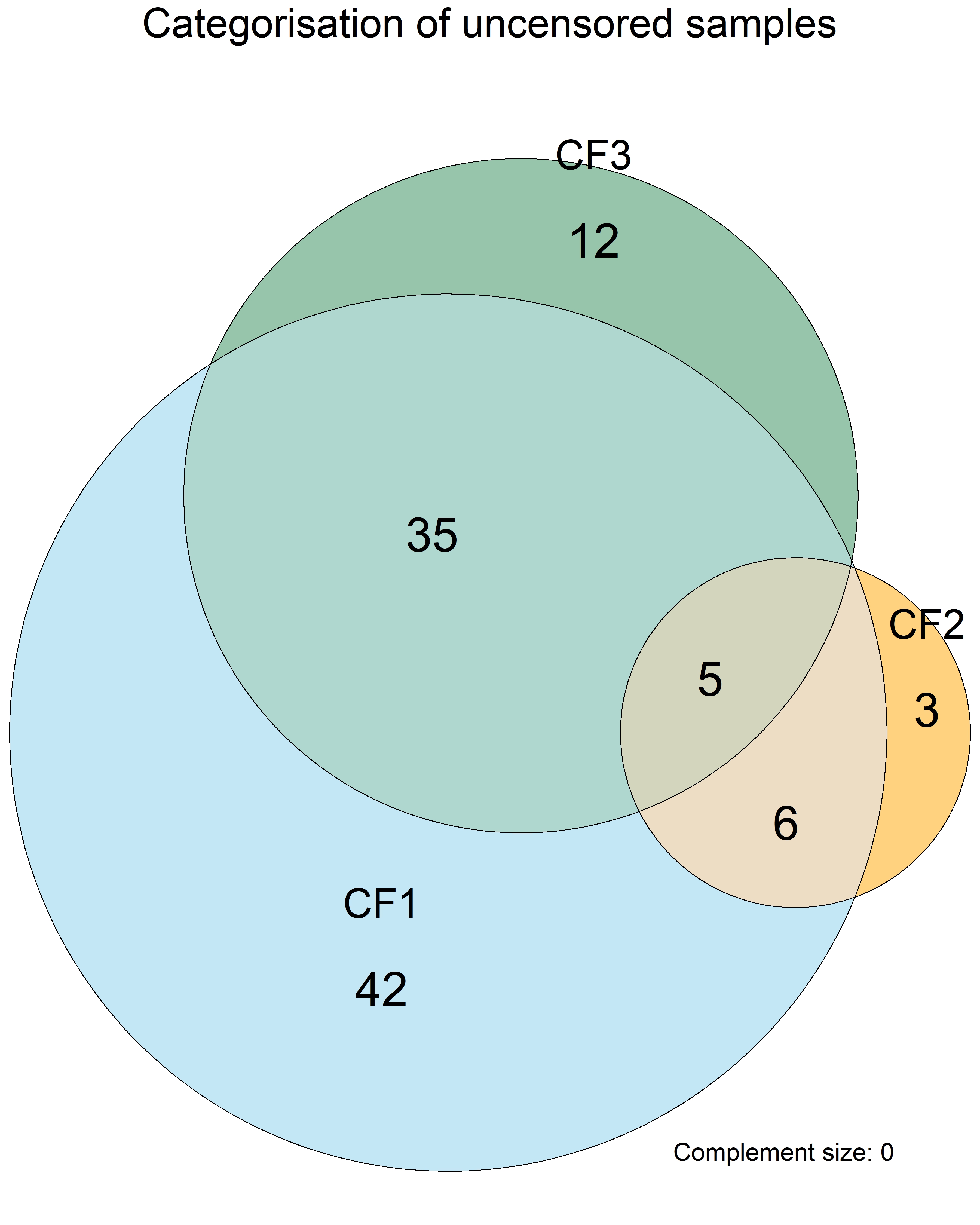}}\\
    \subfloat[]{\includegraphics[width=0.4\linewidth]{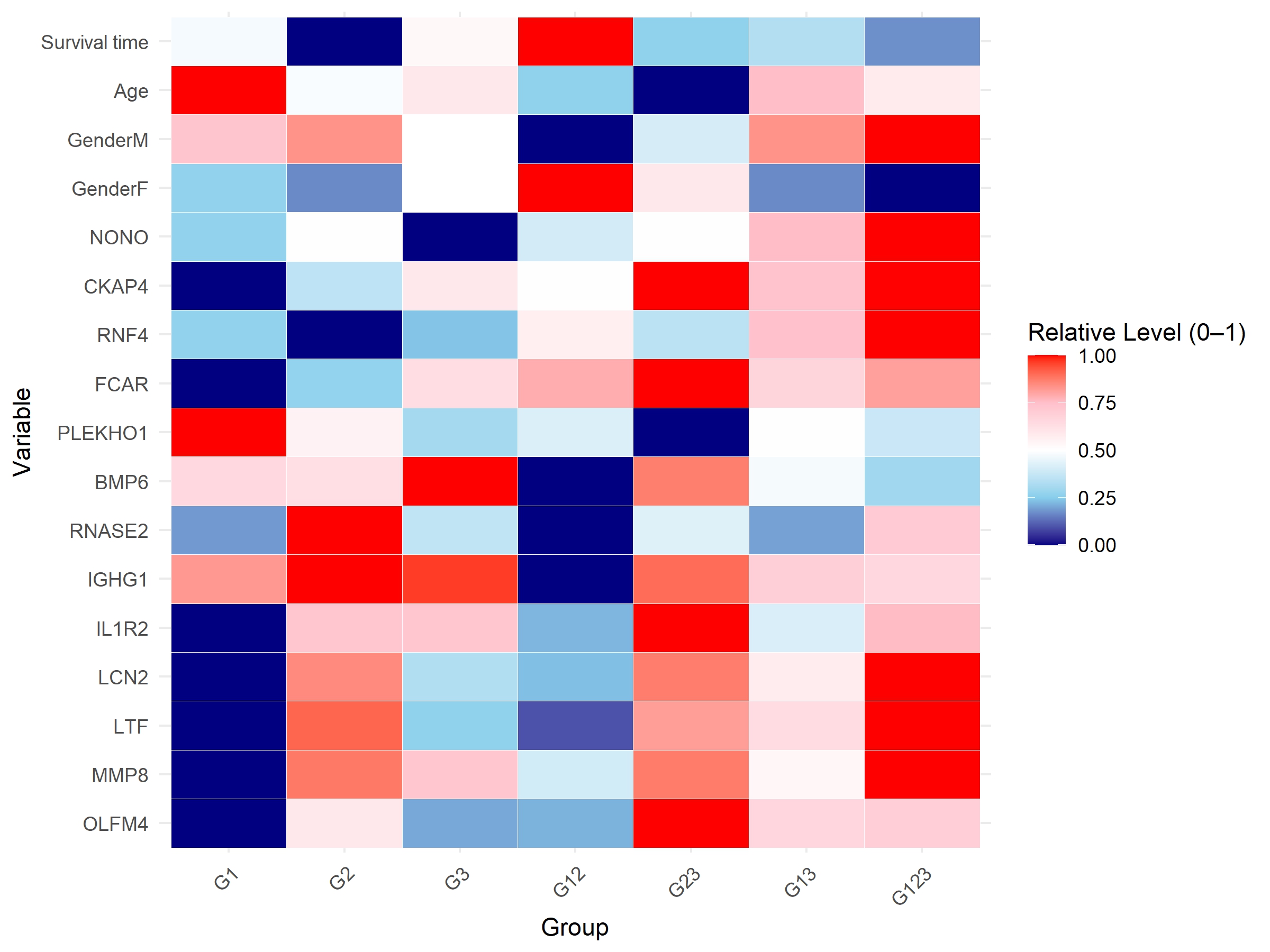}}
    \subfloat[]{\includegraphics[width=0.35\linewidth]{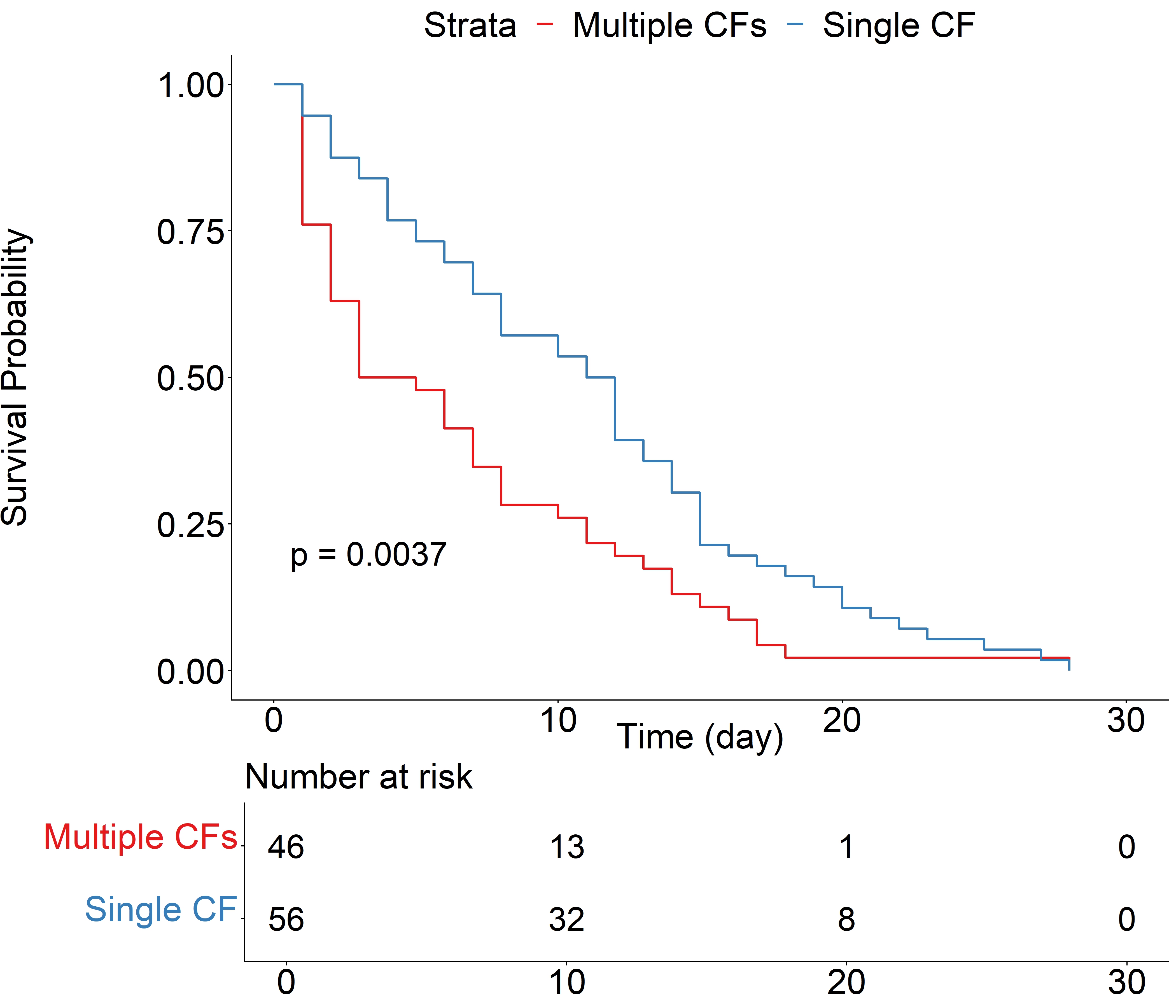}}
    \caption{Sepsis study results: (a) Time-dependent ROC at median survival time (8 days); (b) Time-varying winning probability for failed samples (averaged); (c) Categorisation of failed samples based on IPW attributed survival times shorter the median; (d) Heatmap of relative sample characteristics across categories, obtained by averaging within each category and scaling across all samples; (e) Kaplan-Meier survival curves stratified by single-CF group (G1, G2, G3) and multiple-CFs (G12, G23, G13, G123).}
    \label{fig: sepsis_ROC_WP}
\end{figure}

With tuning parameters $\lambda_1 = 1$ and $\lambda_2 = 0.2$ selected by grid search and specifying three CFs, the resulting competing Weibull model substantially outperformed the Weibull and Cox PH, and disease mixture models (Table \ref{tab: sepsis_est} and Figure \ref{fig: sepsis_ROC_WP}(a)), achieving a C-index of 0.722, an iAUC of 0.737, and an AUC of 0.743 at the median survival time of 8 days, corresponding to around 7\% improvements.

The relative importance of the three CFs for failed samples, measured by the average individualised winning probabilities (Eq.~\eqref{eq: winning probability eta}), is shown in Figure \ref{fig: sepsis_ROC_WP}(b). All the contributions remain stable over time, with CF1 dominating at approximately 60\%, followed by CF3 (30\%) and CF2 (10\%).
For each subject, the attributed survival times for the CFs are computed using the IPW-observed survival time. 
Subjects are assigned to one of the seven subtypes (G1, G2, G3, G12, $\ldots$, G123) if the attributed time for a CF falls below the overall median across all factors and samples; otherwise, they are assigned to a complementary subtype.
Consistent with the dominance of CF1, most patients ($n=42$) belong to subtype G1 (Figure \ref{fig: sepsis_ROC_WP}(c)). Figure \ref{fig: sepsis_ROC_WP}(d) further shows that subtype G2 tends to exhibit shorter survival among single-CF groups, suggesting differential impacts of the CFs on patient outcomes. Patients characterised by multiple-CFs subtypes (G12, G23, G13, and G123) are also associated with significantly shorter survival times ($p=0.0037$; Figure \ref{fig: sepsis_ROC_WP}(e)).

Table \ref{tab: sepsis_est} summarises gene effects on 28-day survival in sepsis patients. Across the four models, most gene effects were consistent with previous studies. Notably, the competing Weibull model reveals heterogeneous effects that conventional approaches fail to capture. In particular, IGHG1 and MMP8 exhibited significant, opposite-direction effects depending on the failure type, whereas the standard models and {the adapted disease mixture model (with merged CF2 \& CF3)} indicated uniform risk (IGHG1) or no association (MMP8).  These divergent effects align with existing literature. IGHG1, a key subclass of immunoglobulin G (IgG), exhibited both positive and negative associations with survival outcomes: upregulation of IgG levels were linked to sepsis patients compared to controls \citep{martinez2021distinguishing} and patients with worse survival outcomes \citep{alagna2021higher}, while low IgG levels were associated with higher sepsis mortality \citep{akatsuka2021low}.  Similarly, elevated MMP8 is associated with poor survival in pediatric sepsis \citep{solan2012novel}, yet its inhibition may be detrimental in juvenile hosts \citep{atkinson2016matrix}. {
This revealed heterogeneous gene effects, highlighting both the limitations of linear additive models in capturing subgroup-specific effects and the inadequacy of a two-group structure (as in the disease-mixture model) for modelling cause competition. Our competing survival model thus provides a more nuanced framework for uncovering individual-level heterogeneity. Moreover, integrating subtype classification with feature heatmaps may further support patient stratification and inform personalised therapeutic strategies.}

\begin{table}[htbp]
    \centering
    \caption{Estimated coefficients with mean (SE), along with C-index and iAUC in the sepsis study, from the competing Weibull model (survival time), Weibull model (survival time), and Cox PH model (hazard). The notation "-" means exclusion from the model.}
    \label{tab: sepsis_est}
    \resizebox{0.95\textwidth}{!}{
    \begin{threeparttable}
    \begin{tabular}{l *{3}{r} r r rr}
    \toprule
\multirow{2}*{Features} &  \multicolumn{3}{c}{Competing Weibull} & \multicolumn{1}{c}{\multirow{2}*{
Weibull}} & \multicolumn{1}{c}{\multirow{2}*{Cox PH}} & \multicolumn{2}{c}{Disease Mixture}\\
\cmidrule{2-4} \cmidrule{7-8}
 &  \multicolumn{1}{c}{CF1} &  \multicolumn{1}{c}{CF2} &  \multicolumn{1}{c}{CF3} &  & &  \multicolumn{1}{c}{State 1 (CF1)} &  \multicolumn{1}{c}{State 0 (CF2 \& CF3)}\\
    \midrule
    Intercept & 6.103 (0.253)   & 10.636 (2.429)   & 9.419 (1.040)  & 5.404 (0.301)   & -  & -& -\\
    \rowcolor[gray]{0.85} Age  & $-$0.698  (0.238)  & -   & -  & $-$0.381 (0.163)   & 0.268 (0.115) & 0.507 (0.186)& -\\
    GenderM   & -    & 2.442 (2.272)   & - & 0.028 (0.287) & $-$0.021 (0.206) & -& $-$0.473 (0.370)\\
    \rowcolor[gray]{0.85} NONO  & 0.179 (0.328) &-   & - &$-$0.095 (0.261)   & 0.076 (0.187) & 0.006 (0.269)& -\\
    CKAP4    & -  &- & 1.558 (0.589) & 0.233 (0.270) & $-$0.167 (0.193) & -& $-$0.528 (0.357)\\
    \rowcolor[gray]{0.85} RNF4 & $-$0.023 (0.306) & - & 0.340  (0.293)  & 0.314 (0.241) & $-$0.226 (0.172) & $-$0.080 (0.253)& $-$0.335 (0.195)\\
    FCAR        & $-$0.305 (0.223) & - & 0.476 (0.438)  & 0.025 (0.239) & $-$0.018 (0.171) & 0.164 (0.192)& $-$0.444 (0.284)\\
    \rowcolor[gray]{0.85} PLEKHO1  & 0.683 (0.241)& $-$1.266 (0.740)  & -    & 0.335 (0.186) & $-$0.234 (0.132) & $-$0.143 (0.203) & $-$0.273 (0.223)\\
    BMP6        & 0.261 (0.219)   & 1.034 (0.911)     & $-$0.539  (0.226)   & $-$0.004 (0.153)  & 0.001 (0.110) & $-$0.320 (0.203)& 0.320 (0.183)\\
    \rowcolor[gray]{0.85} RNASE2  & $-$0.335 (0.210)  & -    & - & $-$0.137 (0.158) & 0.092 (0.113) & 0.076 (0.171)& -\\
    IGHG1        & -  & 1.854 (0.808)      & $-$1.350 (0.357)     & $-$0.296 (0.149)  & 0.207 (0.105) & -& 0.654 (0.194)\\
    \rowcolor[gray]{0.85} IL1R2  & -       & -    & $-$3.653 (0.999) & $-$0.586 (0.254) & 0.417 (0.179) & -& 2.128 (0.623)\\
    LCN2        & -   & 1.142  (1.343)    & -     & $-$0.156 (0.461)  & 0.108 (0.333) & -& 0.071 (0.487)\\
    \rowcolor[gray]{0.85} LTF  & $-$0.878 (0.330) & $-$5.149 (2.720)   & 0.717 (0.478) & $-$0.210 (0.372) & 0.150 (0.268) & 0.710 (0.296)& $-$0.248 (0.408)\\
    MMP8        & 1.487 (0.439)   & -      & $-$2.190 (0.768)     & 0.480 (0.377)  & $-$0.333 (0.269) & $-$0.761 (0.376)& 0.362 (0.462)\\
    \rowcolor[gray]{0.85} OLFM4  & 0.224 (0.301)       & -    & - & 0.033 (0.246) & $-$0.022 (0.177) & $-$0.229 (0.243)& -\\
    Scale ($\widehat{\sigma}$) &1.262 (0.087)   & 0.953 (0.208)   & 1.245 (0.146)   & 1.392 (0.094)   & - & -& -\\ \hline
    C-index        & \multicolumn{3}{c}{0.722}     & 0.654  & 0.653 & \multicolumn{2}{c}{0.662}  \\
    iAUC        &  \multicolumn{3}{c}{0.737}       & 0.658  & 0.658 & \multicolumn{2}{c}{0.677}\\
    \bottomrule
    \end{tabular}
    \end{threeparttable}
    }
\end{table}

\subsection{LUAD patients with GSE72094}\label{sec: luad}
LUAD, the most common histological subtype of lung cancer, remains a leading cause of cancer-related mortality worldwide and is characterised by poor five-year overall survival outcomes \citep{schabath2016differential}. Differential gene expression between tumour and non-tumour tissue may provide important information for diagnostic classification and prognostic stratification across disease stages. In this study, we apply our competing Weibull survival model to investigate the effects of gene expression on overall survival and to compare its predictive performance with conventional survival models. Gene expression and clinical data were obtained from GSE72094 \citep{schabath2016differential}, which profiled tumour samples using the Rosetta/Merck Human RSTA Custom Affymetrix 2.0 microarray (platform GPL15048). Our analysis cohort comprises 398 LUAD patients with complete survival information, and 113 (28.4\%) experienced death. Covariates include patient age and sex, together with expression levels of genes previously implicated in the literature \citep{song2021identification, jiawei2020identification, Zhang2022genomic}: CCNA2, AURKA, AURKB, FEN1, CCND3, NCALD, MACF1, LRC4, NLRC4, PLEKHN1, RASIP1, SPP1, GPT2, SGPL1, and PCOLCE2.

\begin{figure}[H]
    \centering
    \subfloat[]{\includegraphics[width=0.31\linewidth]{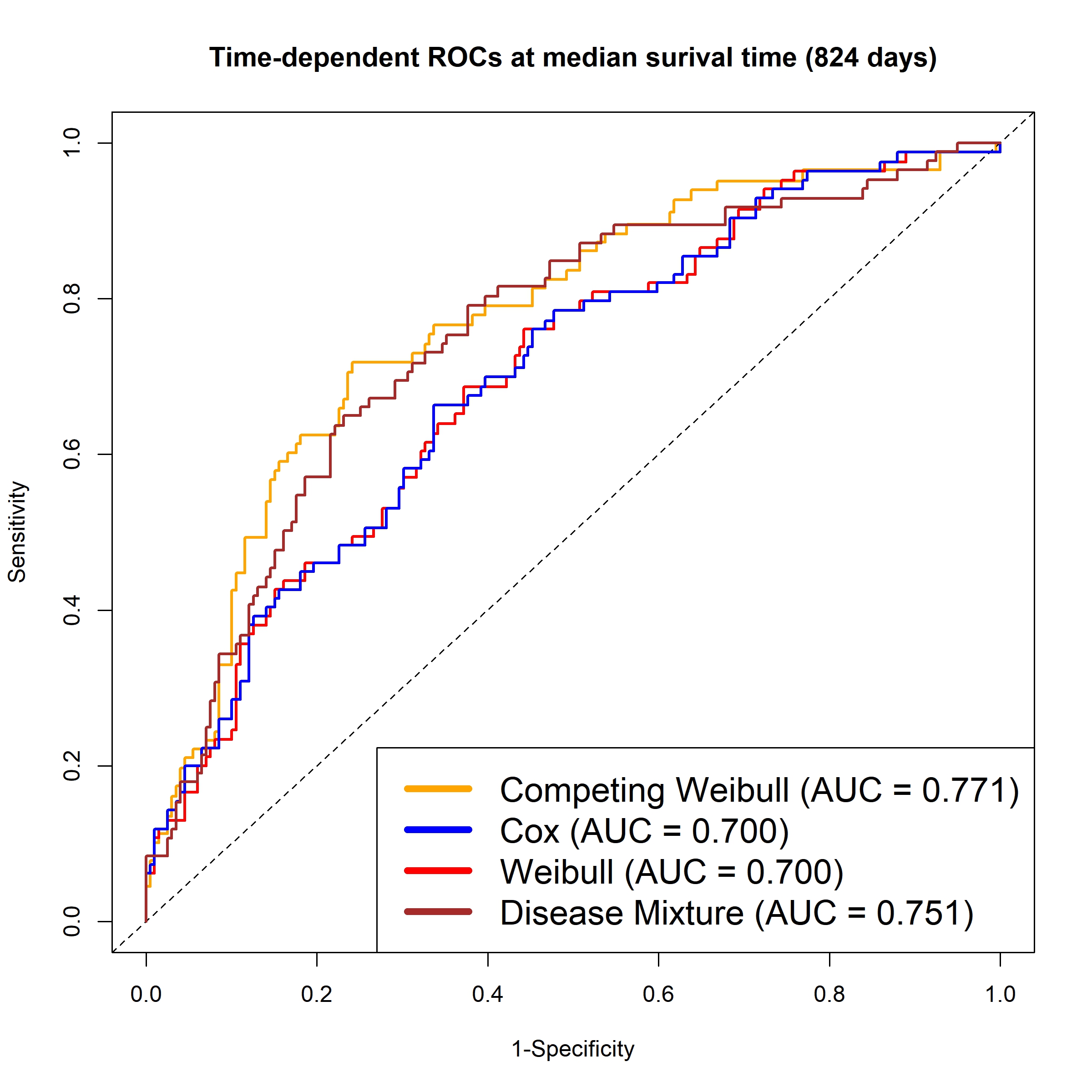}}
    \subfloat[]{\includegraphics[width=0.42\linewidth]{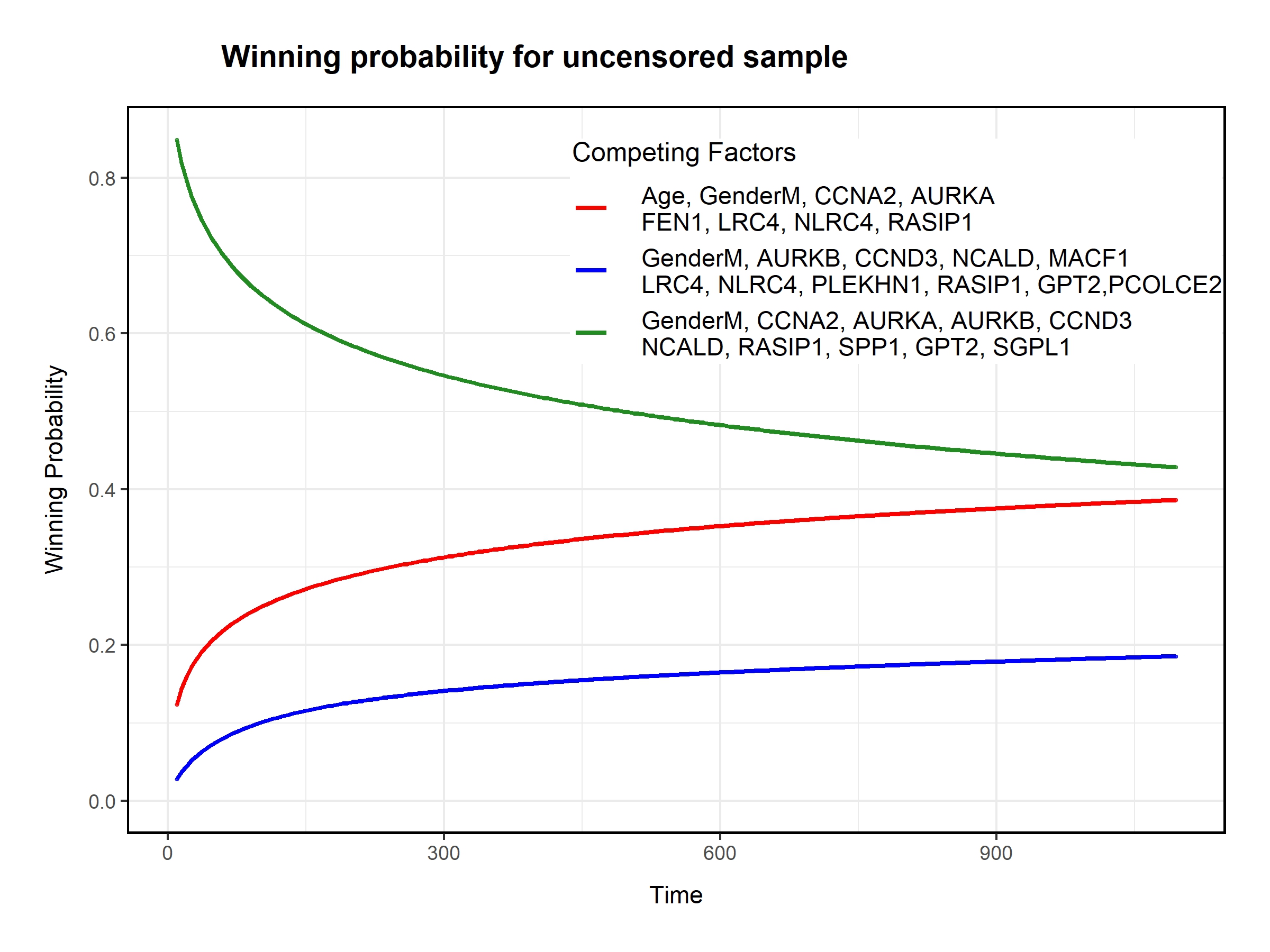}}
    \subfloat[]{\includegraphics[width=0.25\linewidth]{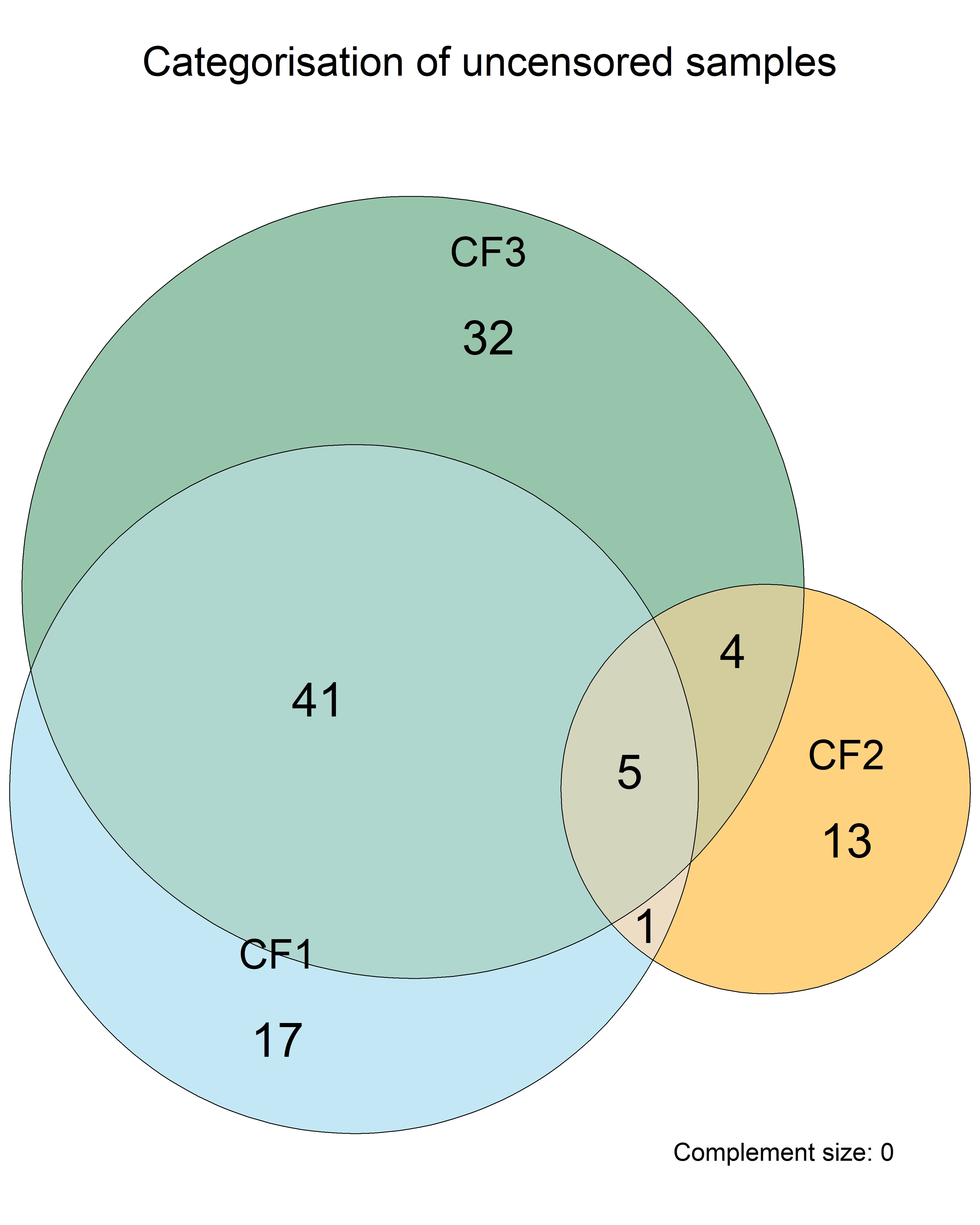}}\\
    \subfloat[]{\includegraphics[width=0.34\linewidth]{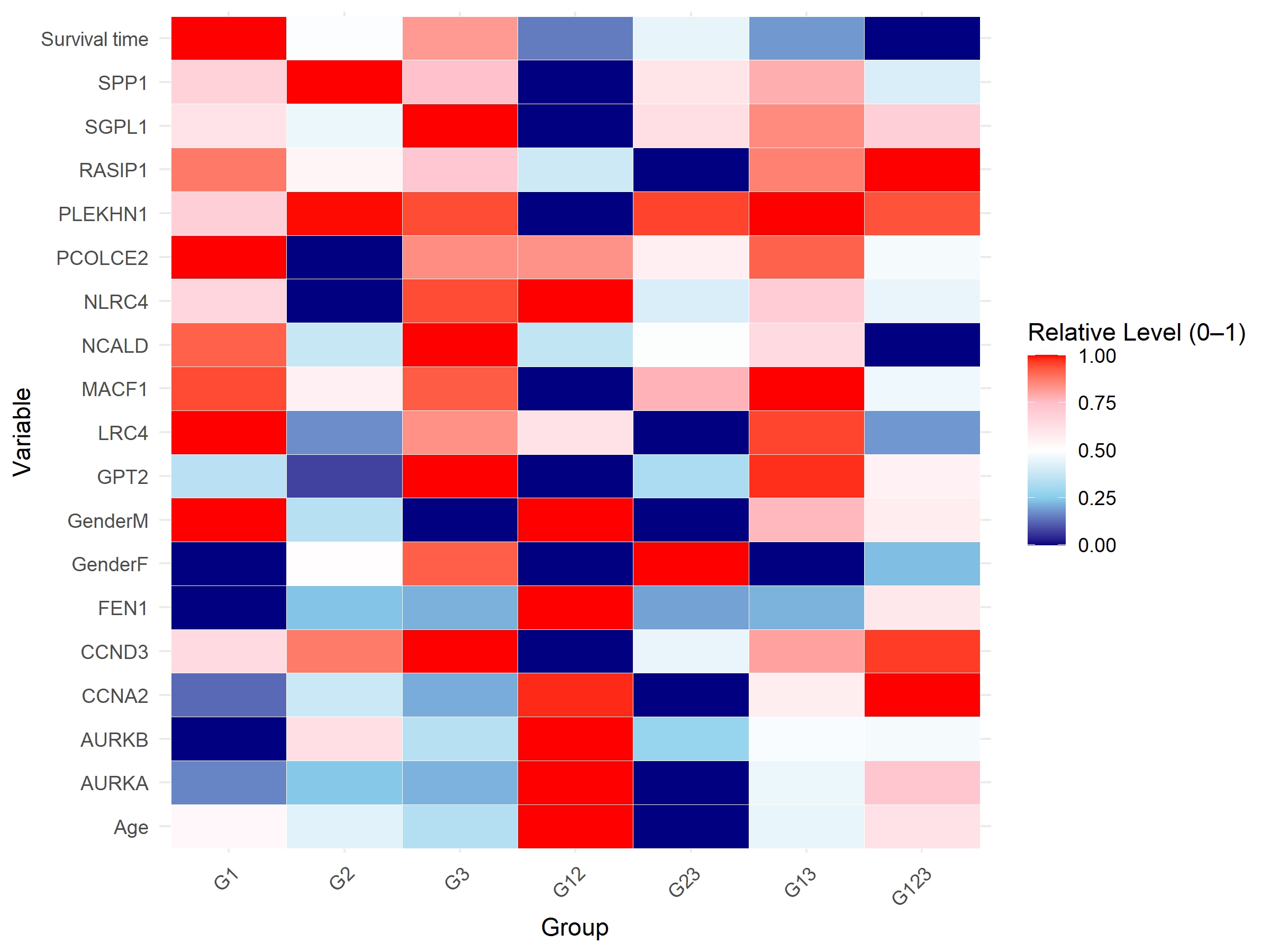}}
    \subfloat[]{\includegraphics[width=0.32\linewidth]{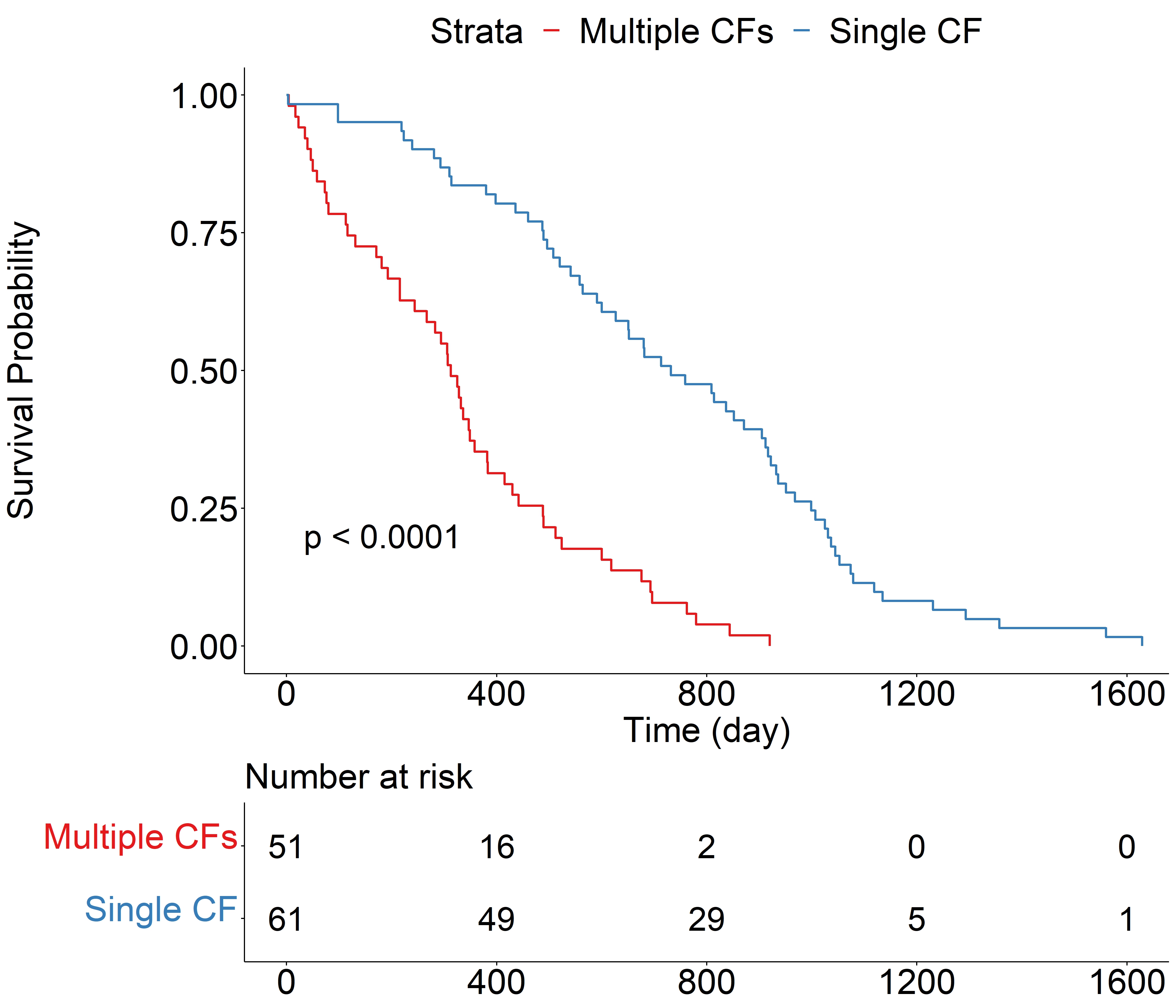}}
    \subfloat[]{\includegraphics[width=0.32\linewidth]{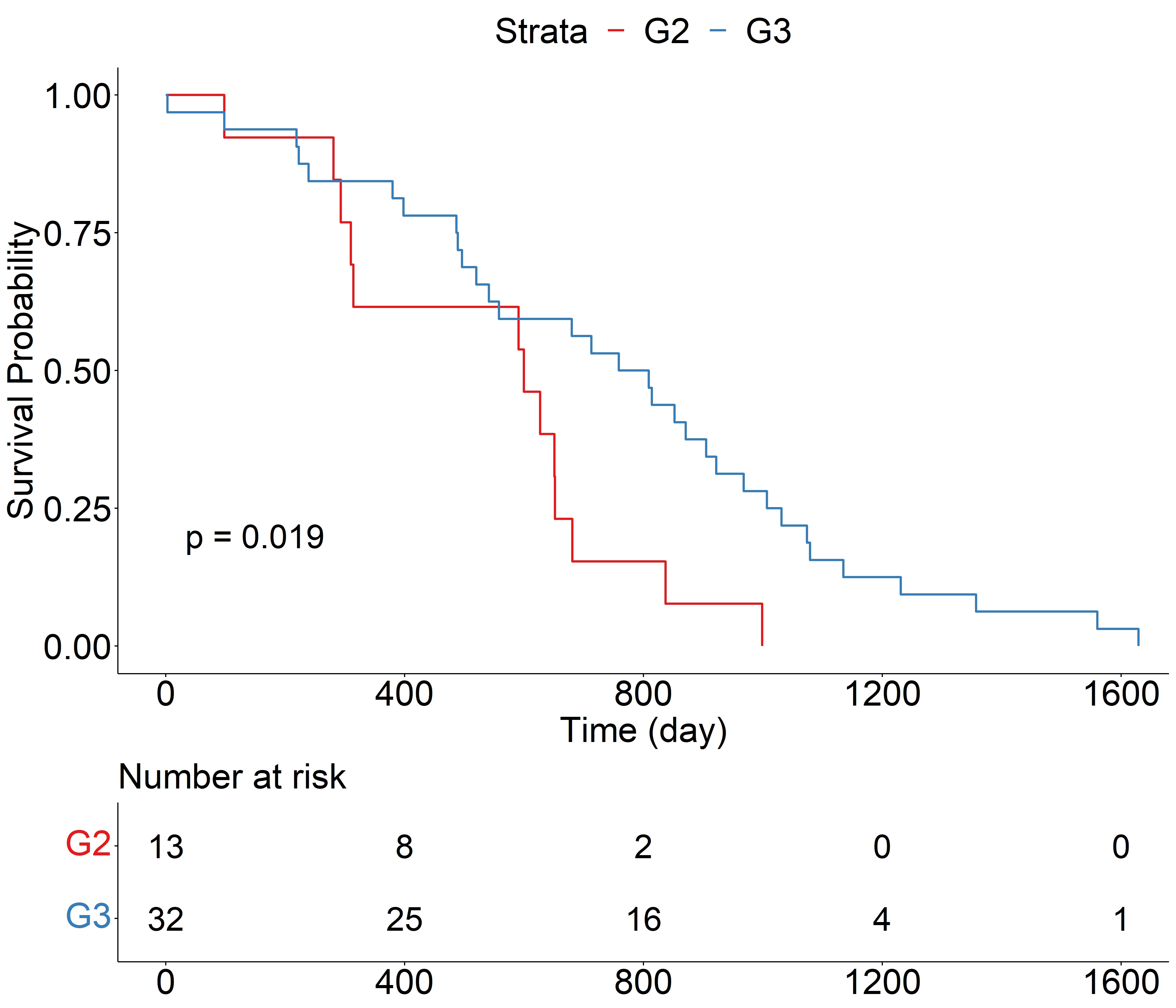}}
    \caption{LUAD study results: (a) Time-dependent ROC at median survival time; (b) Time-varying winning probability for failed samples (averaged); (c) Categorisation of failed samples based on IPW attributed survival times below the median; (d) Heatmap of relative sample characteristics across categories, obtained by averaging within each category and scaling across all categories; (e) Kaplan-Meier survival curves stratified by single-CF group (G1, G2, G3) and multiple-CFs group (G12, G23, G13, G123); (f) Kaplan-Meier survival curves in G2 and G3.}
    \label{fig: LUAD_ROC_WP}
\end{figure}

Using a grid search with selected $\lambda_1 = 1,\ \lambda_2 = 0.1$ and three specified CFs, the competing Weibull model yields a marked improvement in C-index, iAUC, and AUC at the median survival time (824 days), 7\% compared to the Weibull and Cox PH models and 2.5\% to disease mixture model (Table \ref{tab: LUAD_est} and Figure \ref{fig: LUAD_ROC_WP}(a)). The contributions of CFs exhibit a dynamic pattern: The predominant CF3 declines from over 80\% in early follow‑up to 40\% at later times, whereas the contributions of CF1 and CF2 increase modestly over time (Figure \ref{fig: LUAD_ROC_WP}(b)).

The estimated effects are largely consistent across the four models, and the competing specification offers a key advantage by uncovering heterogeneous, context-specific effects (Table \ref{tab: LUAD_est}). It reveals protective effects for NCALD and MACF1, corroborating previous biological evidence \citep{chen2019clinicopathological,lin2025multifaceted}. {Notably, GPT2 exhibits opposing roles depending on the competing failure type, and is supported by the disease mixture model.} It was a risk factor in CF3 but protective in CF2, extending previous studies that reported only its risk association  \citep{WANG2023107415,Zhang2022genomic}. Subtype analysis shows that patients of subtype G2, whose risk was dominated by CF2 and with GPT2 behave protectively, have significantly worse overall survival than those in G3 (log-rank $p=0.019$, Figure \ref{fig: LUAD_ROC_WP}(f)). These findings suggest that GPT2 may function through distinct biological pathways in sepsis, and the therapeutic strategies targeting GPT2 should account for patient-specific contexts.
Moreover, patients with multiple competing failure types experienced poorer outcomes than those with a single dominant type (log-rank $p < 0.0001$, Figure \ref{fig: LUAD_ROC_WP}(e)).

\begin{table}[H]
\centering
\caption{Estimated coefficients with mean (SE), along with C-index and iAUC in the LUAD study, from the competing Weibull model (survival time), Weibull model (survival time), Cox PH model (hazard) and disease mixture model (hazard). The notation "-" means exclusion from the model.}
\resizebox{0.95\textwidth}{!}{
\begin{threeparttable}
\begin{tabular}{lrrrrrrr}
\toprule
\multirow{2}*{Features} &  \multicolumn{3}{c}{Competing Weibull} & \multicolumn{1}{c}{\multirow{2}*{
Weibull}} & \multicolumn{1}{c}{\multirow{2}*{Cox PH}} & \multicolumn{2}{c}{Disease Mixture}\\
\cmidrule{2-4} \cmidrule{7-8}
 &  \multicolumn{1}{c}{CF1} &  \multicolumn{1}{c}{CF2} &  \multicolumn{1}{c}{CF3} &  & &  \multicolumn{1}{c}{State 1 (CF3)} &  \multicolumn{1}{c}{State 0 (CF1 \& CF2)}\\
\midrule
Intercept   & 9.401 (0.674)  & 10.141 (1.154)  & 8.899 (0.240)  & 8.033 (0.159)  & - & -  & -\\
\rowcolor[gray]{0.85} Age   & $-$0.184 (0.152) & -    & -  & $-$0.064 (0.087) &  0.078 (0.101) & -  & 0.344 (0.268) \\
GenderM    & $-$1.500 (0.658) & 0.115 (0.133)   & 0.233 (0.366)   & $-$0.335 (0.171) & 0.392 (0.197) & $-$0.054 (0.240) & 2.617 (1.395) \\
\rowcolor[gray]{0.85} CCNA2  & $-$0.345 (0.195)   & - & 0.183 (0.334) & $-$0.047 (0.170) &  0.039 (0.197) & $-$0.058 (0.234) & 0.912 (0.570) \\
AURKA  & $-$0.121 (0.171)  & -    & $-$0.188 (0.245)  & $-$0.024 (0.123) &  0.025 (0.144) & 0.152 (0.170) & $-$0.629 (0.353)\\
\rowcolor[gray]{0.85} AURKB   & - & 0.239 (0.244)   & $-$0.300 (0.275)  &$-$0.180 (0.170) & 0.231 (0.197)& 0.249 (0.202) & $-$0.666 (0.661)\\
FEN1  & 0.262 (0.205) & - & -   & 0.147 (0.136) & $-$0.182 (0.159) & - &$-$0.033 (0.337)\\
\rowcolor[gray]{0.85} CCND3   & - & 0.687 (0.360)  &  $-$0.276 (0.155)     & $-$0.065 (0.087)  & 0.071 (0.100)  & 0.226 (0.116) & $-$0.467 (0.301)\\
NCALD  & -    & 0.887 (0.227) & $-$0.223 (0.200)  & 0.062 (0.097)  & $-$0.069 (0.112) & 0.069 (0.143) & $-$0.333 (0.200)\\
\rowcolor[gray]{0.85} MACF1    & -  & 1.040 (0.394) & - & 0.184 (0.114)  &  $-$0.223 (0.132) & - & $-$1.412 (0.352)\\
LRC4        & $-$0.216 (0.173)  & $-$0.907 (0.426)  & -    & 0.065 (0.094)  & $-$0.067 (0.108) & - & $-$0.572 (0.290)\\
\rowcolor[gray]{0.85} NLRC4      & 0.317 (0.173)    & $-$0.259 (0.156) & - & 0.113 (0.110) &  $-$0.128 (0.127)& - & $-$0.397 (0.426)\\
PLEKHN1   &  - & $-$0.243  (0.185)  & -  & $-$0.047 (0.085)  &  0.073 (0.099)& - & $-$0.203 (0.252)\\
\rowcolor[gray]{0.85} RASIP1   & $-$0.174 (0.145)  & 1.046 (0.520) & 0.328 (0.194)  & 0.028 (0.098)  &  $-$0.034 (0.114)& $-$0.103 (0.129) & $-$0.093 (0.241)\\
SPP1   & -  & -    & 0.246 (0.173)  & 0.087 (0.090)  &  $-$0.094 (0.103) & $-$0.148 (0.122) & - \\
\rowcolor[gray]{0.85} GPT2   &-  & 1.046 (0.520)  & $-$0.690 (0.153)  & $-$0.180 (0.095)  &  0.206 (0.108)& 0.547 (0.123) & $-$1.007 (0.266)\\
SGPL1   &-  & -     & $-$0.427 (0.158)  & $-$0.142 (0.084)  &  0.166 (0.096)& 0.371 (0.117) & -\\
\rowcolor[gray]{0.85} PCOLCE2   & -  & 1.427  (0.594)  & -  & 0.115 0.096)  &  $-$0.142 (0.112)& - & $-$0.338 (0.297)\\
 Scale ($\widehat{\sigma}$) & 0.638 (0.076)  & 0.421 (0.087)  & 1.028 (0.078)  & 0.866 (0.084)  & - & -& -\\
 \hline
C-index   & \multicolumn{3}{c}{0.741}  &  0.672     & 0.670 & \multicolumn{2}{c}{0.714}\\
iAUC   & \multicolumn{3}{c}{0.767}  &  0.699     & 0.698 & \multicolumn{2}{c}{0.744}\\
\bottomrule
\end{tabular}
\end{threeparttable}
}\label{tab: LUAD_est}
\end{table}

\section{Concluding Remarks \label{sec: Dis}}
We developed a cAFT model for settings where multiple latent processes (comorbidities, chronic diseases) jointly determine the time to an observed event. It provides a principled framework for disentangling heterogeneous sources of risk and quantifying individual-level dominance over time. The model introduces a data-driven, patient-specific, and time-varying "winning probability" to reflect the relative contributions of the latent processes, and is particularly relevant in modern clinical contexts. Patients often present with multiple coexisting conditions that collectively influence prognosis, yet a single event time is observed. From this perspective, survival analysis extends beyond predicting time-to-event to characterising the dynamic interplay among multiple latent processes, thereby offering deeper insights into outcome heterogeneity and supporting personalised therapeutic strategies.

The case studies highlight three key advantages of the proposed model. First, the cAFT model provides a natural framework for biomarker discovery by capturing heterogeneous gene and pathway effects and their interactions. Such heterogeneity across patients is increasingly recognised as central in disease progression \citep{Johansson2023precision}, yet is often attenuated or overlooked in conventional linear additive survival models \citep{zheng2020comparison}. For example, the cAFT model identifies context-dependent effects of IGHG1 and MMP8 in sepsis and GPT2 in LUAD, whereas conventional models typically capture only a single direction or yield non-significant results, potentially due to offsetting influences across the subgroups. Second, the model improves predictive accuracy while remaining interpretable, bridging the gap between black-box machine learning methods and classical survival analysis.
Third, the winning probability provides practical utility for comparing biomarker panels in terms of prognostic performance and for clustering patients into clinically meaningful subtypes, thereby informing biomarker prioritisation, cost-effectiveness considerations, and individualised treatment strategies \citep{Jorgensen2020age, You2018subtype}. Collectively, these features suggest that the cAFT model provides a transparent and flexible framework for studying disease mechanisms and validating biomarkers, enhancing both scientific understanding and clinical decision-making in complex diseases.

Beyond these contributions, the proposed framework offers theoretical, methodological, and applied extensions. On the theoretical side, dependence among the causes could be modelled using copula-based approaches \citep{lo2010copula}, and extensions to longitudinal biomarker data or joint modelling frameworks are also natural directions for future research \citep{han2020statistical}. Semiparametric adaptations using nonparametric MLEs \citep{mao2017efficient} could provide additional flexibility in modelling hazard shapes. On the methodological and applied sides, the quantification of the relative importance of multiple causes can be incorporated within a posterior framework with priors of population-level weights \citep{You2018subtype}. 
{Feature selection into CFs is currently guided by a hybrid strategy that balances fitness and interpretability. Developing fully automated approaches for feature selection, dimensionality reduction, and allocation remains an important direction. As practical guidance, one may consider stepwise selection based on an information criterion that penalises both the number of features and CFs \citep{leblanc2006extreme}, alongside data-driven criteria informed by prior studies \citep{liu2024towards} or incorporate domain knowledge for calibration.}
Broader applications beyond biomedicine are also promising, such as enterprise bankruptcy prediction and financial distress analysis \citep{GEPP2015396}.

\bibliographystyle{apalike}
\bibliography{reference} 
 
\appendix
\renewcommand{\thetable}{S.\arabic{table}}
\setcounter{table}{0}
\renewcommand{\thefigure}{S.\arabic{figure}}
\setcounter{table}{0}
\renewcommand{\thesubsection}{\Alph{subsection}}
\setcounter{subsection}{0}
\renewcommand{\thelemma}{\arabic{lemma}}
\renewcommand{\theproposition}{\arabic{proposition}}
\renewcommand{\theequation}{\arabic{equation}}

\section*{Appendix}\label{sec: Appendix}

In this supplementary material, Appendices A, B and C provide supporting theoretical, application, and simulation results, respectively. Appendix A presents regularity conditions, the unbiased estimator of survival time with numerical approximations, and technical proofs for the competing accelerated failure time (cAFT) models. Appendix B contains additional real-data applications, including Alzheimer’s disease, hepatocellular carcinoma, and breast cancer, along with details on the adapted disease mixture model and diagnostic plots. Appendix C reports further simulation studies, including model comparisons, diagnostic assessments, and scenarios with dependent errors under a competing log-normal model.

\section*{Appendix A}
In this section, we provide the regularity conditions to establish the theoretical results in A.1. The unbiased estimator of survival time, along with the numerical approximation methods, is given in A.2. Technical proofs for the competing Weibull model and a characterisation of key conditions are shown in A.3 and A.4, respectively.

\subsection*{A.1 Regularity Conditions}
We present the following assumptions required for consistency and asymptotic normality.
\begin{enumerate}
    \item[A1.] The parameter space $\Theta$ is compact.
    \item[A2.] The distribution that generates $\vk X_i$ is light-tailed in terms that $\E{\expon{\vk X_i}}<\infty$. No isolated and extreme value of $\vk X_i$ is generated.
    \item[A3.] The proportion of uncensored samples is positive as $n \to \infty$, i.e., $\liminf\limits_{n\to\infty}\sum_{i=1}^n\delta_i/n >0$.
    \item[A4.] Let
    $$
    g(T_i|\vk X_i; \boldsymbol\theta, \delta_i) = \sum_{l=1}^L h_l(T_i; \alpha_l+ \vk X_{il}^\top  \boldsymbol\beta_l, \sigma_l)^{\delta_i} \expon{ \sum_{k=1}^L\log S_k(T_i; \alpha_k+ \vk X_{ik}^\top  \boldsymbol\beta_k, \sigma_k)},$$
    which is the product term in the observed likelihood  $L(\boldsymbol{\theta})$ defined in Eq.(3.1). It holds that
    $\pk{g(T_i|\vk X_i; \boldsymbol\theta,\delta_i)\ne g(T_i|\vk X_i; \boldsymbol\theta_0,\delta_i)}>0, \forall \boldsymbol\theta\in \Theta$ with $\boldsymbol\theta\ne\boldsymbol\theta_0$ (the true parameter).
    \item[A5.] $\boldsymbol\theta_0$ is in the interior of $\Theta$.
    \item[A6.]
    The Fisher information
    \begin{eqnarray*}
        \mathcal I(\boldsymbol\theta)=\E{\frac{\partial}{\partial\boldsymbol\theta}\log  g(T_i|\vk X_i; \boldsymbol\theta,\delta_i)\frac{\partial}{\partial\boldsymbol\theta^\top}\log  g(T_i|\vk X_i; \boldsymbol\theta,\delta_i)}
    \end{eqnarray*} is well-defined and positive definite at $\boldsymbol\theta_0$.
    \item[A7.] $\sigma_l\ge \varsigma>0,\ \forall \ l=1,\ldots,L.$ The noise variances are bounded away from 0.
    \item[A8.] The sequence $\{\boldsymbol\theta^{(m)}, m\ge1\}$ in EM algorithm satisfies  $\norm{\boldsymbol\theta^{(m+1)}-\boldsymbol\theta^{(m)}} \to 0, \ \mbox{as} \ m\to \infty.$
    \item[A9.] The average winning probability of each CF is positive among uncensored samples, i.e., $\liminf\limits_{n\to\infty}(\sum_{i=1}^n\delta_i\eta_{il})/(\sum_{i=1}^n\delta_i) >0, \  l=1\ldots, L.$
   \end{enumerate}

Assumptions \textit{A1} to \textit{A6} are standard regularity conditions in survival analysis requirements to show the consistency and asymptotic normality of MLEs. Assumption \textit{A4} guarantees identifiability of the competing parametric family. The convergence of the EM algorithm further requires Assumptions \textit{A7} to \textit{A8}. In certain competing parametric models (e.g., competing Weibull), the M-step may involve a non-concave objective function $Q(\boldsymbol{\theta}|\boldsymbol\theta^{(m)})$ in the form of a weighted log-likelihood of an accelerated failure time model. Assumption \textit{A9} is imposed to establish convergence of local maximisers to the global maximiser.

\subsection*{A.2 Expected Survival Time}\label{sec: Expected Survival Time}
In this section, we present the expected survival time, which generally does not admit an explicit expression due to heteroscedasticity among the CFs. Specifically, it follows by Eq.(3) that
\begin{eqnarray*}
    \E{T} = \int_0^\infty S(t;\boldsymbol{\theta},\vk X) \d t =
    \int_0^\infty \expon{ \sum\limits_{l=1}^L \log S_l(t;\alpha_l, \boldsymbol{\beta}_l, \sigma_l, \vk X_l) } \d t.
\end{eqnarray*}
A common practical strategy for approximating the complex expectation is to use the importance sampling method \citep{tokdar2010importance} with proposal distributions (e.g.,  exponential and Weibull distributions).

For the competing log-normal model, an alternative expression is discussed in \citet[Section 3.4]{Cui2021Max}. For distributions similar to the competing Weibull case, the tail integral can be approximated using Mill’s ratio over a large real number $M$, as we developed in this study (Proposition \ref{Prop: MR}):
\begin{eqnarray*}
  \int_M^\infty S(t)\d t\approx \frac{1}{\displaystyle\sum\limits_{k=1}^L \zeta_k(M)}S(M) \quad\mbox{with}\ \  \zeta_k(t) :=  \frac{t^{1/\sigma_k-1}}{\sigma_k\expon{\mu_k/ \sigma_k}}.
\end{eqnarray*}

In this study, individualised expected survival times are estimated primarily via importance sampling using the fitted model parameters. These estimated survival times are compared with the observed event times to compute the concordance index (C-index) as well as time-dependent sensitivity and specificity, providing an assessment of model performance (Sections 4 \& 5, Appendices B \& C).

\subsection*{A.3 Technical Proofs for Competing Weibull Model} \label{sec: pfWeibull}
In this section, we primarily consider the competing Weibull model with a likelihood function given by (recall Eq.(4))
\begin{eqnarray*}
    L(\boldsymbol\theta) = \prod_{i=1}^n \left\{\sum\limits_{l=1}^L \frac{t_i^{1/\sigma_l-1}}{ \sigma_l\expon{(\alpha_l+ X_{il}^\top \boldsymbol{\beta}_l)/ \sigma_l}} \right\}^{\delta_i} \exp\left\{- \sum\limits_{l=1}^L \fracl{t_i}{\expon{\alpha_l+ X_{il}^\top \boldsymbol{\beta}_l}}^{1/\sigma_l} \right\}.
\end{eqnarray*}
We will first prove Theorem 1 for the consistency of the MLE for the parameters involved and its asymptotic normality. Then, we show the proof of Theorem 3 by verifying the series of conditions (Assumption A1--A9) assigned to the objective function. Finally, we show Proposition \ref{Prop: MR} for a Mill's ratio to approximate the expected survival time, which is used for the estimate of the C-index and iAUC in this study.

Before we present the proof of Theorem 1, we introduce first Lemmas \ref{Lemma: 2.1}$\sim$\ref{Lemma: 2.3}, which will be used for Lemma \ref{lemma: 5.4} to show the consistency of the MLE, and additional Lemmas \ref{Lemma: 2.4} and \ref{Lemma: 2.5} for its asymptotic normality.
\begin{lemma}\label{Lemma: 2.1}
    Let $\boldsymbol\theta_l=(\sigma_l,\alpha_l, \boldsymbol{\beta}^\top_l)^\top$, $\boldsymbol\theta = (\boldsymbol\theta_1^\top,\ldots,\boldsymbol\theta_L^\top)^\top$, and $X_i=(X_{i1},\ldots,X_{iL})$,  $\widehat{Q}_n(\boldsymbol\theta) = \frac{1}{n}\ell(\boldsymbol\theta) = \frac{1}{n}\sum_{i=1}^n\log  g(T_i|X_i; \boldsymbol\theta,\delta_i)$ and $Q(\boldsymbol\theta) = \mathbb{E}_{\boldsymbol\theta_0}\{\log  g(T_i|X_i; \boldsymbol\theta,\delta_i)\}$, where
\begin{eqnarray*}
    \ell(\boldsymbol\theta) &=& \sum\limits_{i=1}^n \log g(T_i|X_i; \boldsymbol\theta, \delta_i) = \log \left\{f^{\delta_i}(T_i; \boldsymbol\theta,  X_i)S^{1-\delta_i}(T_i; \boldsymbol\theta,  X_i)\right\}\\
   &=& \sum\limits_{i=1}^n \delta_i \log \left\{\sum\limits_{1\le l \le L} \frac{T_i^{1/\sigma_l-1}}{ \sigma_l\expon{(\alpha_l+ X_{il}^\top \boldsymbol{\beta}_l)/ \sigma_l}} \right\}- \sum\limits_{1\le l \le L} \fracl{T_i}{\expon{\alpha_l+ X_{il}^\top  \boldsymbol\beta_l}}^{1/\sigma_l}.
\end{eqnarray*}
Suppose assumptions A1 and A2 hold, then
\begin{eqnarray*}
\sup_{\boldsymbol\theta\in\Theta}|\widehat{Q}_n(\boldsymbol\theta) - Q(\boldsymbol\theta)| \topb 0.
\end{eqnarray*}
\end{lemma}

\begin{lemma}[{\citet[Theorem 2]{Jennrich1969}}]\label{Lemma: Jenrich}
Let $g$ be a function on $\mathcal{X} \times \Theta$ where $\mathcal{X}$ is a Euclidean space and $\Theta$ is a compact subset of a Euclidean space. Let $g(x,\theta)$ be a continuous function of $\theta$ for each $x$ and a measurable function of $x$ for each $\theta$. Assume also that $|g(x,\theta)|\le h(x)$ for all $x$ and $\theta$, where $h$ is integrable with respect to a distribution function $F$ on $\mathcal{X}$. If $x_1, x_2, \ldots$ is a random sample from $F$, then for almost every sequence $(x_i)$,
$$
n^{-1}\sum_{i=1}^n g(x_i, \theta) \to \int g(x,\theta) \d F(x)
$$
uniformly for all $\theta\in\Theta$.
\end{lemma}

{
\makeatletter
\def\proof{\reset@font\rm \trivlist \item[\hskip \parindent
  {\reset@font~~\it Proof of Lemma \ref{Lemma: 2.1}}.]}
\makeatother

\begin{proof}
To apply Lemma \ref{Lemma: Jenrich} with assumptions \textit{A1} and \textit{A2}, we only need to find an upper bound function $h(T_i,X_i)$, such that $\left|\log  g(T_i|X_i; \boldsymbol\theta,\delta_i)\right|\le h(T_i,X_i)$, and  $\E{h(T_i,X_i)} < \infty$ for both $T_i$ and $X_i$.

Firstly, we show that $\log  g(T_i|X_i; \boldsymbol\theta,\delta_i)$ is bounded from above by $h(T_i,X_i)$,
\begin{eqnarray*}
\log  g(T_i|X_i; \boldsymbol\theta,\delta_i)&=& \log \left\{f^{\delta_i}(T_i; \boldsymbol\theta,  X_i)S^{1-\delta_i}(T_i; \boldsymbol\theta,  X_i)\right\} \notag\\
    &\le&   \delta_i \log f(T_i; \boldsymbol\theta,  X_i) \notag\\
   &=&  \delta_i \log \left[\sum\limits_{l=1}^L \frac{T_i^{1/\sigma_l-1}}{ \sigma_l\expon{(\alpha_l+ X_{il}^\top  \boldsymbol\beta_l)/ \sigma_l}} \exp\left\{- \sum\limits_{k=1}^L \fracl{T_i}{\expon{\alpha_k+ X_{ik}^\top  \beta_k}}^{1/\sigma_k}\right\}\right] \notag\\
   &\le& \delta_i \log \left[\sum\limits_{l=1}^L \frac{T_i^{1/\sigma_l-1}}{ \sigma_l\expon{(\alpha_l+ X_{il}^\top  \boldsymbol\beta_l)/ \sigma_l}} \exp\left\{-\fracl{T_i}{\expon{\alpha_l+ X_{il}^\top  \boldsymbol\beta_l}}^{1/\sigma_l}\right\}\right] \notag\\
   &=:& \delta_i \log \left(\sum\limits_{l=1}^L \psi(T_i; \sigma_l, \alpha_l, X_{il}, \boldsymbol\beta_l)\right).
\end{eqnarray*}
Note that $\psi(t; \sigma_l, \alpha_l, X_{il}, \boldsymbol\beta_l), \, t>0,$ is the Weibull density with scale parameter $\expon{\alpha_l+{X}_{il}^\top\boldsymbol\beta_l}$ and shape parameter $1/\sigma_l$. Thus for $0<t<1$, we have $t^{-1}> t^{1/\sigma_l-1}$, and
\begin{eqnarray*}
\delta_i \log \left(\sum\limits_{l=1}^L \psi(t; \sigma_l, \alpha_l, X_{il}, \boldsymbol\beta_l)\right) &\le&   \delta_i \log \left\{\sum\limits_{1\le l \le L} \frac{t^{1/\sigma_l-1}}{ \sigma_l\expon{(\alpha_l+ X_{il}^\top  \boldsymbol\beta_l)/ \sigma_l}} \right\}\\
   &\le& \left|\log \left\{\sum\limits_{1\le l \le L} \frac{t^{-1}}{ \sigma_l\expon{(\alpha_l+ X_{il}^\top  \boldsymbol\beta_l)/ \sigma_l}} \right\}\right| \\
  &\le& \left|\log t\right| + \left|\log \left\{\sum\limits_{1\le l \le L} \frac{1}{ \sigma_l\expon{(\alpha_l+ X_{il}^\top  \boldsymbol\beta_l)/ \sigma_l}} \right\}\right|\\
   &=:& h_1(t,X_i).
\end{eqnarray*}
As for $t\ge1$, when $\sigma_l \ge 1$, the Weibull density is monotonically decreasing, and
\begin{eqnarray*}
 \frac{t^{1/\sigma_l-1}}{ \sigma_l\expon{(\alpha_l+ X_{il}^\top  \boldsymbol\beta_l)/ \sigma_l}} \le  \frac{1}{ \sigma_l\expon{(\alpha_l+ X_{il}^\top  \boldsymbol\beta_l)/ \sigma_l}}, \quad t\ge 1,
\end{eqnarray*}
when $0< \sigma_l < 1$, the Weibull density reaches the max at the mode $t_0 = \expon{\alpha_l+{X}_{il}^\top\boldsymbol\beta_l}(1-\sigma_l)^{\sigma_l}$, with the value
$$
 \psi(t_0; \sigma_l, \alpha_l, X_{il}, \boldsymbol\beta_l)  =   \frac{(1-\sigma_l)^{1-\sigma_l}}{ \sigma_l\expon{\alpha_l+ X_{il}^\top  \boldsymbol\beta_l}} \expon{-(1-\sigma_l)}.
$$
Combine the two cases for $t>1$, we have
\begin{eqnarray*}
&&\delta_i \log \left(\sum\limits_{l=1}^L \psi(t; \sigma_l, \alpha_l, X_{il}, \boldsymbol\beta_l)\right)\\
    &&\quad \le  \delta_i \log \left[\sum\limits_{1\le l \le L} \left\{\frac{1}{ \sigma_l\expon{(\alpha_l+ X_{il}^\top  \boldsymbol\beta_l)/ \sigma_l}}  +\frac{(1-\sigma_l)^{1-\sigma_l}}{ \sigma_l\expon{\alpha_l+ X_{il}^\top  \boldsymbol\beta_l}} \expon{-(1-\sigma_l)}\right\}
   \right]\\
    &&\quad \le  \left|\log \left[\sum\limits_{1\le l \le L} \left\{\frac{1}{ \sigma_l\expon{(\alpha_l+ X_{il}^\top  \boldsymbol\beta_l)/ \sigma_l}}  +\frac{(1-\sigma_l)^{1-\sigma_l}}{ \sigma_l\expon{\alpha_l+ X_{il}^\top  \boldsymbol\beta_l}} \expon{-(1-\sigma_l)}\right\}
   \right]\right|\\
   &&\quad=: h_2(t,X_i).
\end{eqnarray*}
Then the upper bound $h(T_i,X_i)=h_1(T_i,X_i)+h_2(T_i,X_i)$. Under assumption \textit{A1} for the compact $\Theta$ and given $X_{il}$, both $h_2(T_i,X_i)$ and the second term of $h_1(T_i,X_i)$ are constant. For the first term in $h_1(T_i,X_i)$, which is $\left|\log T_i\right|$, we have
\begin{eqnarray*}
  \E{|\log T_i|} &=& \int_0^\infty \left|\log t\right| g(t|X_i; \boldsymbol\theta,\delta_i) \d t\\
   &=&  \int_0^\infty \left|\log t\right| \sum\limits_{l=1}^L \left(\frac{t^{1/\sigma_l-1}}{ \sigma_l\expon{(\alpha_l+ X_{il}^\top  \boldsymbol\beta_l)/ \sigma_l}} \right)^{\delta_i}\exp\left\{- \sum\limits_{k=1}^L \fracl{t}{\expon{\alpha_k+ X_{ik}^\top  \beta_k}}^{1/\sigma_k} \right\} \d t\\
    &\le&  \int_0^\infty  \sum\limits_{l=1}^L \left|\log t\right|\left(\frac{t^{1/\sigma_l-1}}{ \sigma_l\expon{(\alpha_l+ X_{il}^\top  \boldsymbol\beta_l)/ \sigma_l}} \right)^{\delta_i} \exp\left\{- \fracl{t}{\expon{\alpha_l+ X_{il}^\top  \boldsymbol\beta_l}}^{1/\sigma_l} \right\} \d t\\
     &=&  \sum\limits_{l=1}^L \int_0^\infty \left|\log t\right| \left(\frac{t^{1/\sigma_l-1}}{ \sigma_l\expon{(\alpha_l+ X_{il}^\top  \boldsymbol\beta_l)/ \sigma_l}} \right)^{\delta_i} \exp\left\{- \fracl{t}{\expon{\alpha_l+ X_{il}^\top  \boldsymbol\beta_l}}^{1/\sigma_l} \right\} \d t\\
    &=&  \delta_i\sum\limits_{l=1}^L \int_0^\infty \left|\log t\right| \psi(t; \sigma_l, \alpha_l, X_{il}, \boldsymbol\beta_l) \d t\\
    && + (1-\delta_i)\sum\limits_{l=1}^L \int_0^\infty \left|\log t\right| \expon{- \fracl{t}{\expon{\alpha_l+ X_{il}^\top  \boldsymbol\beta_l}}^{1/\sigma_l} } \d t,
\end{eqnarray*}
where both integrals are finite (the first one is finite according to the properties of the Weibull distribution). Hence, $\mathbb{E}_T\{h(T_i,X_i) \} <\infty$ and
under assumption \textit{A2} we have $\mathbb{E}_X\{h(T_i,X_i) \} <\infty$, which indicates that $h(T_i,X_i)$ is a proper upper-bounded function.

Now we show the lower bound. By Jensen's inequality and the concavity of the logarithm function, we have
\begin{eqnarray*}
&&\log \fracl{1}{L} +\log g(T_i|X_i; \boldsymbol\theta, \delta_i)\\
    &&= \log \left[\frac{1}{L} \left(\sum\limits_{l=1}^L \frac{T_i^{1/\sigma_l-1}}{ \sigma_l\expon{(\alpha_l+ X_{il}^\top  \boldsymbol\beta_l)/ \sigma_l}} \right)^{\delta_i}\exp\left\{- \sum\limits_{k=1}^L \fracl{T_i}{\expon{\alpha_k+ X_{ik}^\top  \beta_k}}^{1/\sigma_k} \right\}\right]\\
    &&\ge \log \left[\frac{1}{L^2} \sum\limits_{l=1}^L \left(\frac{T_i^{1/\sigma_l-1}}{ \sigma_l\expon{(\alpha_l+ X_{il}^\top  \boldsymbol\beta_l)/ \sigma_l}} \right)^{\delta_i}\exp\left\{- \sum\limits_{k=1}^L \fracl{T_i}{\expon{\alpha_k+ X_{ik}^\top  \beta_k}}^{1/\sigma_k} \right\}\right]\\
&&\ge\log \frac{1}{L} + \frac{1}{L}  \sum\limits_{l=1}^L \log \left[\left(\frac{T_i^{1/\sigma_l-1}}{ \sigma_l\expon{(\alpha_l+ X_{il}^\top  \boldsymbol\beta_l)/ \sigma_l}} \right)^{\delta_i}\exp\left\{- \sum\limits_{k=1}^L \fracl{T_i}{\expon{\alpha_k+ X_{ik}^\top  \beta_k}}^{1/\sigma_k} \right\}\right]\\
&& =: \log \frac{1}{L} +\frac{1}{L}  \sum\limits_{l=1}^L g_l (T_i|X_i; \boldsymbol\theta_l, \delta_i).
\end{eqnarray*}
Clearly, we have
\begin{eqnarray*}
|g_l (T_i|X_i; \boldsymbol\theta_l, \delta_i)|&=&  \abs{\delta_i \log \left\{\frac{T_i^{1/\sigma_l-1}}{ \sigma_l\expon{(\alpha_l+ X_{il}^\top  \boldsymbol\beta_l)/ \sigma_l}} \right\} - \sum\limits_{k=1}^L \fracl{T_i}{\expon{\alpha_k+ X_{ik}^\top  \beta_k}}^{1/\sigma_k}}\\
&\le& |(1/\sigma_l-1)\log T_i| + |\log \sigma_l| + \left|\frac{\alpha_l+ X_{il}^\top  \boldsymbol\beta_l}{\sigma_l}\right|
 + \sum\limits_{k=1}^L \fracl{T_i}{\expon{\alpha_k+ X_{ik}^\top  \boldsymbol\beta_k}}^{1/\sigma_k}\\
 &=:& G(T_i,X_i).
\end{eqnarray*}
Next, we show that $\E{G(T_i,X_i)}\le \infty$ for both $T_i$ and $X_i$.
Under assumptions \textit{A1} and \textit{A2}, $\E{G(T_i,X_i)}\le \infty$ for both $T_i$ and $X_i$, where
$\mathbb{E}\{|\log T_i|\}<\infty$ by previous calculations, and $\mathbb{E}\{ T_i^{1/\sigma_k}\}<\infty$ by
\begin{eqnarray*}
  \E{T_i^{1/\sigma_k}} &\le&  \delta_i\sum\limits_{l=1}^L \int_0^\infty t^{1/\sigma_k} \psi(t; \sigma_l, \alpha_l, X_{il}, \boldsymbol\beta_l) \d t\\
    && + (1-\delta_i)\sum\limits_{l=1}^L \int_0^\infty t^{1/\sigma_k} \exp\left\{- \fracl{t}{\expon{\alpha_l+ X_{il}^\top  \boldsymbol\beta_l}}^{1/\sigma_l} \right\} \d t,
\end{eqnarray*}
where both integrals are finite, by the properties of the Weibull distribution.

Consequently, Lemma \ref{Lemma: 2.1} is proved since all the conditions in Lemma \ref{Lemma: Jenrich} have been verified.
\end{proof}
}

\begin{lemma}\label{Lemma: 2.3}
    Suppose that assumptions A1 and A4 hold, there exists $\boldsymbol\theta_0\in \Theta$ such that $\forall \epsilon >0$,
    $$
    \sup_{\boldsymbol\theta \notin B(\boldsymbol\theta_0,\epsilon)} Q(\boldsymbol\theta) < Q(\boldsymbol\theta_0).
    $$
\end{lemma}

\begin{proof} It follows by Jensen's inequality that
\begin{eqnarray*}
   Q(\boldsymbol\theta)- Q(\boldsymbol\theta_0) &=& \mathbb{E}_{\boldsymbol\theta_0} \left\{\log \frac{g(T_i|X_i; \boldsymbol\theta,\delta_i)}{g(T_i|X_i; \boldsymbol\theta_0,\delta_i)}\right\}\\
    &\le& \log \mathbb{E}_{\boldsymbol\theta_0}\{g(T_i|X_i; \boldsymbol\theta,\delta_i)/g(T_i|X_i; \boldsymbol\theta_0,\delta_i)\}=0.
\end{eqnarray*}
Under assumption \textit{A4}, the inequality is strict, and $\boldsymbol\theta_0$ is the unique {maximiser} of $Q(\boldsymbol\theta)$ over $\Theta$. Noting that $\Theta$ is compact (assumption \textit{A1}) and $Q(\boldsymbol\theta)$ is continuous, Lemma \ref{Lemma: 2.3} follows.
\end{proof}

\begin{lemma}[{Extremum Consistency Theorem, \citet[Theorem 2.1]{newey1994large}}]\label{lemma: 5.4}
If there is a function $Q(\boldsymbol\theta)$, such that
\begin{enumerate}
    \item $Q(\boldsymbol\theta)$ is uniquely maximised at $\boldsymbol\theta_0$,
    \item $\Theta$ is compact,
    \item $Q(\boldsymbol\theta)$ is continuous,
    \item $\widehat{Q}_n(\boldsymbol\theta)$ converges uniformly in probability to $Q(\boldsymbol\theta)$,
\end{enumerate}
then
$$
\widehat{\boldsymbol\theta}_n \topb \boldsymbol\theta_0.
$$
\end{lemma}

For asymptotic normality, we first need to show
\begin{eqnarray}\label{eq: first-derivative}
   \E{\left|\frac{\partial}{\partial \boldsymbol\theta}\log g(T_i|X_i; \boldsymbol\theta, \delta_i)\right|} < \infty
\end{eqnarray}
and \begin{eqnarray}\label{eq: second-derivative}
  \E{\left|\frac{\partial^2}{\partial \boldsymbol\theta\partial\boldsymbol\theta^\top}\log g(T_i|X_i; \boldsymbol\theta, \delta_i)\right|} < \infty.
\end{eqnarray} To show Eq.\eqref{eq: first-derivative}, we rewrite $\log g$ (recall Lemma \ref{Lemma: 2.1}) as
\begin{eqnarray*}
  \log  g(T_i|X_i; \boldsymbol\theta, \delta_i)
  &=& {\delta_i} \log \sum\limits_{l=1}^L \left\{\frac{T_i^{1/\sigma_l-1}}{ \sigma_l\expon{(\alpha_l+ X_{il}^\top  \boldsymbol\beta_l)/ \sigma_l}} \right\} {- \sum\limits_{k=1}^L \fracl{T_i}{\expon{\alpha_k+ X_{ik}^\top  \boldsymbol\beta_k}}^{1/\sigma_k} }\\
  &=& {\delta_i} \log \sum\limits_{l=1}^L \left(\frac{1}{T_i\sigma_l} S_{il}^{1/\sigma_l} \right) {- \sum\limits_{k=1}^L S_{ik}^{1/\sigma_k} },
\end{eqnarray*}
where $S_{il} = {T_i}/{\expon{\alpha_l+ X_{il}^\top  \boldsymbol\beta_l}}>0$, and $\E{|S_{il}|}<\infty$, $\E{|\log S_{il}|}<\infty$ by previous calculations under assumptions \textit{A1} and \textit{A2}. In the following, we consider only the case with the non-censoring case (that is, all $\delta_i$'s equal 1).

First, we have
\begin{eqnarray} \label{eq: first-d-sigma}
    \left|\frac{\partial \log  g(T_i|X_i; \boldsymbol\theta, 1)}{\partial \sigma_l}\right| &=& \left| \frac{-\frac{S_{il}^{1/\sigma_l}(\log S_{il}+\sigma_l)}{T_i \sigma_l^3}}{\sum\limits_{l=1}^L \left(\frac{1}{T_i\sigma_l} S_{il}^{1/\sigma_l} \right)}  + \frac{S_{il}^{1/\sigma_l}\log S_{il}}{\sigma_l^2}\right|\\
    &\le&
    \frac{\left|\frac{S_{il}^{1/\sigma_l}(\log S_{il}+\sigma_l)}{T_i \sigma_l^3}\right|}{ \frac{1}{T_i\sigma_l} S_{il}^{1/\sigma_l} }  + \left|\frac{S_{il}^{1/\sigma_l}\log S_{il}}{\sigma_l^2}\right|\notag\\
    &\le&
    \left|\frac{\log S_{il}+\sigma_l}{ \sigma_l^2}\right| + \frac{S_{il}^{1/\sigma_l}}{\sigma_l^2}\left|\log S_{il}\right|.\notag
\end{eqnarray}
Now, we consider the partial derivative w.r.t. $\alpha_l$.
 \begin{eqnarray}
    \label{eq: first-d-alpha}&&\left|\frac{\partial \log  g(T_i|X_i; \boldsymbol\theta, 1)}{\partial \alpha_l}\right|=  \left|\frac{\partial \log  g(T_i|X_i; \boldsymbol\theta, 1)}{\partial S_{il}} \cdot \frac{\partial S_{il}}{\partial \alpha_l}\right|
   \notag \\
    &&= \left|\frac{-\frac{1}{T_i\sigma^2_l} S_{il}^{1/\sigma_l}}{\sum\limits_{l=1}^L \left(\frac{1}{T_i\sigma_l} S_{il}^{1/\sigma_l} \right)} +  \frac{S_{il}^{1/\sigma_l}}{\sigma_l}\right|\le  \left|\frac{\frac{1}{T_i\sigma^2_l} S_{il}^{1/\sigma_l} }{\frac{1}{T_i\sigma_l} S_{il}^{1/\sigma_l}}\right| +  \left|\frac{S_{il}^{1/\sigma_l}}{\sigma_l}\right|
    \le  \frac{1+ S_{il}^{1/\sigma_l}}{\sigma_l}.
\end{eqnarray}
Similarly,
\begin{eqnarray}\label{eq: first-d-beta}
\left|\frac{\partial \log  g(T_i|X_i; \boldsymbol\theta, 1)}{\partial  \beta_l}\right|  =  \left|\frac{\partial \log  g(T_i|X_i; \boldsymbol\theta, 1)}{\partial \alpha_l} \cdot X_{il} \right| \le  \left|X_{il} \right|  \frac{1+ S_{il}^{1/\sigma_l}}{\sigma_l}.
\end{eqnarray}
Again, under assumptions \textit{A1} and \textit{A2}, all first-order derivatives are bounded by integrable functions.

Next, we show Eq.\eqref{eq: second-derivative}. We have second-order derivative w.r.t. $\sigma_l$ and $\sigma_k$. Recalling Eq.\eqref{eq: first-d-sigma}, we have for $l\neq k$
\begin{eqnarray*}
    \left|\frac{\partial^2 \log  g(T_i|X_i; \boldsymbol\theta, 1)}{\partial \sigma_l\partial \sigma_k} \right| &=& \left| \frac{\partial}{\partial\sigma_k} \left\{\frac{-\frac{S_{il}^{1/\sigma_l}(\log S_{il}+\sigma_l)}{T_i \sigma_l^3}}{\sum\limits_{l=1}^L \left(\frac{1}{T_i\sigma_l} S_{il}^{1/\sigma_l} \right)} +\frac{S_{il}^{1/\sigma_l}\log S_{il}}{\sigma_l^2}\right\}\right| \\
    &=&  \left|-\frac{\left(-\frac{S_{il}^{1/\sigma_l}(\log S_{il}+\sigma_l)}{T_i \sigma_l^3}\right) \left(-\frac{S_{ik}^{1/\sigma_k}(\log S_{ik}+\sigma_k)}{T_i \sigma_l^3}\right)}{\left(\sum\limits_{l=1}^L \left(\frac{1}{T_i\sigma_l} S_{il}^{1/\sigma_l} \right)\right)^2}\right|\\
    &\le& \frac{\left|\left(-\frac{S_{il}^{1/\sigma_l}(\log S_{il}+\sigma_l)}{T_i \sigma_l^3}\right) \left(-\frac{S_{ik}^{1/\sigma_k}(\log S_{ik}+\sigma_k)}{T_i \sigma_l^3}\right)\right|}{\left(\frac{1}{T_i\sigma_l} S_{il}^{1/\sigma_l} \right) \left(\frac{1}{T_i\sigma_k} S_{ik}^{1/\sigma_k} \right)}
    \\
&\le&  \left|\frac{\log S_{il}+\sigma_l}{ \sigma_l^2}\right|\cdot \left|\frac{\log S_{ik}+\sigma_k}{ \sigma_k^2}\right|
\end{eqnarray*}
and for $l=k$,
\begin{eqnarray*}
    \left|\frac{\partial^2 \log  g(T_i|X_i; \boldsymbol\theta, 1)}{\partial \sigma_l^2} \right| &=&  \left| \frac{\partial}{\partial\sigma_l} \left\{\frac{-\frac{S_{il}^{1/\sigma_l}(\log S_{il}+\sigma_l)}{T_i \sigma_l^3}}{\sum\limits_{l=1}^L \left(\frac{1}{T_i\sigma_l} S_{il}^{1/\sigma_l} \right)} +\frac{S_{il}^{1/\sigma_l}\log S_{il}}{\sigma_l^2}\right\}\right|
    \end{eqnarray*}
    with
    \begin{eqnarray*}
        \left|\frac{\partial}{\partial\sigma_l}\frac{S_{il}^{1/\sigma_l}\log S_{il}}{\sigma_l^2}\right| = \left|-\frac{S_{il}^{1/\sigma_l}((\log S_{il})^2+2\sigma_l\log S_{il})}{\sigma_l^4}\right| \le \left|\frac{S_{il}^{1/\sigma_l}(\log S_{il})^2}{\sigma_l^4}\right| +\left|\frac{2S_{il}^{1/\sigma_l}\log S_{il}}{\sigma_l^3}\right|
    \end{eqnarray*}
    and
    \begin{eqnarray*}
        && \left| \frac{\partial}{\partial\sigma_l}\frac{-\frac{S_{il}^{1/\sigma_l}(\log S_{il}+\sigma_l)}{T_i \sigma_l^3}}{\sum\limits_{l=1}^L \left(\frac{1}{T_i\sigma_l} S_{il}^{1/\sigma_l} \right)} \right|= \left|-\frac{1}{T_i} \frac{-\frac{S_{il}^{1/\sigma_l}(\log S_{il})^2}{\sigma_l^5}-\frac{4S_{il}^{1/\sigma_l}\log S_{il}}{\sigma_l^4}-\frac{2S_{il}^{1/\sigma_l}}{\sigma_l^3}}{\sum\limits_{l=1}^L \left(\frac{1}{T_i\sigma_l} S_{il}^{1/\sigma_l} \right)} -\frac{\left(-\frac{S_{il}^{1/\sigma_l}(\log S_{il}+\sigma_l)}{T_i \sigma_l^3}\right)^2}{\left(\sum\limits_{l=1}^L \left(\frac{1}{T_i\sigma_l} S_{il}^{1/\sigma_l} \right)\right)^2}\right|\\
        & & \le \frac{(\log S_{il})^2}{\sigma_l^4}
    + \left|\frac{4\log S_{il}}{\sigma_l^3}\right|
    +\left|{\frac{2}{\sigma_l^2}}\right| +  \left(\frac{\log S_{il}+\sigma_l}{ \sigma_l^2}\right)^2.
    \end{eqnarray*}
Next, we deal with the second-order derivative w.r.t. $\alpha_l$ and $\alpha_k$. Recalling Eq.\eqref{eq: first-d-alpha}, we have for $l\neq k$
\begin{eqnarray*}
  & \left| \frac{\partial^2 \log  g(T_i|X_i; \boldsymbol\theta, 1)}{\partial \alpha_l\partial \alpha_k} \right|
    = \left|\frac{\partial}{\partial\alpha_k} \frac{\left(-\frac{1}{T_i\sigma^2_l} S_{il}^{1/\sigma_l} \right)}{\sum\limits_{l=1}^L \left(\frac{1}{T_i\sigma_l} S_{il}^{1/\sigma_l} \right)} +  \frac{\partial}{\partial\alpha_k}\frac{S_{il}^{1/\sigma_l}}{\sigma_l}\right| 
     =\frac{\left(-\frac{1}{T_i\sigma^2_l} S_{il}^{1/\sigma_l} \right)\left(-\frac{1}{T_i\sigma^2_k} S_{ik}^{1/\sigma_k} \right) }{\left\{\sum\limits_{l=1}^L \left(\frac{1}{T_i\sigma_l} S_{il}^{1/\sigma_l} \right)\right\}^2} \le \frac1{\sigma_l\sigma_k}
\end{eqnarray*}
and for $l=k$
\begin{eqnarray*}
  \left| \frac{\partial^2 \log  g(T_i|X_i; \boldsymbol\theta, 1)}{\partial \alpha_l^2} \right| \le  \left|\frac{\partial}{\partial\alpha_l}\frac{-\frac{1}{T_i\sigma^2_l} S_{il}^{1/\sigma_l} }{\sum\limits_{l=1}^L \left(\frac{1}{T_i\sigma_l} S_{il}^{1/\sigma_l} \right)}\right| + \left|\frac{\partial}{\partial\alpha_l} \frac{S_{il}^{1/\sigma_l}}{\sigma_l}\right| =: I + I\!I,
\end{eqnarray*}
where $I\!I = S_{il}^{1/\sigma_l}/\sigma_l^2$ and
\begin{eqnarray*}
   & I =\left| \frac{\frac{1}{T_i\sigma_l^3}{S_{il}^{1/\sigma_l}}   }{\sum\limits_{l=1}^L \left(\frac{1}{T_i\sigma_l} S_{il}^{1/\sigma_l} \right)}- \frac{\left(-\frac{1}{T_i\sigma^2_l} S_{il}^{1/\sigma_l} \right)^2   }{\left\{\sum\limits_{l=1}^L \left(\frac{1}{T_i\sigma_l} S_{il}^{1/\sigma_l} \right)\right\}^2}\right|
 \le  \left|\frac{\frac{1}{T_i\sigma_l^3}{S_{il}^{1/\sigma_l}}   }{\sum\limits_{l=1}^L \left(\frac{1}{T_i\sigma_l} S_{il}^{1/\sigma_l} \right)}\right|
    + \left|\frac{\left\{-\frac{1}{T_i\sigma^2_l} S_{il}^{1/\sigma_l} \right\}^2   }{\left\{\sum\limits_{l=1}^L \left(\frac{1}{T_i\sigma_l} S_{il}^{1/\sigma_l} \right)\right\}^2}\right| \le  \frac{2}{\sigma_l^2}.
\end{eqnarray*}
Further, we deal with the second-order derivative w.r.t. $\beta_l$ and $\beta_k$. Recalling Eqs.\eqref{eq: first-d-alpha} and \eqref{eq: first-d-beta}, we have for $l\neq k$
\begin{eqnarray*}
    \left|\frac{\partial^2 \log  g(T_i|X_i; \boldsymbol\theta, 1)}{\partial  \beta_l\partial  \beta_k} \right|
    &\le & \left|\frac{\partial}{\partial\beta_k} \frac{\left(-\frac{1}{T_i\sigma^2_l} S_{il}^{1/\sigma_l}X_{il} \right)}{\sum\limits_{l=1}^L \left(\frac{1}{T_i\sigma_l} S_{il}^{1/\sigma_l} \right)}\right| + \left| \frac{\partial}{\partial\beta_k}\frac{S_{il}^{1/\sigma_l}X_{il}}{\sigma_l}\right|\\
     &=&\left|-\frac{\left(-\frac{1}{T_i\sigma^2_l} S_{il}^{1/\sigma_l}X_{il} \right)\left(-\frac{1}{T_i\sigma^2_k} S_{ik}^{1/\sigma_k}X_{ik} \right) }{\left(\sum\limits_{l=1}^L \left(\frac{1}{T_i\sigma_l} S_{il}^{1/\sigma_l} \right)\right)^2}\right| \le \frac{|X_{il} X_{ik}|}{\sigma_l\sigma_k}
\end{eqnarray*}
and for $l=k$
\begin{eqnarray*}
    \left|\frac{\partial^2 \log  g(T_i|X_i; \boldsymbol\theta, 1)}{\partial  \beta_l^2} \right| &=& \left|\frac{\frac{1}{T_i\sigma_l^3}{S_{il}^{1/\sigma_l}} X^2_{il}  }{\sum\limits_{l=1}^L \left(\frac{1}{T_i\sigma_l} S_{il}^{1/\sigma_l} \right)}- \frac{\left(-\frac{1}{T_i\sigma^2_l} S_{il}^{1/\sigma_l}X_{il} \right)^2   }{\left\{\sum\limits_{l=1}^L \left(\frac{1}{T_i\sigma_l} S_{il}^{1/\sigma_l} \right)\right\}^2}
     -  \frac{S_{il}^{1/\sigma_l}X^2_{il}}{\sigma_l^2}\right|\\
    &\le&  \left|\frac{\frac{1}{T_i\sigma_l^3}{S_{il}^{1/\sigma_l}} X^2_{il}  }{\sum\limits_{l=1}^L \left(\frac{1}{T_i\sigma_l} S_{il}^{1/\sigma_l} \right)}\right|
    + \left|\frac{\left(-\frac{1}{T_i\sigma^2_l} S_{il}^{1/\sigma_l}X_{il} \right)^2   }{\left\{\sum\limits_{l=1}^L \left(\frac{1}{T_i\sigma_l} S_{il}^{1/\sigma_l} \right)\right\}^2}\right|
     +  \left|\frac{S_{il}^{1/\sigma_l}X^2_{il}}{\sigma_l^2}\right|\\
    &\le&  \frac{2X^2_{il}}{\sigma_l^2} +  \frac{S_{il}^{1/\sigma_l}X^2_{il}}{\sigma_l^2}.
\end{eqnarray*}
Lastly, for other second-order derivatives w.r.t. $(\sigma_l, \alpha_k)$ and $(\sigma_l, \beta_k)$, we recall Eq.\eqref{eq: first-d-sigma} for $l\neq k$
\begin{eqnarray*}
       \left|\frac{\partial^2 \log  g(T_i|X_i; \boldsymbol\theta, 1)}{\partial \sigma_l\partial \alpha_k} \right|
    &=& \left|\frac{\partial}{\partial\alpha_k} \frac{\left(-\frac{S_{il}^{1/\sigma_l}(\log S_{il}+\sigma_l)}{T_i \sigma_l^3}\right)}{\sum\limits_{l=1}^L \left(\frac{1}{T_i\sigma_l} S_{il}^{1/\sigma_l} \right)}  +\frac{\partial}{\partial\alpha_k}  \frac{S_{il}^{1/\sigma_l}\log S_{il}}{\sigma_l^2}\right|\\
       &=& \left|-\frac{\left(-\frac{S_{il}^{1/\sigma_l}(\log S_{il}+\sigma_l)}{T_i \sigma_l^3}\right) \left(-\frac{1}{T_i\sigma^2_k} S_{ik}^{1/\sigma_k} \right)}{\left\{\sum\limits_{l=1}^L \left(\frac{1}{T_i\sigma_l} S_{il}^{1/\sigma_l} \right)\right\}^2}\right| \le \frac{1}{\sigma_k}\left|\frac{\log S_{il}+\sigma_l}{ \sigma_l^2} \right|
\end{eqnarray*}
and
\begin{eqnarray*}
      \left|\frac{\partial^2 \log  g(T_i|X_i; \boldsymbol\theta, 1)}{\partial \sigma_l\partial \beta_k} \right|
       &\le& \frac{|X_{ik}|}{\sigma_k}\left|\frac{\log S_{il}+\sigma_l}{ \sigma_l^2} \right|.
\end{eqnarray*}
Now, for $l=k$
\begin{eqnarray*}
 \left|\frac{\partial^2 \log  g(T_i|X_i; \boldsymbol\theta, 1)}{\partial \sigma_l\partial \alpha_l} \right|
    &=& \left|\frac{\partial}{\partial\alpha_l} \frac{\left(-\frac{S_{il}^{1/\sigma_l}(\log S_{il}+\sigma_l)}{T_i \sigma_l^3}\right)}{\sum\limits_{l=1}^L \left(\frac{1}{T_i\sigma_l} S_{il}^{1/\sigma_l} \right)}  +\frac{\partial}{\partial\alpha_l}  \frac{S_{il}^{1/\sigma_l}\log S_{il}}{\sigma_l^2}\right| =: |I + I\!I|
    \end{eqnarray*}
with
\begin{eqnarray*}
 I\!I   &= &-\frac{S_{il}^{1/\sigma_l}\log S_{il}}{\sigma_l^3}-\frac{S_{il}^{1/\sigma_l}}{\sigma_l^2} \le \frac{S_{il}^{1/\sigma_l}|\log S_{il}|}{\sigma_l^3}+\frac{S_{il}^{1/\sigma_l}}{\sigma_l^2},\\
  I  &=&  \frac{\frac{\partial}{\partial\alpha_l}\left(-\frac{S_{il}^{1/\sigma_l}(\log S_{il}+\sigma_l)}{T_i \sigma_l^3}\right)}{\sum\limits_{l=1}^L \left(\frac{1}{T_i\sigma_l} S_{il}^{1/\sigma_l} \right)} - \frac{\left(-\frac{S_{il}^{1/\sigma_l}(\log S_{il}+\sigma_l)}{T_i \sigma_l^3}\right)\left(-\frac{1}{T_i\sigma^2_l} S_{il}^{1/\sigma_l} \right)}{\left(\sum\limits_{l=1}^L \left(\frac{1}{T_i\sigma_l} S_{il}^{1/\sigma_l} \right)\right)^2} \\
   &=& \frac{\frac{1}{T_i \sigma_l^3} \left(\frac{S_{il}^{1/\sigma_l}}{\sigma_l} \log S_{il} + S_{il}^{1/\sigma_l}       \right)}{\sum\limits_{l=1}^L \left(\frac{1}{T_i\sigma_l} S_{il}^{1/\sigma_l} \right)} + \frac{\frac{S_{il}^{1/\sigma_l}}{T_i \sigma_l^3}}{\sum\limits_{l=1}^L \left(\frac{1}{T_i\sigma_l} S_{il}^{1/\sigma_l} \right)} - \frac{\left(-\frac{S_{il}^{1/\sigma_l}(\log S_{il}+\sigma_l)}{T_i \sigma_l^3}\right)\left(-\frac{1}{T_i\sigma^2_l} S_{il}^{1/\sigma_l} \right)}{\left\{\sum\limits_{l=1}^L \left(\frac{1}{T_i\sigma_l} S_{il}^{1/\sigma_l} \right)\right\}^2},
\end{eqnarray*}
which is bounded by
\begin{eqnarray*}
  &&\left|\frac{\frac{S_{il}^{1/\sigma_l}\log S_{il}}{T_i \sigma_l^4}}{\sum\limits_{l=1}^L \left(\frac{1}{T_i\sigma_l} S_{il}^{1/\sigma_l} \right)}  \right|
    +\left|\frac{\frac{2S_{il}^{1/\sigma_l}}{T_i \sigma_l^3} }{\sum\limits_{l=1}^L \left(\frac{1}{T_i\sigma_l} S_{il}^{1/\sigma_l} \right)}\right|  + \left|\frac{-\frac{S_{il}^{1/\sigma_l}(\log S_{il}+\sigma_l)}{T_i \sigma_l^3}}{\sum\limits_{l=1}^L \left(\frac{1}{T_i\sigma_l} S_{il}^{1/\sigma_l} \right)}\right| \cdot
    \left|\frac{-\frac{1}{T_i\sigma^2_l} S_{il}^{1/\sigma_l}}{\sum\limits_{l=1}^L \left(\frac{1}{T_i\sigma_l} S_{il}^{1/\sigma_l} \right)}\right| \\
     && \le\left|{\frac{\log S_{il}}{ \sigma_l^3} }\right|
    +\frac{2}{ \sigma_l^2}  + \left|\frac{\log S_{il}+\sigma_l}{\sigma_l^3}\right|.
\end{eqnarray*}
Next,
\begin{eqnarray*}
   \left|\frac{\partial^2 \log  g(T_i|X_i; \boldsymbol\theta, 1)}{\partial \sigma_l\partial  \beta_l} \right|
    = \left|X_{il}\cdot\left\{\frac{\partial}{\partial\alpha_l} \frac{\left(-\frac{S_{il}^{1/\sigma_l}(\log S_{il}+\sigma_l)}{T_i \sigma_l^3}\right)}{\sum\limits_{l=1}^L \left(\frac{1}{T_i\sigma_l} S_{il}^{1/\sigma_l} \right)} +\frac{\partial}{\partial\alpha_l}  \frac{S_{il}^{1/\sigma_l}\log S_{il}}{\sigma_l^2}\right\}\right| = |X_{il}\cdot(I + I\!I)| \\
    \le  |X_{il}| \left(\left|{\frac{\log S_{il}}{ \sigma_l^3} }\right|
    +\frac{2}{ \sigma_l^2}  + \left|\frac{\log S_{il}+\sigma_l}{\sigma_l^3}\right| + \frac{S_{il}^{1/\sigma_l}|\log S_{il}|}{\sigma_l^3}+\frac{S_{il}^{1/\sigma_l}}{\sigma_l^2}\right).
\end{eqnarray*}
Finally, we show the second order derivatives w.r.t. $\alpha_l$ and $\beta_k$. Recalling Eq.\eqref{eq: first-d-alpha}, we have for $l\neq k$
\begin{eqnarray*}
    & \left|\frac{\partial^2 \log  g(T_i|X_i; \boldsymbol\theta, 1)}{\partial \alpha_l\partial \beta_k} \right|
    \le \left|\frac{\partial}{\partial\beta_k} \frac{-\frac{1}{T_i\sigma^2_l} S_{il}^{1/\sigma_l} }{\sum\limits_{l=1}^L \left(\frac{1}{T_i\sigma_l} S_{il}^{1/\sigma_l} \right)}\right| +  \left|\frac{\partial}{\partial\beta_k}\frac{S_{il}^{1/\sigma_l}}{\sigma_l}\right| 
     =\left|-\frac{\left(-\frac{1}{T_i\sigma^2_l} S_{il}^{1/\sigma_l} \right)\left(-\frac{1}{T_i\sigma^2_k} S_{ik}^{1/\sigma_k} X_{ik}\right) }{\left\{\sum\limits_{l=1}^L \left(\frac{1}{T_i\sigma_l} S_{il}^{1/\sigma_l} \right)\right\}^2}\right| \le \frac{|X_{ik}|}{\sigma_k\sigma_l}
\end{eqnarray*}
and for $l=k$
\begin{eqnarray*}
       &\left|\frac{\partial^2 \log  g(T_i|X_i; \boldsymbol\theta, 1)}{\partial \alpha_l\partial  \beta_l} \right|
    = \left| \frac{\frac{1}{T_i\sigma_l^3}{S_{il}^{1/\sigma_l}}  X_{il} }{\sum\limits_{l=1}^L \left(\frac{1}{T_i\sigma_l} S_{il}^{1/\sigma_l} \right)}- \frac{X_{il}\left(-\frac{1}{T_i\sigma^2_l} S_{il}^{1/\sigma_l} \right)^2   }{\left(\sum\limits_{l=1}^L \left(\frac{1}{T_i\sigma_l} S_{il}^{1/\sigma_l} \right)\right)^2}
     -  \frac{S_{il}^{1/\sigma_l}X_{il}}{\sigma_l^2}\right|
\le  \frac{2|X_{il}|}{\sigma_l^2} + \left|\frac{S_{il}^{1/\sigma_l}X_{il}}{\sigma_l^2}\right|.
\end{eqnarray*}
Note that to show $\E{(\log S_{il})^2}<\infty$ is equivalent to show $\E{(\log T_i)^2}<\infty$. It follows by elementary calculation and the properties of Weibull density that
\begin{eqnarray*}
  \E{(\log T_i)^2} &\le& \delta_i\sum\limits_{l=1}^L \int_0^\infty \left(\log t\right)^2 \psi(t; \sigma_l, \alpha_l, X_{il}, \boldsymbol\beta_l) \d t\\
    && + (1-\delta_i)\sum\limits_{l=1}^L \int_0^\infty \left(\log t\right)^2 \exp\left\{- \fracl{t}{\expon{\alpha_l+ X_{il}^\top  \boldsymbol\beta_l}}^{1/\sigma_l} \right\} \d t<\infty,
\end{eqnarray*}
and we obtain Eq.\eqref{eq: second-derivative}.
\QED

The next two lemmas are used for the proof of the asymptotic normality of the MLE for the parameters involved.
\begin{lemma}\label{Lemma: 2.4} Under assumptions A1$\sim$A6,
    $$\sqrt{n}\frac{\partial}{\partial \boldsymbol\theta}\widehat{Q}_n(\boldsymbol\theta_0)\todis \mathcal{N}(0,I(\boldsymbol\theta_0)).$$
\end{lemma}

\begin{proof}
Under assumptions \textit{A1}$\sim$\textit{A6} and Eq.\eqref{eq: first-derivative}, we have $\E{\left|\frac{\partial }{\partial \boldsymbol\theta}\log  g(T_i|X_i; \boldsymbol\theta_0, \delta_i)\right|}<\infty$. By the dominated convergence theorem,
\begin{eqnarray*}
\vk0 &=&\frac{\partial }{\partial \boldsymbol\theta}Q(\boldsymbol\theta_0)= \frac{\partial }{\partial \theta}\E{\log  g(T_i|X_i; \boldsymbol\theta_0, \delta_i)} = \E{\frac{\partial }{\partial \boldsymbol\theta}\log  g(T_i|X_i; \boldsymbol\theta_0, \delta_i)} \\
 &=& \E{\frac{1}{n}\sum_{i=1}^n\frac{\partial}{\partial \boldsymbol\theta}\log  g(T_i|X_i; \boldsymbol\theta_0, \delta_i)}= \E{\frac{\partial}{\partial \boldsymbol\theta}\widehat{Q}_n(\boldsymbol\theta_0)}.
\end{eqnarray*}
As the Fisher information matrix $\mathcal I(\boldsymbol\theta_0)$ is well-defined at $\boldsymbol\theta_0$, {the claim follows} by central limit theorem and the proved Eq.\eqref{eq: second-derivative}. 
\end{proof}

\begin{lemma}\label{Lemma: 2.5} Under assumptions A1$\sim$A6, for any sequence $\widetilde{\boldsymbol\theta}_n \topb \boldsymbol\theta_0$,
$$\frac{\partial^2}{\partial \boldsymbol\theta\partial \boldsymbol\theta^\top}\widehat{Q}_n(\widetilde{\boldsymbol\theta}_n) + \mathcal I(\boldsymbol\theta_0) \topb 0.$$
\end{lemma}

\begin{proof}
By assumptions \textit{A1}$\sim$\textit{A5}, we have all bounded first- and second-order derivatives, i.e., $\E{\left|\frac{\partial}{\partial \boldsymbol\theta}\log g(T_i|X_i; \boldsymbol\theta, \delta_i)\right|} < \infty$ and $\E{\left|\frac{\partial^2}{\partial \boldsymbol\theta\partial\boldsymbol\theta^\top}\log g(T_i|X_i; \boldsymbol\theta, \delta_i)\right|}$ $ < \infty$. Applying the dominated convergence theorem and Lemma \ref{Lemma: Jenrich} obtains that for  $\boldsymbol\theta\in\Theta$,
$$
\frac{\partial^2}{\partial \boldsymbol\theta\partial \boldsymbol\theta^\top}\widehat{Q}_n(\boldsymbol\theta) \topb \frac{\partial^2}{\partial \boldsymbol\theta\partial \boldsymbol\theta^\top}{Q}(\boldsymbol\theta).
$$
Further, it follows by assumptions \textit{A3} and \textit{A6} that
$$
\sup_{\boldsymbol\theta\in\Theta}\left|\frac{\partial^2}{\partial \boldsymbol\theta\partial \boldsymbol\theta^\top}\widehat{Q}_n(\boldsymbol\theta)-(-\mathcal I(\boldsymbol\theta)) \right| = \sup_{\boldsymbol\theta\in\Theta}\left|\frac{\partial^2}{\partial \boldsymbol\theta\partial \boldsymbol\theta^\top}\widehat{Q}_n(\boldsymbol\theta)-\frac{\partial^2}{\partial \boldsymbol\theta\partial \boldsymbol\theta^\top}{Q}(\boldsymbol\theta) \right| \topb 0.
$$
Then for any sequence  $\widetilde{\boldsymbol\theta}_n \topb \boldsymbol\theta_0$, by the continuity of $\mathcal I(\boldsymbol\theta)$ at $\boldsymbol\theta_0$,
\begin{eqnarray*}
    \left|\frac{\partial^2}{\partial \boldsymbol\theta\partial \boldsymbol\theta^\top}\widehat{Q}_n(\widetilde{\boldsymbol\theta}_n) +\mathcal I(\boldsymbol\theta_0) \right| &\le&
    \left|\frac{\partial^2}{\partial \boldsymbol\theta\partial \boldsymbol\theta^\top}\widehat{Q}_n(\widetilde{\boldsymbol\theta}_n)+I(\widetilde{\boldsymbol\theta}_n) \right|+\left|\mathcal I(\boldsymbol\theta_0)- \mathcal I(\widetilde{\boldsymbol\theta}_n)\right|\\
    &\le& \sup_{\boldsymbol\theta\in\Theta}\left|\frac{\partial^2}{\partial \boldsymbol\theta\partial \boldsymbol\theta^\top}\widehat{Q}_n(\boldsymbol\theta)+\mathcal I(\boldsymbol\theta) \right| +\left|\mathcal I(\boldsymbol\theta_0)- \mathcal I(\widetilde{\boldsymbol\theta}_n)\right| \topb 0.
\end{eqnarray*}

\end{proof}

Now, combining all the results and lemmas, we are ready to prove the consistency and asymptotic normality.

{
\makeatletter
\def\proof{\reset@font\rm \trivlist \item[\hskip \parindent
  {\reset@font~~\it Proof of Theorem 1.}]}
\makeatother

\begin{proof}
 Under assumptions \textit{A1}$\sim$\textit{A4}, Lemmas \ref{Lemma: 2.1} and \ref{Lemma: 2.3} hold, ensuring that all conditions in Lemma \ref{lemma: 5.4} are satisfied, thereby the consistency is proved.

Under assumptions \textit{A1}$\sim$\textit{A6}, the MLE  $\widehat{\boldsymbol\theta}_n \topb \boldsymbol\theta_0$. As $\boldsymbol\theta_0$ is in an interior point of $\Theta$, by mean value theorem,
$$
\frac{\partial}{\partial \boldsymbol\theta}\widehat{Q}_n(\widehat{\boldsymbol\theta}_n)-\frac{\partial}{\partial \boldsymbol\theta}\widehat{Q}_n(\boldsymbol\theta_0)=\frac{\partial^2}{\partial \boldsymbol\theta\partial \boldsymbol\theta^\top}\widehat{Q}_n(\widetilde{\boldsymbol\theta}_n)(\widehat{\boldsymbol\theta}_n-\boldsymbol\theta_0),
$$
where $\widetilde{\boldsymbol\theta}_n$ lies on the segment between $\widehat{\boldsymbol\theta}_n$ and $\boldsymbol\theta_0$. Thus,
$$
\frac{\partial}{\partial \boldsymbol\theta}\widehat{Q}_n(\boldsymbol\theta_0) + \frac{\partial^2}{\partial \boldsymbol\theta\partial \boldsymbol\theta^\top}\widehat{Q}_n(\widetilde{\boldsymbol\theta}_n)(\widehat{\boldsymbol\theta}_n-\boldsymbol\theta_0) = 0,
$$
where $|\widetilde{\boldsymbol\theta}_n-\boldsymbol\theta_0|\le |\widehat{\boldsymbol\theta}_n-\boldsymbol\theta_0| \topb 0$ as $n\to \infty$. Applying Lemmas \ref{Lemma: 2.4} and \ref{Lemma: 2.5},
$$
\sqrt{n}(\widehat{\boldsymbol\theta}_n-\boldsymbol\theta_0)=\left\{-\frac{\partial^2}{\partial \boldsymbol\theta\partial \boldsymbol\theta^\top}\widehat{Q}_n(\widetilde{\boldsymbol\theta}_n)\right\}^{-1} \sqrt{n}\frac{\partial}{\partial \boldsymbol\theta}\widehat{Q}_n(\boldsymbol\theta_0) \todis \mathcal{N}(0,\mathcal I(\boldsymbol\theta_0)^{-1}).
$$
\end{proof}}

{
\makeatletter
\def\proof{\reset@font\rm \trivlist \item[\hskip \parindent
  {\reset@font~~\it Proof of Theorem 3.}]}
\makeatother

\begin{proof}
Recall that for the objective function $Q(\boldsymbol{\theta}|\boldsymbol\theta^{(m)})$ in the $m$-th step of M-step is separated into groups ($l=1,\ldots, L$) with
\begin{eqnarray*}
    Q_l(\boldsymbol{\theta}|\boldsymbol\theta^{(m)}) =
    \sum\limits_{i=1}^n   \delta_i \eta_{il}^{(m)} \left\{\left(\frac{1}{\sigma_l}-1\right)\log T_i- \log \sigma_l- \frac{\alpha_l+\vk X_{il}^\top \boldsymbol{\beta}_l}{\sigma_l}\right\}- \fracl{T_i}{\expon{\alpha_l+\vk X_{il}^\top \boldsymbol{\beta}_l}}^{1/\sigma_l}.
\end{eqnarray*}
To study the asymptotic behavior of $Q_l(\boldsymbol{\theta}|\boldsymbol\theta^{(m)})$ is equivalent to the cause-specific analysis that all samples have survival time $\widetilde{T}_i\sim Weibull(1/\sigma_l, \exp(\alpha_l+\vk{X_{il}^\top\beta_l}) )$ with corresponding weights. For censored subjects, the weight is 1. Each observed failure is represented by two pseudo-observations: a pseudo-failure with weight $\eta_{il}^{(m)}$ and a pseudo-censored observation with weight $1-\eta_{il}^{(m)}$.
As $\delta_i$ and $\eta^{(m)}_{il}$ are already given/estimated as constants in data/E-step, it's equivalent to consider the asymptotic behaviour of the following function for simplicity:
$$q_{n}(\boldsymbol\theta) = \frac{1}{n}\sum_{i=1}^n\delta_{i}\left(\frac{1}{\sigma}-1\right)\log t_i- \delta_{i}\log\sigma-\delta_{i}\frac{X_i\beta}{\sigma}- \expon{\frac{\log t_i-{ X_i\beta}}{\sigma}}$$
with $\boldsymbol\theta=(\sigma, \beta)$, and $\delta_{i}$ being the failure indicator.

The function $q_{n}(\vk\theta)$ is the same as a Weibull survival log-likelihood. Under regularity conditions of assumptions \textit{A2}, \textit{A3} and \textit{A9},
the score function, denoted by $\mathcal{U}(\boldsymbol\theta_0)$, is a multivariate normal distribution with a mean of 0 and variance of the Fisher information matrix $\mathcal{I}(\boldsymbol\theta_0)$ \citep{kalbfleisch2011statistical}, at the true parameter $\boldsymbol\theta_0$.

Now we consider the approach outlined in \citet[Proof of Theorem 2]{fan2025dynamic} and \cite{smith1985maximum} to show $q_n(\boldsymbol\theta)$ has a sequence of local maximisers converging to the true parameter $\boldsymbol\theta_0=(\sigma_0, \beta_0)$. The above illustration is equivalent to showing that there is a sequence of $\Delta_n\to0$, $n\Delta_n \to \infty$ as $n\to \infty$, the following function
$$
h_n(\boldsymbol x):=h_n(y,z) = \frac{1}{\Delta_n^2}q_n(\sigma_0+\Delta_ny, \beta_0+\Delta_nz)
$$
has local maximiser $(\widehat{y},\widehat{z})$ within the open set $\norm{(y,z)} < 1$. By showing the existence of a local maximiser in the open set, the true parameter $\boldsymbol\theta_0$ acts like a local maximiser as $n\to \infty$, and the open set converges to the true parameter $\boldsymbol\theta_0$  as $n\to \infty$.

By \citet[Lemma 5]{smith1985maximum} as well as the illustration of boundary in \citet[Proof of Theorem 2]{fan2025dynamic}, the proof of the existence of local maximiser $(\widehat{y},\widehat{z})$ in the open set, is also equivalent to showing that, for $\boldsymbol x=(y,z)^{{\top}}$,
\begin{eqnarray}\label{eq: hn}
\boldsymbol x^\top\frac{\partial h_n(\boldsymbol x)}{\partial \boldsymbol x} = y\frac{\partial h_n(y,z)}{\partial y}+ z\frac{\partial h_n(y,z)}{\partial z}
\end{eqnarray}
is negative at the boundary $\norm{\boldsymbol x}=1$ as $n\to \infty$.

We firstly consider the term ${\partial h_n(y,z)}/{\partial y}$ in Eq.\eqref{eq: hn} with the mean value theorem, we have
\begin{eqnarray*}
    \frac{\partial h_n(y,z)}{\partial y} &=& \frac{1}{\Delta_n}\frac{\partial q_n(\sigma, \beta_0+\Delta_nz)}{\partial \sigma}\Big|_{\sigma = \sigma_0+\Delta_ny}\\
    &=& \frac{1}{\Delta_n}\frac{\partial q_n(\sigma, \beta_0)}{\partial \sigma}\Big|_{\sigma = \sigma_0} + \frac{1}{\Delta_n}\frac{\partial^2 q_n(\sigma, \beta_c)}{\partial \sigma^2}\Big|_{\sigma = \sigma_c}(\sigma-\sigma_0)\\
    && + \frac{1}{\Delta_n}\frac{\partial^2 q_n(\sigma, \beta_c)}{\partial \sigma\partial\beta}\Big|_{\sigma = \sigma_c,\beta=\beta_c}(\beta-\beta_0)\\
    &=& \frac{1}{\Delta_n}\frac{\partial q_n(\sigma, \beta_0)}{\partial \sigma}\Big|_{\sigma = \sigma_0} + \frac{\partial^2 q_n(\sigma, \beta_c)}{\partial \sigma^2}\Big|_{\sigma = \sigma_c}y + \frac{\partial^2 q_n(\sigma, \beta)}{\partial \sigma\partial\beta}\Big|_{\sigma = \sigma_c,\beta=\beta_c}z,
\end{eqnarray*}
where $|\sigma_c-\sigma_0|<\Delta_ny,|\beta_c-\beta_0|<\Delta_nz$ are in the neighbouring open set of $(\sigma_0,\beta_0)$.

It is easy to check the first part
$$
\frac{1}{\Delta_n}\frac{\partial q_n(\sigma, \beta_0)}{\partial \sigma}\Big|_{\sigma = \sigma_0} = \frac{1}{n\Delta_n}\sum_{i=1}^n \mathcal{U}_i(\sigma_0, \beta_0) \to 0,
$$
by $\E{\mathcal{U}(\boldsymbol\theta_0)}=0$ and $n\Delta_n \to \infty$. For the second and third parts, by the existence of the positive definite Fisher information matrix $\mathcal{I}(\boldsymbol\theta_0)$, we will derive that
$$
\frac{\partial^2 q_n(\sigma, \beta)}{\partial \sigma^2}-\frac{\partial^2 q_n(\sigma, \beta_0)}{\partial \sigma^2}\Big|_{\sigma = \sigma_0} \topb 0,\ \frac{\partial^2 q_n(\sigma, \beta)}{\partial \sigma\partial\beta} - \frac{\partial^2 q_n(\sigma, \beta)}{\partial \sigma\partial\beta}\Big|_{\sigma = \sigma_0,\beta=\beta_0} \topb 0
$$
for $\norm{\boldsymbol\theta - \boldsymbol\theta_0}\le \Delta_n$.

The first-order derivatives are
\begin{eqnarray*}
    \frac{\partial q_n(\sigma, \beta)}{\partial \beta} &=& \frac{1}{n}\sum_{i=1}^n\frac{X_i}{\sigma}\expon{\frac{\log t_i-{X_i \beta}}{\sigma}}- \delta_i\frac{X_i}{\sigma},\\
    \frac{\partial q_n(\sigma, \beta)}{\partial \sigma} &=& \frac{1}{n}\sum_{i=1}^n\delta_i\left(-\frac{1}{\sigma}-\frac{\log t_i-{X_i \beta}}{\sigma^2}\right)+\frac{\log t_i-{X_i \beta}}{\sigma^2}\expon{\frac{\log t_i-{X_i \beta}}{\sigma}},
\end{eqnarray*}
and the second-order derivatives are
\begin{eqnarray*}
    \frac{\partial^2 q_n(\sigma, \beta)}{\partial \beta^2} &=& \frac{1}{n}\sum_{i=1}^n-\frac{X_i^2}{\sigma^2}\expon{\frac{\log t_i-{X_i \beta}}{\sigma}}, \\
   \frac{\partial^2 q_n(\sigma, \beta)}{\partial \beta\partial\sigma} &=& \frac{1}{n}\sum_{i=1}^n \frac{X_i}{\sigma^2}\left\{\delta_i-\expon{\frac{\log t_i-{X_i \beta}}{\sigma}}\right\}-\frac{X_i(\log t_i-{X_i \beta})}{\sigma^3}\expon{\frac{\log t_i-{X_i \beta}}{\sigma}},\\
\frac{\partial^2 q_n(\sigma, \beta)}{\partial \sigma^2} &=& \frac{1}{n}\sum_{i=1}^n\delta_i\left(\frac{1}{\sigma^2}+\frac{2(\log t_i-X_i \beta)}{\sigma^3}\right)-\frac{2(\log t_i-X_i \beta)}{\sigma^3}\expon{\frac{\log t_i-{X_i \beta}}{\sigma}}\\
&&-\frac{1}{n}\sum_{i=1}^n\frac{(\log t_i-X_i \beta)^2}{\sigma^4}\expon{\frac{\log t_i-{X_i \beta}}{\sigma}}.
\end{eqnarray*}
A direct calculation yields:
\begin{eqnarray*}
&&   \frac{\partial^2 q_n(\sigma, \beta)}{\partial \sigma^2}-\frac{\partial^2 q_n(\sigma, \beta_0)}{\partial \sigma^2}\Big|_{\sigma = \sigma_0}=: I_1+ I_2+I_3 \\
&&   =  \frac{1}{n}\sum_{i=1}^n\delta_i\left(\frac{1}{\sigma^2}-\frac{1}{\sigma_0^2}+\frac{2(\log t_i-X_i \beta)}{\sigma^3}-\frac{2(\log t_i-X_i \beta_0)}{\sigma_0^3}\right)\\
&& -\frac{1}{n}\sum_{i=1}^n\frac{2(\log t_i-X_i \beta)}{\sigma^3}\expon{\frac{\log t_i-{X_i \beta}}{\sigma}}-\frac{2(\log t_i-X_i \beta_0)}{\sigma_0^3}\expon{\frac{\log t_i-{X_i \beta_0}}{\sigma_0}}\\
&&-\frac{1}{n}\sum_{i=1}^n\frac{(\log t_i-X_i \beta)^2}{\sigma^4}\expon{\frac{\log t_i-{X_i \beta}}{\sigma}}-\frac{(\log t_i-X_i \beta_0)^2}{\sigma_0^4}\expon{\frac{\log t_i-{X_i \beta_0}}{\sigma_0}}.
\end{eqnarray*}
The first part
$$
I_1 = \frac{1}{n}\sum_{i=1}^n\delta_i\left\{(\sigma_0-\sigma)\left(\frac{\sigma_0+\sigma}{\sigma_0^2\sigma^2}+\frac{2(\log t_i- X_i\beta_0)(\sigma^2+\sigma\sigma_0+\sigma_0^2)}{\sigma^3\sigma_0^3}\right)-(\beta-\beta_0)\frac{X_i}{\sigma^3}\right\},
$$
since $|\sigma-\sigma_0|\le\Delta_n, |\beta-\beta_0|\le\Delta_n$ and the rest part of $I_1$ is finite, the first part converges to 0 as $\Delta_n\to 0$. The second and third parts also converge to 0, with
\begin{eqnarray*}
I_2&=& -\frac{1}{n}\sum_{i=1}^n\frac{2(\log t_i-X_i \beta)}{\sigma^3}\expon{\frac{\log t_i-{X_i \beta}}{\sigma}}-\frac{2(\log t_i-X_i \beta_0)}{\sigma_0^3}\expon{\frac{\log t_i-{X_i \beta}}{\sigma}}\\
&&\ -\frac{1}{n}\sum_{i=1}^n\frac{2(\log t_i-X_i \beta_0)}{\sigma_0^3}\expon{\frac{\log t_i-{X_i \beta}}{\sigma}}-\frac{2(\log t_i-X_i \beta_0)}{\sigma_0^3}\expon{\frac{\log t_i-{X_i \beta_0}}{\sigma_0}}\\
&=& -\frac{1}{n}\sum_{i=1}^n\left\{ \frac{2(\log t_i-X_i \beta)}{\sigma^3}-\frac{2(\log t_i-X_i \beta_0)}{\sigma_0^3}\  \right\}\expon{\frac{\log t_i-{X_i \beta}}{\sigma}}\\
&&\ -\frac{1}{n}\sum_{i=1}^n\frac{2(\log t_i-X_i \beta_0)}{\sigma_0^3}\expon{\frac{\log t_i-{X_i \beta_0}}{\sigma_0}}\left\{\expon{\frac{\log t_i-{X_i \beta}}{\sigma}-\frac{\log t_i-{X_i \beta_0}}{\sigma_0}}-1\right\},
\end{eqnarray*}
and
\begin{eqnarray*}
&&I_3= -\frac{1}{n}\sum_{i=1}^n\left\{ \frac{2(\log t_i-X_i \beta)^2}{\sigma^4}-\frac{2(\log t_i-X_i \beta_0)^2}{\sigma_0^4}\  \right\}\expon{\frac{\log t_i-{X_i \beta}}{\sigma}}\\
&&-\frac{1}{n}\sum_{i=1}^n\frac{2(\log t_i-X_i \beta_0)^2}{\sigma_0^4}\expon{\frac{\log t_i-{X_i \beta_0}}{\sigma_0}}\left\{\expon{\frac{\log t_i-{X_i \beta}}{\sigma}-\frac{\log t_i-{X_i \beta_0}}{\sigma_0}}-1\right\}.
\end{eqnarray*}
Using the factorisation in $I_1$ leads to
$$
\frac{\partial^2 q_n(\sigma, \beta)}{\partial \sigma^2}-\frac{\partial^2 q_n(\sigma, \beta_0)}{\partial \sigma^2}\Big|_{\sigma = \sigma_0} \topb 0.
$$
For the next part of
\begin{eqnarray*}
&& \frac{\partial^2 q_n(\sigma, \beta)}{\partial\sigma\partial \beta}-\frac{\partial^2 q_n(\sigma, \beta)}{\partial\sigma\partial \beta}\Big|_{\sigma = \sigma_0,\beta=\beta_0} = \frac{1}{n}\sum_{i=1}^n \frac{X_i}{\sigma^2}\delta_i-\frac{X_i}{\sigma_0^2}\delta_i\\
&&-\frac{1}{n}\sum_{i=1}^n\frac{X_i}{\sigma^2}\expon{\frac{\log t_i-{X_i \beta}}{\sigma}}-\frac{X_i}{\sigma_0^2}\expon{\frac{\log t_i-{X_i \beta_0}}{\sigma_0}}\\
&&-\frac{1}{n}\sum_{i=1}^n\frac{X_i(\log t_i-{X_i \beta})}{\sigma^3}\expon{\frac{\log t_i-{X_i \beta}}{\sigma}}-\frac{X_i(\log t_i-{X_i \beta_0})}{\sigma_0^3}\expon{\frac{\log t_i-{X_i \beta_0}}{\sigma_0}}.
\end{eqnarray*}
It is easy to check that all parts go to 0 as $n\to \infty$ by calculating
\begin{eqnarray*}
&&\frac{\partial^2 q_n(\sigma, \beta)}{\partial\sigma\partial \beta}-\frac{\partial^2 q_n(\sigma, \beta)}{\partial\sigma\partial \beta}\Big|_{\sigma = \sigma_0,\beta=\beta_0}= \frac{1}{n}\sum_{i=1}^n (\sigma_0-\sigma)\frac{X_i\delta_i(\sigma_0+\sigma)}{\sigma^2\sigma_0^2}\\
&&-\frac{1}{n}\sum_{i=1}^n(\sigma_0-\sigma)\frac{X_i(\sigma_0+\sigma)}{\sigma^2\sigma_0^2}\expon{\frac{\log t_i-{X_i \beta}}{\sigma}}\\
&&-\frac{1}{n}\sum_{i=1}^n\frac{X_i}{\sigma_0^2}\expon{\frac{\log t_i-{X_i \beta_0}}{\sigma_0}}\left\{\expon{\frac{\log t_i-{X_i \beta}}{\sigma}-\frac{\log t_i-{X_i \beta_0}}{\sigma_0}}-1\right\}\\
&& -\frac{1}{n}\sum_{i=1}^n\left\{ \frac{X_i(\log t_i-X_i \beta)}{\sigma^3}-\frac{X_i(\log t_i-X_i \beta_0)}{\sigma_0^3}\  \right\}\expon{\frac{\log t_i-{X_i \beta}}{\sigma}}\\
&&-\frac{1}{n}\sum_{i=1}^n\frac{X_i(\log t_i-X_i \beta_0)}{\sigma_0^3}\expon{\frac{\log t_i-{X_i \beta_0}}{\sigma_0}}\left\{\expon{\frac{\log t_i-{X_i \beta}}{\sigma}-\frac{\log t_i-{X_i \beta_0}}{\sigma_0}}-1\right\},
\end{eqnarray*}
and using the same factorisation in $I_1$ leads to
$$
\frac{\partial^2 q_n(\sigma, \beta)}{\partial \sigma\partial\beta} - \frac{\partial^2 q_n(\sigma, \beta)}{\partial \sigma\partial\beta}\Big|_{\sigma = \sigma_0,\, \beta=\beta_0} \topb 0.
$$

Applying the same strategy to the term ${\partial h_n(y,z)}/{\partial z}$ also leads to a similar result. Combined with Eq.\eqref{eq: hn}, we have
$$
\boldsymbol x\frac{\partial h_n(\boldsymbol x)}{\partial \boldsymbol x} = y\frac{\partial h_n(y,z)}{\partial y}+ z\frac{\partial h_n(y,z)}{\partial z} \topb \boldsymbol x^\top \mathcal{H}(\boldsymbol\theta_0)\boldsymbol x =-\boldsymbol x^\top \mathcal{I}(\boldsymbol\theta_0)\boldsymbol x
$$
with Hessian matrix $\mathcal{H}(\boldsymbol\theta_0)$ and  positive definite Fisher information $\mathcal{I}(\boldsymbol\theta_0)$. Consequently, Eq.\eqref{eq: hn} is thus negative at the boundary $\norm{\boldsymbol x}=1$, which completes the proof.
\end{proof}}

\begin{proposition}[Mill's ratio]\label{Prop: MR}
    Let $\gamma(t)=\expon{- \sum\limits_{k=1}^L t^{1/\sigma_k}/\expon{\mu_k/\sigma_k} }, t>0.$ For  large $M>0$ and $\sigma_* = \min(\sigma_1, \ldots, \sigma_L)$, we have
    \begin{eqnarray*}
 \frac{\gamma(M)}{\sum\limits_{k=1}^L \zeta_k(M)}\left\{1-
      \frac{1/\sigma_*}{\sum\limits_{k=1}^L M\zeta_k(M)}\right\} \le \int_M^\infty  \gamma(t) \d t\le  \frac{\gamma(M)}{\sum\limits_{k=1}^L \zeta_k(M)} \left\{1+ \frac{1}{\sum\limits_{k=1}^L M\zeta_k(M)-1}\right\},
\end{eqnarray*}
where $\zeta_k(t) =  t^{1/\sigma_k-1}/(\sigma_k\expon{\mu_k/ \sigma_k})$.
\end{proposition}

\begin{proof} Recalling that for $\gamma(t)=\expon{- \sum\limits_{k=1}^L t^{1/\sigma_k}/\expon{\mu_k/\sigma_k} }$, the derivative
$$
\gamma'(t) = -\gamma(t){\sum\limits_{k=1}^L \frac{ t^{1/\sigma_k-1}}{\sigma_k\expon{\mu_k/ \sigma_k}}} =: -\gamma(t)\sum\limits_{k=1}^L \zeta_k(t),
$$
where $\zeta_k(t)>0,t>0$, and $t\cdot\zeta_k(t)=  t^{1/\sigma_k}/({\sigma_k\expon{\mu_k/ \sigma_k}})$ is a monotonically increasing function in $t>0$.
For the upper bound, we have for large $M>0$,
\begin{eqnarray*}
  \left\{\sum\limits_{k=1}^L M\zeta_k(M)-1\right\}\int_M^\infty  \gamma(t) \d t&\le& \int_M^\infty  -\left\{1-\sum\limits_{k=1}^L\frac{ t^{1/\sigma_k}}{\sigma_k\expon{\mu_k/ \sigma_k}}\right\}  \gamma(t) \d t\\
  &=&{-t\cdot \gamma(t)\bigg|_{M}^{\infty}}=M\cdot \gamma(M),
\end{eqnarray*}
yielding that, for large $M>0$ such that $\sum_{l=1}^M \zeta_k(M) >1/M$
\begin{eqnarray*}
 \int_M^\infty  \gamma(t) \d t\le \frac{\gamma(M)}{\sum\limits_{k=1}^L \zeta_k(M)-\frac{1}{M}} =\frac{\gamma(M)}{\sum\limits_{k=1}^L \zeta_k(M)}\left\{1+ \frac{1}{\sum\limits_{k=1}^L M\zeta_k(M)-1}\right\}.
\end{eqnarray*}

Next, we will show the lower bound. Recall $\sigma_*= \min(\sigma_1,\ldots,\sigma_L)$, we consider
\begin{eqnarray*}
&&\Bigg\{1+\frac{ {1}/{\sigma_*}  }{\sum\limits_{k=1}^LM\zeta_k(M)  } \Bigg\}\int_M^\infty  \gamma(t) \d t
\ge \int_M^\infty \Bigg\{1+\frac{{1}/{\sigma_*}}{\sum\limits_{k=1}^Lt\zeta_k(t)  } \Bigg\} \gamma(t) \d t\\
&& = \int_M^\infty \Bigg\{1+\frac{\frac{1}{\sigma_*}\sum\limits_{k=1}^Lt\zeta_k(t)}{\left(\sum\limits_{k=1}^Lt\zeta_k(t)\right)^2} \Bigg\} \gamma(t) \d t\ge\int_M^\infty \Bigg\{1+\frac{\sum\limits_{k=1}^L \frac{1}{\sigma_k}t\zeta_k(t)}{\left(\sum\limits_{k=1}^Lt\zeta_k(t)\right)^2} \Bigg\} \gamma(t) \d t\\
&& =\int_M^\infty \Bigg\{1+\frac{\sum\limits_{k=1}^L \frac{1}{\sigma_k} \frac{ \zeta_k(t)}{t}}{\left(\sum\limits_{k=1}^L\zeta_k(t)  \right)^2} \Bigg\} \gamma(t) \d t\ge\int_M^\infty \Bigg\{1+\frac{\sum\limits_{k=1}^L (\frac{1}{\sigma_k}-1) \frac{ \zeta_k(t)}{t}}{\left(\sum\limits_{k=1}^L\zeta_k(t)  \right)^2} \Bigg\} \gamma(t) \d t\\
&&  = \int_{M}^{\infty} \frac{-\gamma'(t){\sum\limits_{k=1}^L \zeta_k(t)}}{\left\{\sum\limits_{k=1}^L\zeta_k(t)  \right\}^2}
  +\frac{\gamma({t})\sum\limits_{k=1}^L (\frac{1}{\sigma_k}-1) \frac{ \zeta_k(t)}{t}}{\Bigg\{\sum\limits_{k=1}^L\zeta_k(t)  \Bigg\}^2} \d t= -\frac{\gamma(t)}{\sum\limits_{k=1}^L \zeta_k(t) }\bigg|_{M}^{\infty}=\frac{\gamma(M)}{\sum\limits_{k=1}^L \zeta_k(M)}.
\end{eqnarray*}

Thus,
\begin{eqnarray*}
     \int_M^\infty  \gamma(t) \d t&\ge& \frac{\gamma(M)}{\sum\limits_{k=1}^L \zeta_k(M)} \left\{1+
      \frac{1}{\sigma_*M\sum\limits_{k=1}^L \zeta_k(M)}\right\}^{-1}
     \ge \frac{\gamma(M)}{\sum\limits_{k=1}^L \zeta_k(M)} \left\{1-
      \frac{1}{\sigma_*M\sum\limits_{k=1}^L \zeta_k(M)}\right\}
\end{eqnarray*}
using $1/(1+x) \ge 1-x$ for $x>0$. We complete the proof of Proposition \ref{Prop: MR}.
\end{proof}

\subsection*{A.4 Characterisation of Competing Log-normal Model}\label{sec: pfln}
For the competing log-normal model, the likelihood function is given by
\begin{eqnarray*}
    L(\boldsymbol\theta) &=& \prod_{i=1}^n f^{\delta_i}(t_i; \boldsymbol\theta,  X_i) S^{1-\delta_i}(t_i; \boldsymbol\theta,  X_i)  \\
     &=& \prod_{i=1}^n \left\{\sum\limits_{ l =1}^L \frac{1}{\sigma_lt_i}\phi\fracl{ \log t_i - \alpha_l- X_{il}^\top \boldsymbol{\beta}_l}{\sigma_l}\prod_{k \ne l} \bar{\Phi}\fracl{ \log t_i - \alpha_k- X_{ik}^\top \boldsymbol{\beta}_k}{\sigma_k} \right\}^{\delta_i}\\
     &&\times\left\{\prod_{ l =1}^L \bar{\Phi}\fracl{ \log t_i - \alpha_l- X_{il}^\top \boldsymbol{\beta}_l}{\sigma_l}\right\}^{1-\delta_i},
\end{eqnarray*}
leading thus to the log-likelihood
\begin{eqnarray*}
    \ell(\boldsymbol\theta)  &=& \sum\limits_{i=1}^n \delta_i \log \left\{\sum\limits_{ l =1}^L \frac{1}{\sigma_lt_i}\phi\fracl{ \log t_i - \alpha_l- X_{il}^\top \boldsymbol{\beta}_l}{\sigma_l}\prod_{k \ne l} \bar{\Phi}\fracl{\log t_i - \alpha_k- X_{ik}^\top \boldsymbol{\beta}_k}{\sigma_k} \right\} \\
    &&+ \sum\limits_{i=1}^n(1-\delta_i) \sum\limits_{ l =1}^L \log \left\{ \bar{\Phi}\fracl{ \log t_i - \alpha_l- X_{il}^\top \boldsymbol{\beta}_l}{\sigma_l} \right\}\\
     &=& \sum\limits_{i=1}^n \delta_i \log \left\{\sum\limits_{l=1}^L\frac{ \frac{1}{\sigma_lt_i}\phi\fracl{ \log t_i - \alpha_l- X_{il}^\top \boldsymbol{\beta}_l}{\sigma_l}}{ \bar{\Phi}\fracl{ \log t_i - \alpha_l- X_{il}^\top \boldsymbol{\beta}_l}{\sigma_l} }\right\}+ \sum\limits_{i=1}^n\sum_{l=1}^L \log\left\{\bar{\Phi}\fracl{ \log t_i - \alpha_l- X_{il}^\top \boldsymbol{\beta}_l}{\sigma_l} \right\}\\
     &=:& \sum\limits_{i=1}^n\log g(T_i|X_i; \boldsymbol\theta,\delta_i),
\end{eqnarray*}
which is differentiable with respect to $\boldsymbol\theta$.

In the following, we only need to show that $|\log  g(T_i|X_i; \boldsymbol\theta,\delta_i)|$ is bounded by $G(T_i,X_i)$ with $\E{h(T_i,X_i)}<\infty$, and
\begin{eqnarray*}
   \E{\left|\frac{\partial}{\partial \boldsymbol\theta}\log g(T_i|X_i; \boldsymbol\theta, \delta_i)\right|} < \infty ,\quad
  \E{\left|\frac{\partial^2}{\partial \boldsymbol\theta\partial\boldsymbol\theta^\top}\log g(T_i|X_i; \boldsymbol\theta, \delta_i)\right|} < \infty.
\end{eqnarray*}
Firstly, we show the upper bound by
\begin{eqnarray*}
   \log g(T_i|X_i; \boldsymbol\theta,\delta_i) &=&   \delta_i \log \left\{\sum\limits_{ l =1}^L \frac{1}{\sigma_lT_i}\phi\fracl{ \log T_i - \alpha_l- X_{il}^\top \boldsymbol{\beta}_l}{\sigma_l}\prod_{k \ne l} \bar{\Phi}\fracl{\log T_i - \alpha_k- X_{ik}^\top \boldsymbol{\beta}_k}{\sigma_k} \right\} \\
    &&+ (1-\delta_i) \sum\limits_{ l =1}^L \log \left\{ \bar{\Phi}\fracl{ \log T_i - \alpha_l- X_{il}^\top \boldsymbol{\beta}_l}{\sigma_l} \right\}\\
    &\le& \delta_i \log \left(\sum\limits_{ l =1}^L \frac{1}{\sigma_lT_i}\phi(0) \right) = \delta_i \log \left(\sum\limits_{ l =1}^L \frac{1}{T_i\sqrt{2\pi}\sigma^2_l} \right).
\end{eqnarray*}
By Jensen's inequality and the concavity of the logarithm function, we have the lower bound
\begin{eqnarray*}
&&\log \fracl{1}{L} +\log g(T_i|X_i; \boldsymbol\theta, \delta_i)\\
    &&= \log \left[\frac{1}{L} \left\{\sum\limits_{ l =1}^L \frac{1}{\sigma_lT_i}\phi\fracl{ \log T_i - \alpha_l- X_{il}^\top \boldsymbol{\beta}_l}{\sigma_l}\prod_{k \ne l} \bar{\Phi}\fracl{ \log T_i - \alpha_k- X_{ik}^\top \boldsymbol{\beta}_k}{\sigma_k} \right\}^{\delta_i}\right]\\
    && \quad + \log\left[\left\{\prod_{ l =1}^L \bar{\Phi}\fracl{ \log T_i - \alpha_l- X_{il}^\top \boldsymbol{\beta}_l}{\sigma_l}\right\}^{1-\delta_i}\right],\\
&&\ge \log \left[\frac{1}{L^2} \sum\limits_{l =1}^L \left\{\frac{1}{\sigma_lT_i}\phi\fracl{ \log T_i - \alpha_l- X_{il}^\top \boldsymbol{\beta}_l}{\sigma_l}\prod_{k \ne l} \bar{\Phi}\fracl{ \log T_i - \alpha_k- X_{ik}^\top \boldsymbol{\beta}_k}{\sigma_k} \right\}^{\delta_i}\right]\\
&& \quad +  (1-\delta_i) \sum\limits_{ l =1}^L \log \left\{ \bar{\Phi}\fracl{ \log T_i - \alpha_l- X_{il}^\top \boldsymbol{\beta}_l}{\sigma_l} \right\},\\
&&\ge \log \frac{1}{L} +  \frac{1}{L}\sum\limits_{l=1}^L \delta_i\log \left\{ \frac{1}{\sigma_lT_i}\phi\fracl{ \log T_i - \alpha_l- X_{il}^\top \boldsymbol{\beta}_l}{\sigma_l}\prod_{k \ne l} \bar{\Phi}\fracl{\log T_i - \alpha_k- X_{ik}^\top \boldsymbol{\beta}_k}{\sigma_k}\right\}\\
&& \quad+ (1-\delta_i) \sum\limits_{ l =1}^L \log \left\{ \bar{\Phi}\fracl{ \log T_i - \alpha_l- X_{il}^\top \boldsymbol{\beta}_l}{\sigma_l} \right\}\\
&& = \log \frac{1}{L} +\frac{1}{L}\sum\limits_{l=1}^Lg_l (T_i|X_i; \boldsymbol\theta_l, \delta_i).
\end{eqnarray*}
To show the existence of a bounded function, it is sufficient to show that for the second term in the formula above (i.e., the case $\delta_i=1$)
\begin{eqnarray*}
&& \abs{ \log \left\{ \frac{1}{\sigma_lT_i}\phi\fracl{ \log T_i - \alpha_l- X_{il}^\top \boldsymbol{\beta}_l}{\sigma_l}\right\}+ \sum_{k \ne l}  \log\bar{\Phi}\fracl{\log T_i - \alpha_k- X_{ik}^\top \boldsymbol{\beta}_k}{\sigma_k}}\\
&&\le \left|\log(\sqrt{2\pi}\sigma_l^2T_i)\right|+\frac{ (\log T_i - \alpha_l- X_{il}^\top \boldsymbol{\beta}_l)^2}{2\sigma_l^2} + \sum_{k \ne l}  \left|\log\bar{\Phi}\fracl{\log T_i - \alpha_k- X_{ik}^\top \boldsymbol{\beta}_k}{\sigma_k}\right|\\
 &&=: G(T_i,X_i)
\end{eqnarray*}
with $\E{G(T_i,X_i)|}<\infty$. Under the assumptions \textit{A1} and \textit{A2}, the proof is straightforward using Mill's ratio for Gaussian distribution and Jensen's inequality, as outlined in the proof of \citet{Cui2021Max}.

Defining $S_{il} = (\log T_i - \alpha_l- X_{il}^\top \boldsymbol{\beta}_l)/\sigma_l$, the first-order derivatives are bounded by
\begin{eqnarray*}
    \left|\frac{\partial \log  g(T_i|X_i; \boldsymbol\theta, 1)}{\partial (\sigma^2_l)}\right| &\le& \left|\frac{\frac{S_{il}^2-1}{2\sigma^3_l}\phi(S_{il})\prod\limits_{k \ne l} \bar{\Phi}(S_{ik}) }{\sum\limits_{ l =1}^L \frac{1}{\sigma_l}\phi(S_{il})\prod\limits_{k \ne l} \bar{\Phi}(S_{ik}) }\right| + \left|\frac{\sum_{r\ne l}\frac{1}{\sigma_r}\phi(S_{ir})\frac{S_{il}}{2\sigma^2_l}\phi(S_{il})\prod_{s \ne l,r} \bar{\Phi}(S_{is}) }{\sum\limits_{l =1}^L \frac{1}{\sigma_l}\phi(S_{il})\prod\limits_{k \ne l} \bar{\Phi}(S_{ik}) }\right|, \\
    &\le& \left|\frac{S_{il}^2-1}{2\sigma^2_l}\right| + \left|\frac{S_{il}}{2\sigma_l}\sum_{r\ne l}\frac{\phi(S_{ir})}{\bar{\Phi}(S_{ir})}\right|\\
    \left|\frac{\partial \log  g(T_i|X_i; \boldsymbol\theta, 1)}{\partial \alpha_l}\right| &\le& \left|\frac{\frac{S_{il}}{\sigma^2_l}\phi(S_{il})\prod\limits_{k \ne l} \bar{\Phi}(S_{ik}) }{\sum\limits_{ l =1}^L \frac{1}{\sigma_l}\phi(S_{il})\prod\limits_{k \ne l} \bar{\Phi}(S_{ik}) }\right| + \left|\frac{\sum_{r\ne l}\frac{1}{\sigma_l\sigma_r}\phi(S_{ir})\phi(S_{il})\prod_{s \ne l,r} \bar{\Phi}(S_{is}) }{\sum\limits_{ l =1}^L \frac{1}{\sigma_l}\phi(S_{il})\prod\limits_{k \ne l} \bar{\Phi}(S_{ik}) }\right|\\
    &\le& \left|\frac{S_{il}}{\sigma_l} \right| + \left|\sum_{r\ne l}\frac{\phi(S_{ir})}{\sigma_r\bar{\Phi}(S_{ir})}\right|.
\end{eqnarray*}
By the symmetry of the standard normal density, we have $\phi(x)/\bar{\Phi}(x) = \phi(-x)/\Phi(-x) \le C(1 + |x|)$ for all $x \in \R$ and some constant $C$, following an argument analogous to that in \citet{Cui2021Max}. This establishes the boundedness condition, and the existence of second-order derivatives follows similarly (details omitted).
\begin{eqnarray*}
    \left|\frac{\partial \log^2  g(T_i|X_i; \boldsymbol\theta, 1)}{\partial (\alpha_l)^2}\right|
    &\le& \left|\frac{S^2_{il}-1}{\sigma^2_l} \right| + \left|\frac{S_{il}}{\sigma_l} \right|\sum_{r\ne l}\left|\frac{C(1+S_{ir})}{\sigma_r}\right| + \left(\frac{1}{\sigma_l}+\sum_{r\ne l}\left|\frac{C(1+S_{ir})}{\sigma_r}\right|\right)^2,\\
    \left|\frac{\partial \log^2  g(T_i|X_i; \boldsymbol\theta, 1)}{\partial (\alpha_l)\partial (\alpha_j)}\right|
    &\le& \left|\frac{S_{i l}}{\sigma_l\sigma_j}  C\left(1+S_{i j}\right)\right|+\left|\frac{S_{ij}}{\sigma_j^2} C\left(1+S_{i j}\right)\right|+\sum_{r \neq l, j} \frac{C\left(1+\left|S_{i r}\right|\right)\left(1+\left|S_{i j}\right|\right)}{\sigma_r\sigma_j}  \\
&& +\left(\frac{\left|S_{i l}\right|}{\sigma_l}+\sum_{r \neq l} \frac{1}{\sigma_r} C\left(1+\left|S_{i r}\right|\right)\right)\left(\frac{\left|S_{i j}\right|}{\sigma_j}+\sum_{r \neq j} \frac{1}{\sigma_r} C\left(1+\left|S_{i r}\right|\right)\right),\\
    \left|\frac{\partial \log^2  g(T_i|X_i; \boldsymbol\theta, 1)}{\partial (\sigma_l^2)\partial (\sigma^2_l)}\right|
    & \leq& \frac{1}{4 \sigma_l^4}\left|S_{i l}^4-6 S_{i l}^2+3\right|+\frac{1}{4 \sigma_l^3}\left|S_{i l}^3-3 S_{i l}\right| \sum_{r \neq l} \frac{1}{\sigma_r} C\left(1+\left|S_{i r}\right|\right) \\
&& +\left(\frac{\left|S_{i l}^2-1\right|}{2 \sigma_l^2}+\frac{\left|S_{i l}\right|}{2 \sigma_l} \sum_{r \neq l} \frac{1}{\sigma_r} C\left(1+\left|S_{i r}\right|\right)\right)^2 .\\
\end{eqnarray*}
\begin{eqnarray*}
\left|\frac{\partial \log^2  g(T_i|X_i; \boldsymbol\theta, 1)}{\partial (\sigma_l^2)\partial (\sigma^2_j)}\right|
    & \leq& \frac{1}{2 \sigma_l^2}\left|S_{i l}^2-1\right| \frac{\left|S_{i j}\right|}{2 \sigma_j^2} C\left(1+\left|S_{i j}\right|\right)+\frac{1}{2 \sigma_j^2}\left|S_{i j}^2-1\right| \frac{\left|S_{i l}\right|}{2 \sigma_l} C\left(1+\left|S_{i j}\right|\right) \\
&& +\sum_{r \neq l, j} \frac{1}{\sigma_r} C\left(1+\left|S_{i r}\right|\right) \frac{\left|S_{i j}\right|}{2 \sigma_j^2} C\left(1+\left|S_{i j}\right|\right) \frac{\left|S_{i l}\right|}{2 \sigma_l} \\
&& +\left(\frac{\left|S_{i l}^2-1\right|}{2 \sigma_l^2}+\frac{\left|S_{i l}\right|}{2 \sigma_l} \sum_{r \neq l} \frac{1}{\sigma_r} C\left(1+\left|S_{i r}\right|\right)\right)\\
&&\times\left(\frac{\left|S_{i j}^2-1\right|}{2 \sigma_j^2}+\frac{\left|S_{i j}\right|}{2 \sigma_j} \sum_{r \neq j} \frac{1}{\sigma_r} C\left(1+\left|S_{i r}\right|\right)\right)\\
 \left|\frac{\partial \log^2  g(T_i|X_i; \boldsymbol\theta, 1)}{\partial (\sigma_l^2)\partial (\alpha_j)}\right|
& \leq& \frac{1}{2 \sigma_l^3}\left|S_{i l}^3-3 S_{i l}\right|+\frac{1}{2 \sigma_l^2}\left|S_{i l}^2-1\right| \sum_{r \neq l} \frac{1}{\sigma_r} C\left(1+\left|S_{i r}\right|\right) \\
&& +\left(\frac{1}{2 \sigma_l^2}\left|S_{i l}^2-1\right|+\sum_{r \neq l} \frac{\left|S_{i l}\right|}{2 \sigma_l} \frac{1}{\sigma_r} C\left(1+\left|S_{i r}\right|\right)\right)\\
&&\times\left(\frac{\left|S_{i l}\right|}{\sigma_l}+\sum_{r \neq l} \frac{\left|S_{i l}\right|}{\sigma_l} \frac{1}{\sigma_r} C\left(1+\left|S_{i r}\right|\right)\right) .
\end{eqnarray*}

\section*{Appendix B}

In this section, we apply the competing parametric survival model to three additional real-world applications: (1) evaluating biomarker combinations in Alzheimer’s disease (B.1); (2) examining hepatocellular carcinoma (B.2); and (3) analysing recurrence-free time of breast cancer patients (B.3). The modifications applied to the disease mixture model \citep{moreno2017survival} to enable a fair comparison and the diagnostic plots for all applications are presented in B.4.

\subsection*{B.1 Alzheimer's Disease with ADNI}
Alzheimer’s disease is a neurodegenerative disorder, usually characterised by prominent amnestic cognitive impairment and short-term memory difficulties, and is the most common cause of dementia \citep{knopman2021alzheimer}. Definitive diagnosis of Alzheimer’s disease is only possible through autopsy, while clinical diagnoses are categorised as "possible" or "probable" based on available findings. Although the amyloid beta hypothesis is the predominant explanation, the exact causes of Alzheimer’s disease remain poorly understood \citep{burns2009alzheimer}. As a result, ongoing research focuses on developing diagnostic algorithms to identify the most informative biomarker combinations and to determine their optimal ordering for clinical situations  \citep{chetelat2020amyloid}.

We collated the data from four phases of the ADNI study (ADNI-1, ADNI-GO, ADNI-2, and ADNI-3) \citep{mueller2005alzheimer}, a longitudinal study designed to detect early Alzheimer’s disease signs and track disease progression using biomarkers. Since its launch in 2004, ADNI has recruited approximately 2,000 participants aged 55--90 across 57 sites in the United States and Canada, including individuals with dementia due to Alzheimer’s disease, those with mild memory problems known as mild cognitive impairment \citep[MCI;][]{petersen2004mild}, and healthy elderly controls.

To ensure data completeness, our analysis includes 751 participants diagnosed with MCI at baseline, of whom 289 progressed to dementia during follow-up. Eleven features, measured at baseline, were selected (Table \ref{tab: var}) based on prior evidence from relevant studies \citep{li2017prediction,li2018prognostic, Zhou2024joint,Kang2022joint}, and preliminary analyses using Weibull and Cox proportional hazards (PH) models across all available features (with a significance level $<10\%$). These features span demographic characteristics, genetic information, cognitive assessments, neuropathological markers, and brain imaging measures.

\begin{table}[H]
\centering
\caption{Description and summary of the features selected by Weibull and Cox PH models (with a significance level $<10\%$) in the Alzheimer's disease study.}
\resizebox{\linewidth}{!}{
\begin{tabular}{*{3}{clc}}
\toprule
Features & Description &  Mean (range)/Counts (percentage) \\
\midrule
\multirow{2}*{APOE$\epsilon$4} & Number of the apolipoprotein E$\epsilon$4 alleles (0, 1, or 2). Proceeded  & One: 291 (38.7\%) \\
 &\cellcolor[gray]{0.85}into two indicators (APOE1, APOE2) &\cellcolor[gray]{0.85} Two: 84 (11.2\%) \\
\midrule
ADAS13 & Score from 13-item Alzheimer's Disease Assessment Scale (ADAS) at baseline & 16.19 (0.67 to 39.67) \\
\cmidrule{2-3}
RAVLT & \cellcolor[gray]{0.85}Total number of words recalled across the 5 learning trials of the& \cellcolor[gray]{0.85} 34.62  \\
(immediate) &   Rey Auditory Verbal Learning Test (RAVLT) at baseline & (11 to 68) \\
\cmidrule{2-3}
RAVLT & \cellcolor[gray]{0.85}The difference between the number of words recalled in two trials & \cellcolor[gray]{0.85} 4.21  \\
(learning) &   (Trial 5 $-$ Trial 1) of the RAVLT at baseline & ($-$3 to 12) \\
\cmidrule{2-3}
{FAQ} & \cellcolor[gray]{0.85}Total score of Functional Activities Questionnaire (FAQ) at baseline \cellcolor[gray]{0.85} & \cellcolor[gray]{0.85} 3.17 (0 to 24)  \\
 \midrule
GenderM & Participant gender (1 for male, 0 for female) & Male: 443 (52.6\%) \\
\cmidrule{2-3}
Education & \cellcolor[gray]{0.85}Participant's years of education at baseline& \cellcolor[gray]{0.85} 15.94 (6 to 20) \\
\cmidrule{2-3}
Age & Participant age (in years) at baseline & 72.51 (55 to 88) \\
\midrule
{Hippocampus} & \cellcolor[gray]{0.85}Total volume of the bilateral hippocampus at baseline, measured at baseline (in mm$^3$) & \cellcolor[gray]{0.85} 6,829.76 (3,281 to 9,929) \\
\cmidrule{2-3}
\multirow{2}*{ICV} & Intracranial volume (ICV) at baseline derived using UCSF's methodology, & {1,536,780 } \\
 &   \cellcolor[gray]{0.85}measured at baseline (in mm$^3$) from T1-weighted MRI scans  & \cellcolor[gray]{0.85} (1,164,940 to 21,10,290) \\
\cmidrule{2-3}
\multirow{2}*{MidTemp} &Total volume of the bilateral middle temporal gyrus at baseline using  & 19,737.63  \\
 & \cellcolor[gray]{0.85}UCSF's methodology, measured at baseline (in mm$^3$)  &\cellcolor[gray]{0.85}  (10,472 to 28,682) \\
\bottomrule
\end{tabular} }
\label{tab: var}
\end{table}

\begin{figure}[H]
    \centering
    \subfloat[]{\includegraphics[width=0.31\linewidth]{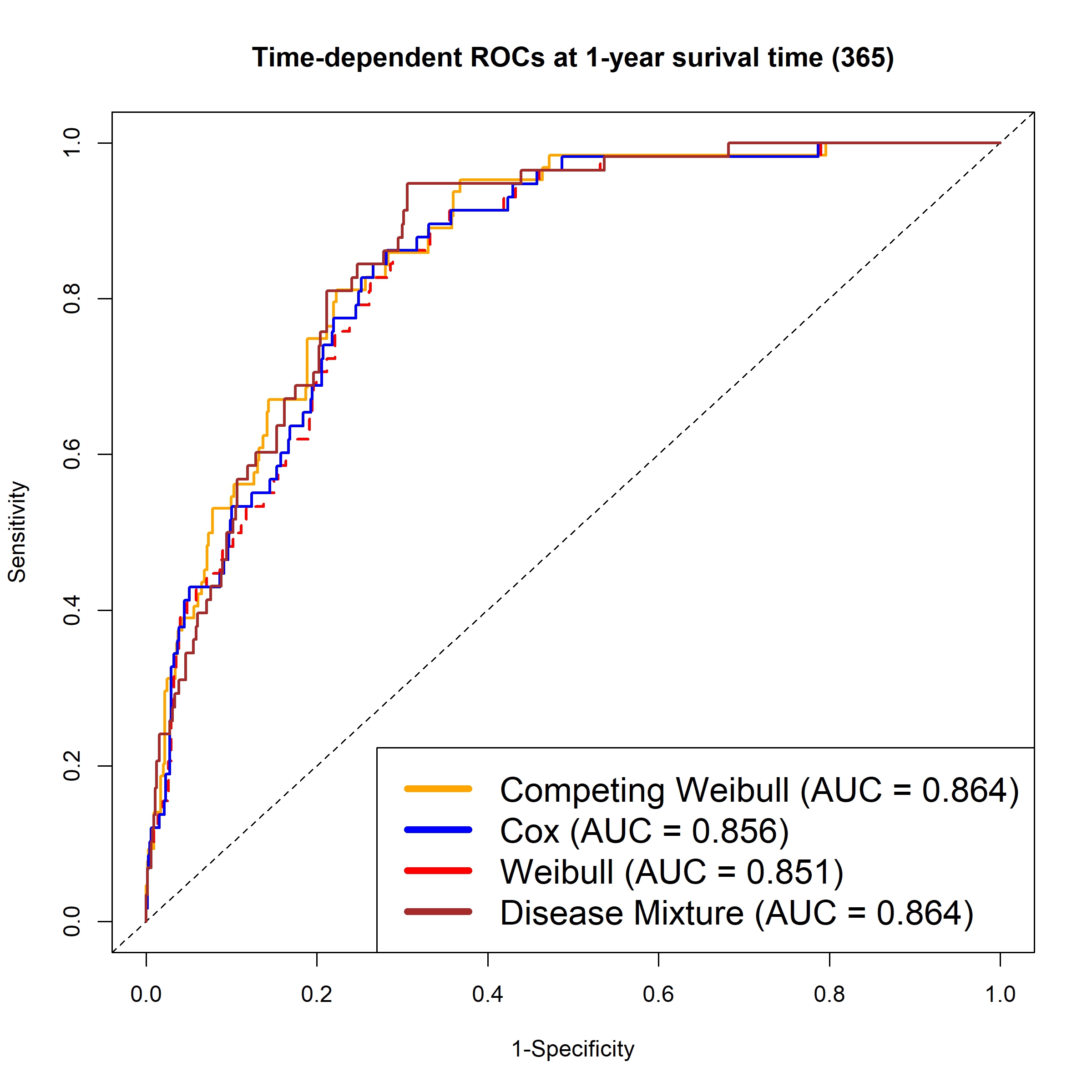}}
    \subfloat[]{\includegraphics[width=0.42\linewidth]{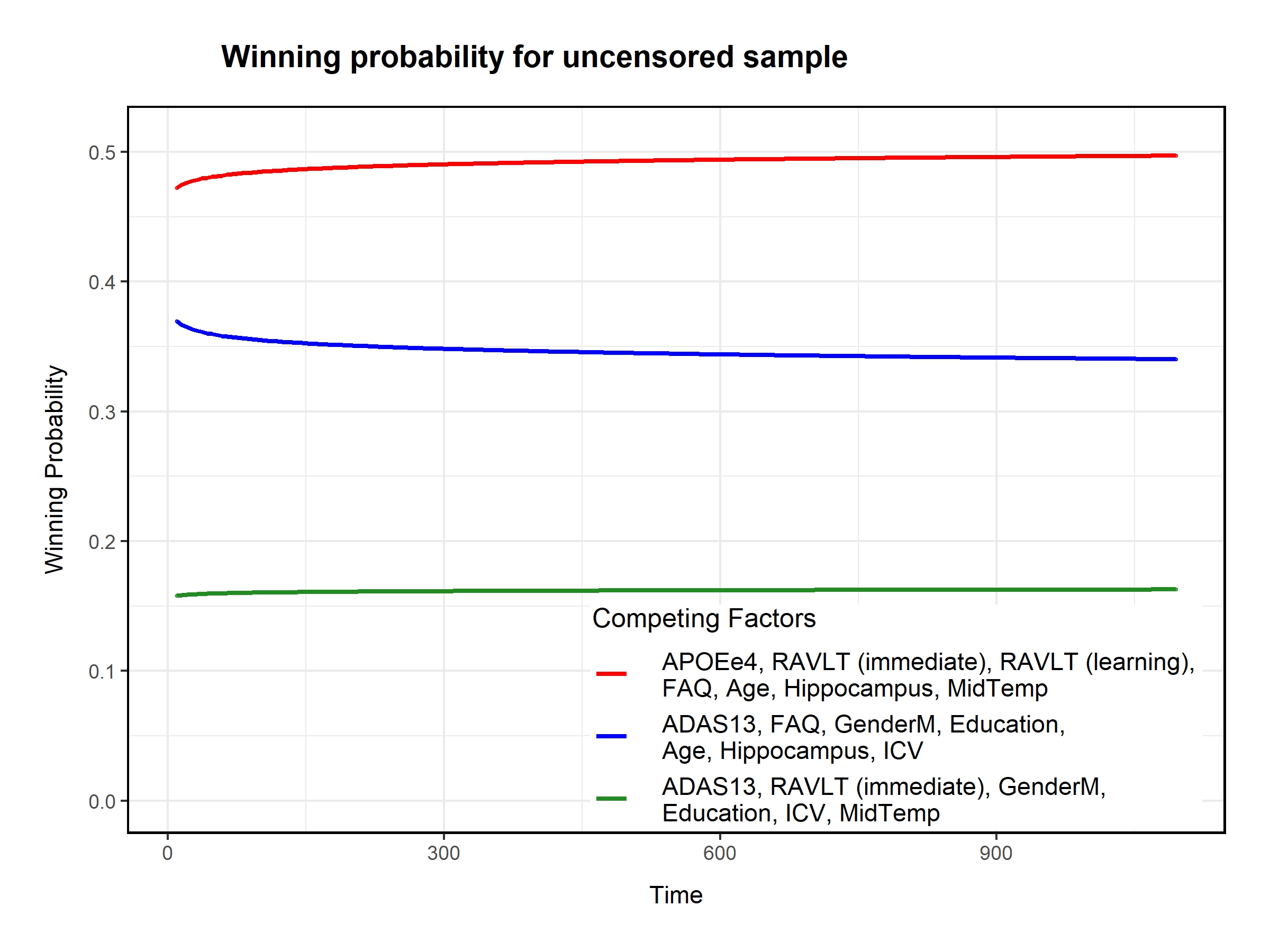}}
    \subfloat[]{\includegraphics[width=0.25\linewidth]{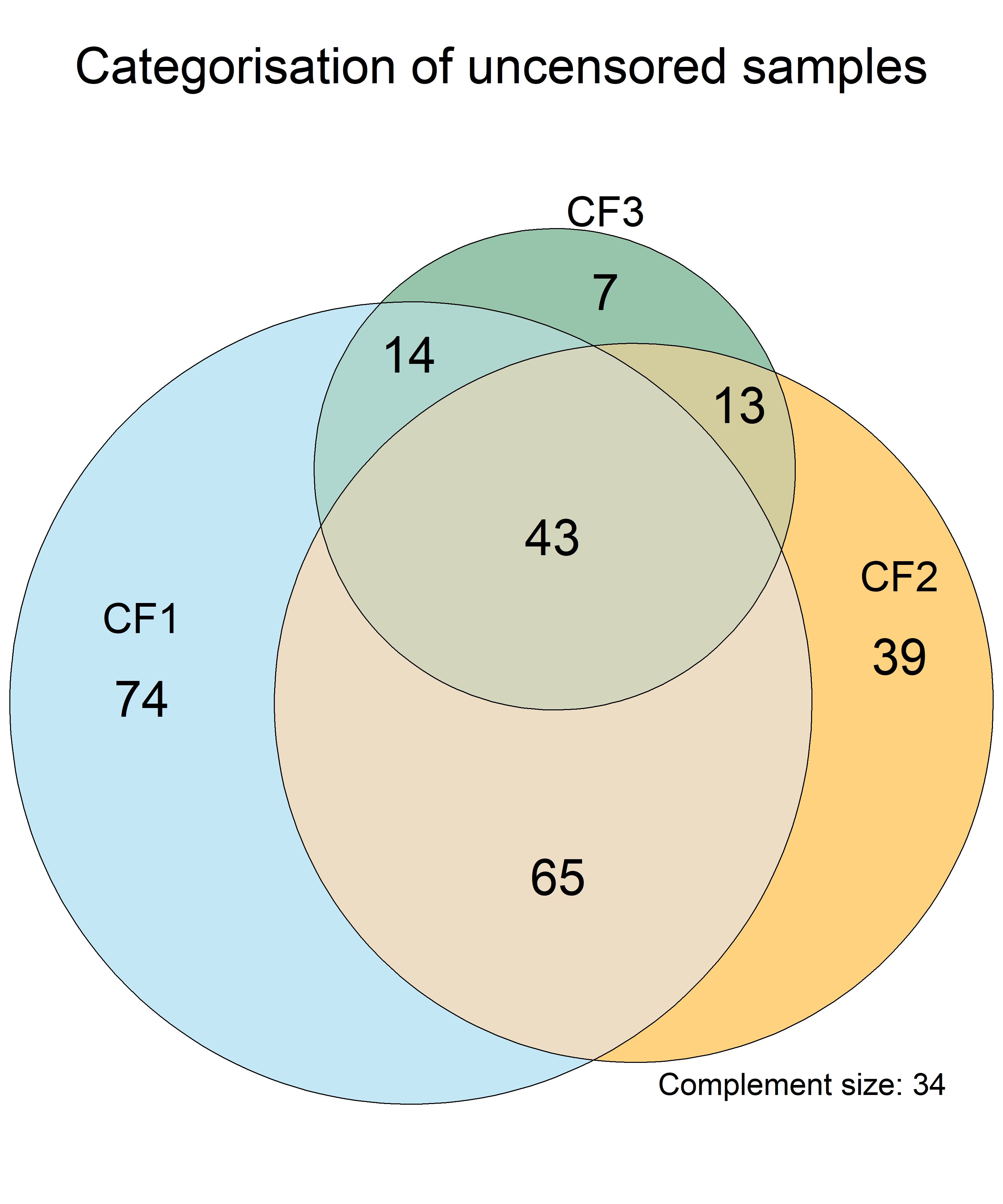}} \\
    \subfloat[]{\includegraphics[width=0.4\linewidth]{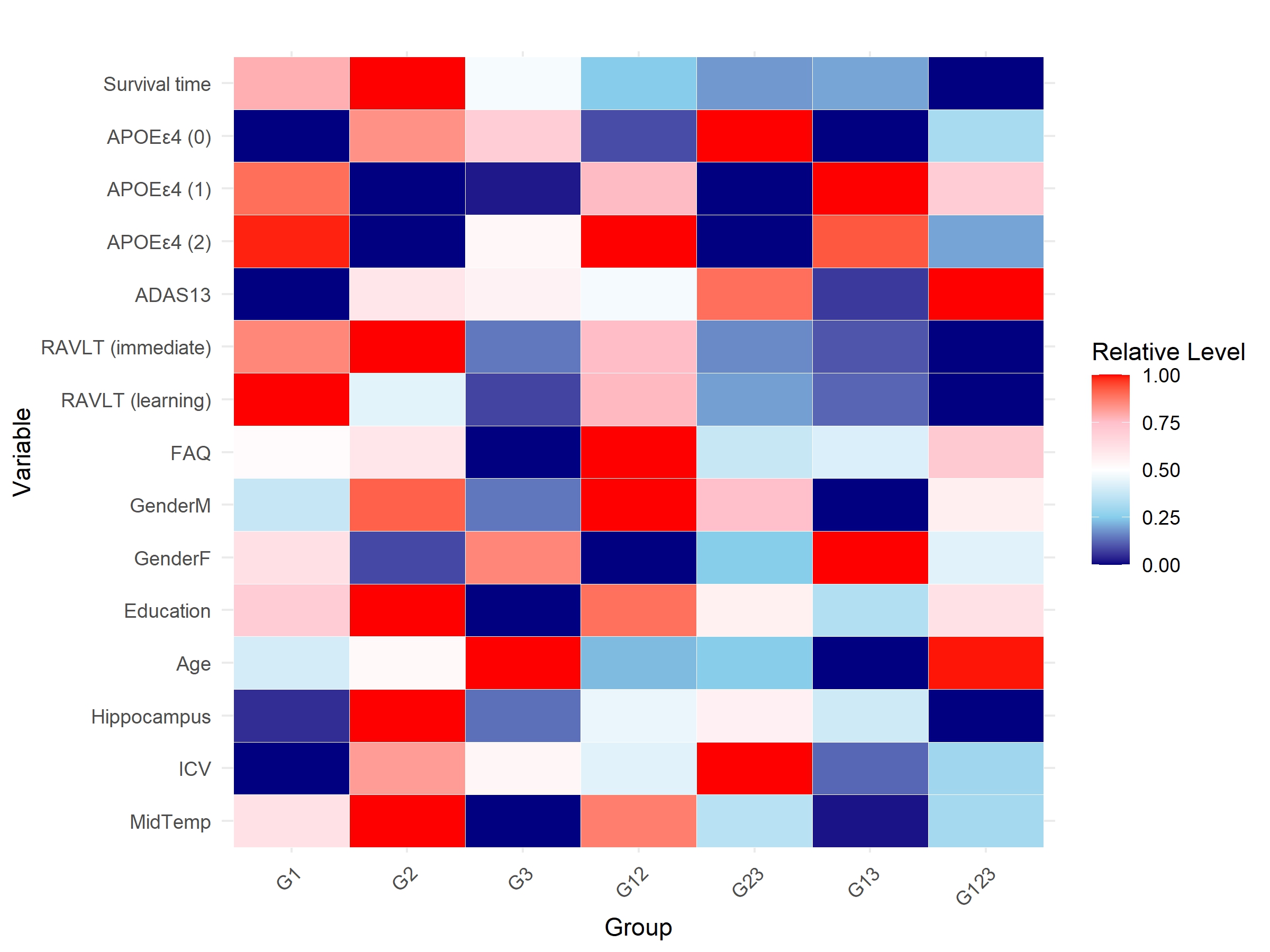}}
    \subfloat[]{\includegraphics[width=0.35\linewidth]{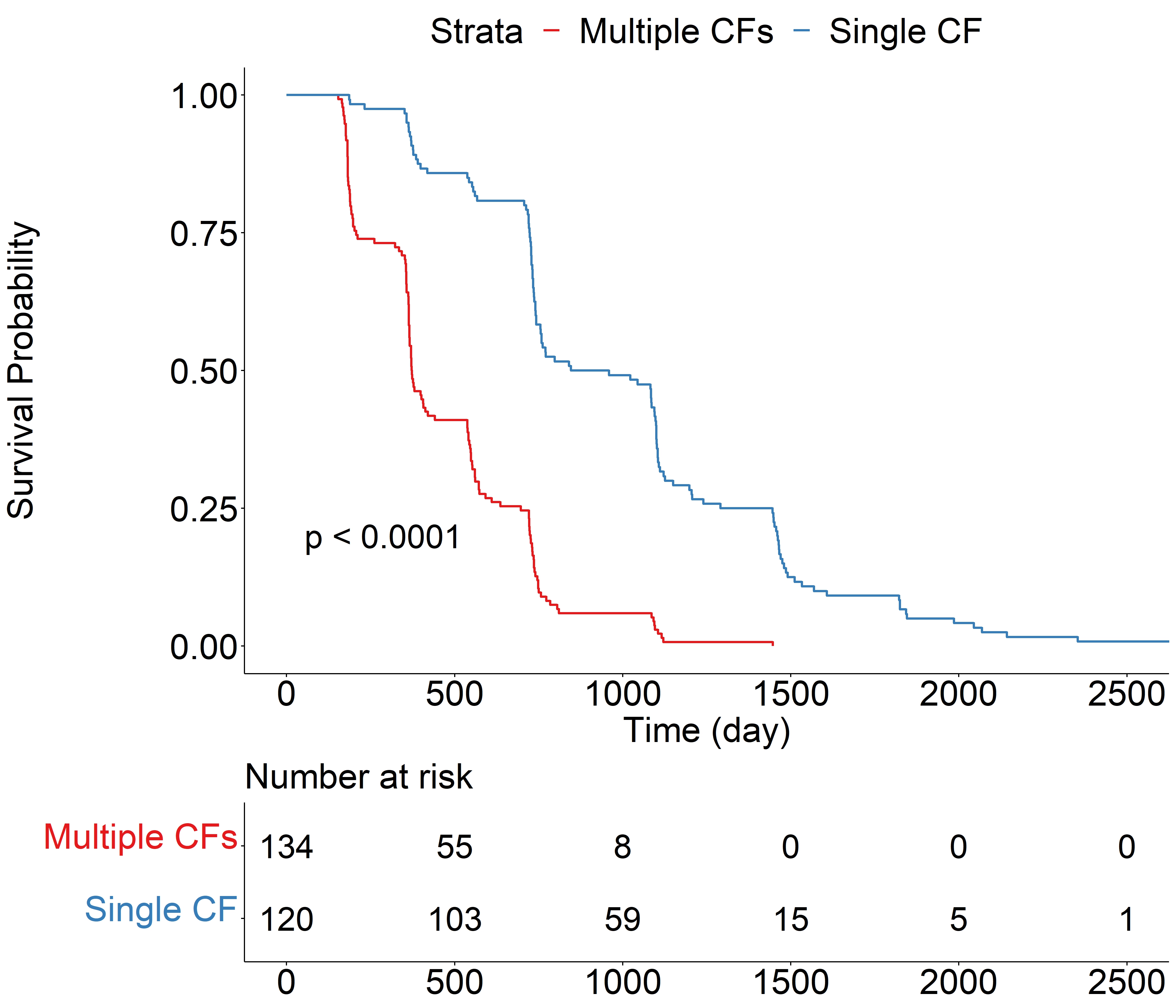}}
    \caption{Alzheimer’s disease study results: (a) Time-dependent ROC curves at the 1-year survival horizon; (b) Time-varying winning probabilities for uncensored samples; (c) Categorisation of uncensored samples based on IPW attributed survival times below the median; (d) Heatmap of relative sample characteristics across categories, obtained by averaging within each category and scaling across all categories (e.g., G123 denotes the 43 samples classified by all three CFs); (e) Kaplan-Meier survival curves stratified by single-CF group (G1, G2, G3) and multiple-CFs (G12, G23, G13, G123).}
    \label{fig: ADNI_ROC_WP}
\end{figure}

We employ a man-machine collaborative strategy \citep{zhang2021five} to determine the feature allocation to three predetermined competing factors (CFs). Guided by preliminary analyses, the first CF incorporates biomarkers with strong significance, while the remaining features are grouped to capture alternative patterns that may characterise smaller subpopulations. The resulting CFs are as follows: (1) APOE$\epsilon$4, RAVLT (immediate), RAVLT (learning), FAQ, Age, Hippocampus, MidTemp; (2) ADAS13, FAQ, Gender, Education, Age, Hippocampus, ICV; (3) ADAS13, RAVLT (immediate), Gender, Education, ICV, MidTemp. Tuning parameters were selected as $\lambda_1=1$ and $\lambda_2=0.2$ via grid search over $\lambda_1\in[0,2]$ and $\lambda_2\in[0,1]$. This configuration achieved better performance in both C-index and iAUC (Table \ref{tab: ADNI_est}), with stronger discrimination for shorter survival times, as shown by the time-dependent ROC curves at the 1-year survival time (Figure \ref{fig: ADNI_ROC_WP}(a)).

\begin{table}[H]
\centering
\caption{Estimated coefficients with mean (SE), along with C-index and iAUC, from the competing Weibull model (survival time), Weibull model (survival time), Cox PH model (hazard) and disease mixture model (hazard) in Alzheimer's disease study. The notation "-" means exclusion from the model.}
\resizebox{0.95\textwidth}{!}{
\begin{threeparttable}
\begin{tabular}{lrrrrrrr}
\toprule
\multirow{2}*{Features} &  \multicolumn{3}{c}{Competing Weibull} & \multicolumn{1}{c}{\multirow{2}*{
Weibull}} & \multicolumn{1}{c}{\multirow{2}*{Cox PH}} & \multicolumn{2}{c}{Disease Mixture}\\
\cmidrule{2-4} \cmidrule{7-8}
 &  \multicolumn{1}{c}{CF1} &  \multicolumn{1}{c}{CF2} &  \multicolumn{1}{c}{CF3} &  & &  \multicolumn{1}{c}{State 1 (CF1)} &  \multicolumn{1}{c}{State 0 (CF2 \& CF3)}\\
\midrule
Intercept   & 9.302 (0.209)  & 8.642 (0.175)  & 9.431 (0.405)  & 8.240 (0.093)  & - & -& - \\
\rowcolor[gray]{0.85} APOE1    & $-$1.143 (0.225) & -              & -              & $-$0.488 (0.088) &  0.661 (0.135) & 1.286 (0.201)& -\\
APOE2    & $-$1.356 (0.255) & -     & -   & $-$0.683 (0.122) & 0.937 (0.189) & 1.486 (0.249)& -\\
\rowcolor[gray]{0.85} ADAS13  & -     & $-$0.772 (0.091) & $-$0.322 (0.174) & $-$0.352 (0.057) &  0.537 (0.089)& -& 1.651 (0.226)\\
RAVLT (immediate)  & 0.569 (0.108)  & -    & 0.458 (0.235)  & 0.379 (0.071)  &  $-$0.574 (0.108) & $-$0.869 (0.121)& 0.232 (0.229)\\
\rowcolor[gray]{0.85} RAVLT (learning)   & $-$0.180 (0.096) & -   & -  &$-$0.095 (0.054) & 0.187 (0.084)& 0.255 (0.103)& -\\
FAQ  & $-$0.292 (0.049) & $-$0.261 (0.061) & -   & $-$0.235 (0.031) & 0.358 (0.048) & 0.406 (0.058)& 0.277 (0.119)\\
\rowcolor[gray]{0.85} GenderM   & - & 0.297 (0.242)  &  0.492 (0.489)     & 0.228 (0.104)  & $-$0.326 (0.161)  & -& $-$0.357 (0.377)\\
Education  & -    & $-$0.155 (0.097) & 0.264 (0.129)  & $-$0.023 (0.039)  & 0.032 (0.059) & -& 0.058 (0.126)\\
\rowcolor[gray]{0.85} Age    & 0.191 (0.081)  & $-$0.041 (0.118) & - & 0.082 (0.046)  &  $-$0.141 (0.070) & $-$0.304 (0.085)& 0.169 (0.152)\\
Hippocampus        & 0.503 (0.090)  & 0.064 (0.116)  & -    & 0.256 (0.056)  & $-$0.353 (0.085) & $-$0.625 (0.104)& 0.346 (0.179)\\
\rowcolor[gray]{0.85} ICV                & -              & $-$0.195 (0.119) & $-$0.644 (0.229) & $-$0.197 (0.051) &  0.327 (0.079)& -& 0.742 (0.169)\\
MidTemp   & 0.117 (0.089)  & -              & 0.948 (0.181)  & 0.229 (0.056)  &  $-$0.365 (0.084)& $-$0.278 (0.093)& $-$0.601 (0.197)\\
\rowcolor[gray]{0.85} Scale ($\widehat{\sigma}$) & 0.625 (0.038)  & 0.641 (0.044)  & 0.628 (0.070)  & 0.653 (0.044)  & - & -& -\\ \hline
C-index   &  \multicolumn{3}{c}{0.850} & 0.844  & 0.845 &  \multicolumn{2}{c}{0.848}\\
iAUC   & \multicolumn{3}{c}{0.903}  & 0.895  &  0.896 & \multicolumn{2}{c}{0.897}\\
\bottomrule
\end{tabular}
\end{threeparttable}
}\label{tab: ADNI_est}
\end{table}

The estimated coefficients (Table \ref{tab: ADNI_est}) across the three CFs in the competing Weibull model (for survival time) are largely consistent with those from the Weibull model (for survival time), the Cox PH model (for hazard), and the disease mixture model (for hazard). A minor discrepancy is observed in the effect of age within CF2, though none of these estimates is statistically significant at the 5\% level. Another notable difference arises in the effect of education: CF3 shows a significant positive association, indicating that more years of education are linked to longer survival time, whereas both the Weibull, Cox PH model, and adapted disease mixture model (with merged CF2 \& CF3) yield inconsistent but non-significant estimates ($p>0.50$). One potential reason is that education can positively and negatively affect the progression of the disease, and as a result, it becomes insignificant in the fitted classical models without differentiating disease subtypes. Further analysis shows that when fewer biomarkers are included in the Weibull and Cox models, the effect of education changes direction, becomes significant, and aligns with the CF3 results. This pattern is consistent with systematic reviews indicating that, although higher education is generally associated with reduced dementia risk, the relationship varies across populations and settings \citep{sharp2011relationship}. These results indicate that conventional linear additive models may misspecify heterogeneous effects across populations.

Figure \ref{fig: ADNI_ROC_WP}(b) illustrates the relative importance of the CFs over time. CF1 consistently emerges as the dominant contributor, with a stable winning probability around 50\%, though not overwhelming the outcome. The importance of CF2 gradually declines to approximately 35\%, still representing a meaningful combination of biomarkers. CF3 remains the least influential, contributing around 15\% throughout. The dominance of CF1 over time also informs cost-effective strategies, suggesting that prioritising measurements within this group could capture most of the predictive information and reduce diagnostic costs without substantially compromising model performance.

Figure \ref{fig: ADNI_ROC_WP}(c) categorises failure samples into {seven} subtypes. For each sample, the attributed survival times for the CFs are estimated using the IPW-observed survival time. A sample is assigned to a subtype based on whether the attributed time for a specific CF is shorter than the median of all attributed times across all CFs and samples, which serves as the cut-off. In the ADNI cohort, most high-risk individuals are identified by abnormalities in CF1 or CF2, while seven samples exhibit high risk solely attributable to CF3.

Figure \ref{fig: ADNI_ROC_WP}(d) displays the sample characteristics of the seven subtypes defined by the Venn diagram, computed by averaging within each category and scaling across all categories. Among the single-CF groups, the CF3 subtype exhibits the shortest survival time, suggesting its relevance for investigating outcomes in individuals with more severe symptoms. {We note that education levels in CF3 are lower than those in CF1 and CF2, which may explain why patients in this group had the shortest survival time.} The combination of CF1 and CF2 appears particularly informative for early detection and prediction.  Predictor profiles also align with CF definitions, e.g., G1 shows a relatively high proportion of individuals carrying one or two APOE$\epsilon$4 alleles. {In addition, patients assigned to subtypes with multiple CFs experience significantly poorer survival (log-rank $p < 0.0001$; Figure \ref{fig: ADNI_ROC_WP}(e)).}

\subsection*{B.2 Hepatocellular Carcinoma Patients with GSE76427}\label{sec: HCC}
Liver cancer is the third leading cause of cancer-related deaths worldwide, with HCC accounting for the majority of cases \citep{kinsey2024management}. We analysed overall survival in HCC patients using five critical differentially expressed genes (CCDC107, CXCL12, GIGYF1, GMNN, and IFFO1) previously identified between HCC and control samples \citep{liu2023five}. We collect the data from GSE76427, which includes clinical information (overall survival, age, gender, etc.) and mRNA expression profiles from 115 primary tumour samples generated on the GPL10558 [Illumina HumanHT-12 V4.0 expression beadchip] platform \citep{grinchuk2018tumor}, among whom 23 (20\%) experienced an event.

\begin{figure}[H]
    \centering
    \subfloat[]{\includegraphics[width=0.35\linewidth]{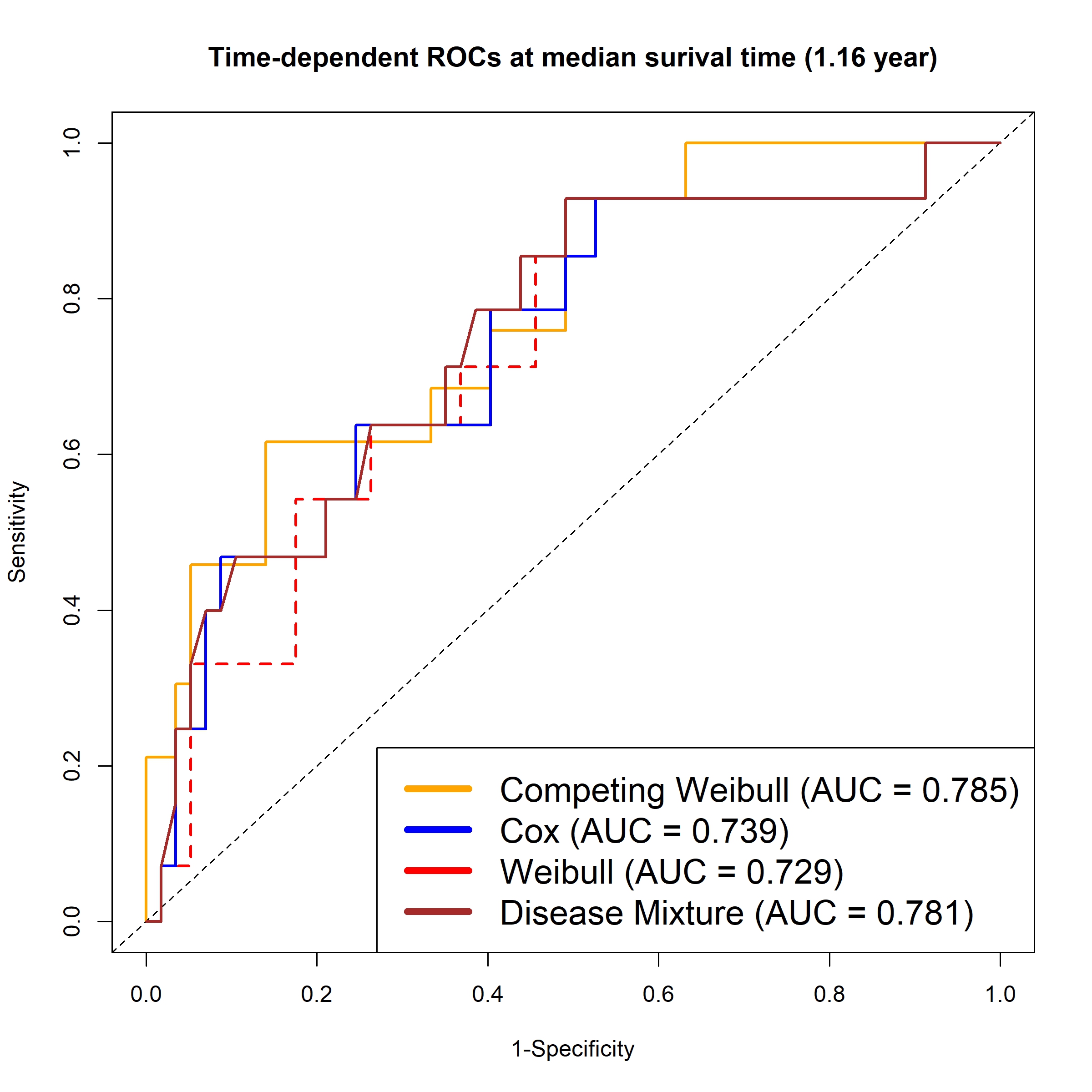}}
    \subfloat[]{\includegraphics[width=0.48\linewidth]{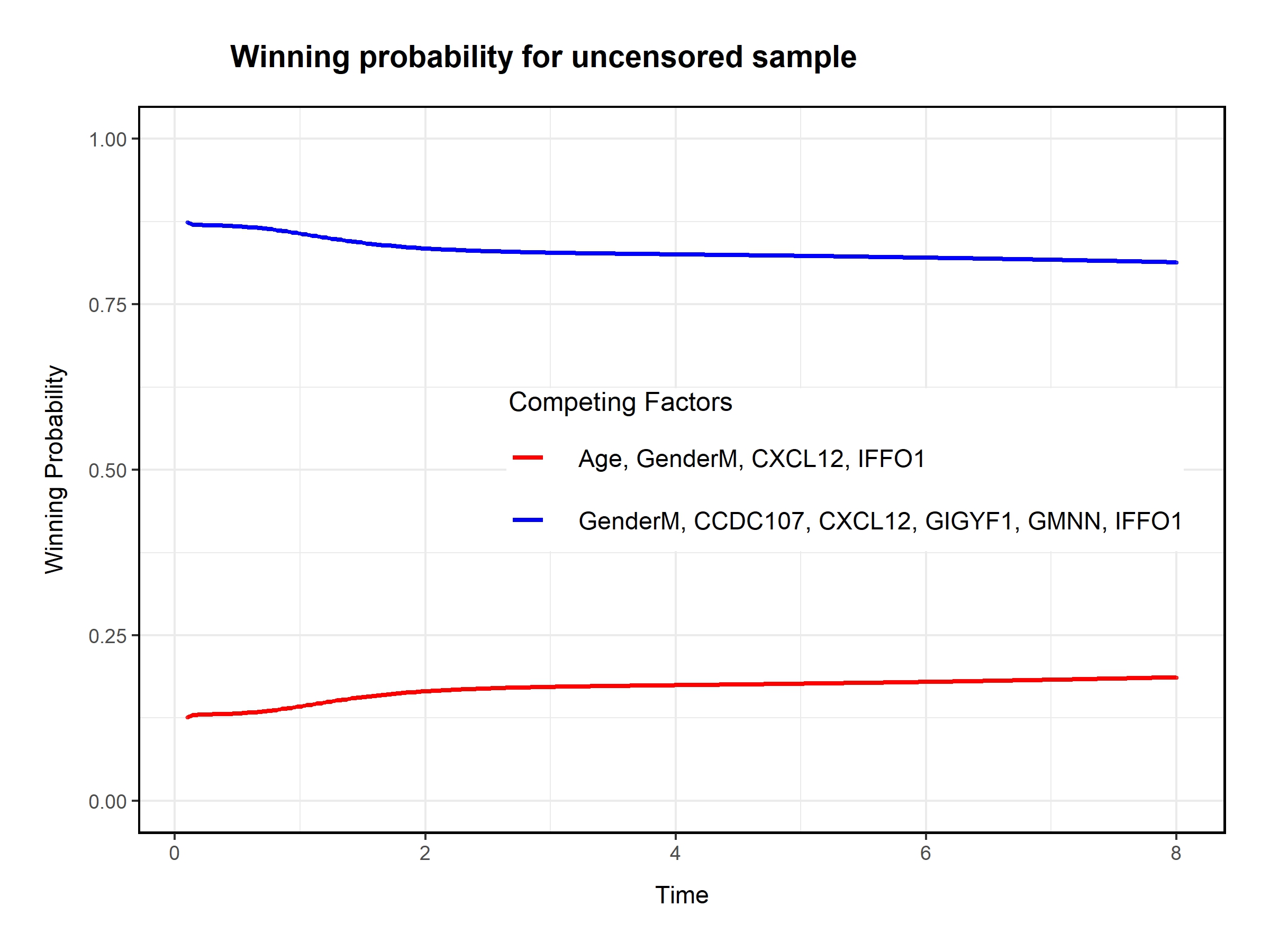}}
    \caption{HCC study results: (a) Time-dependent ROCs at median survival time (1.16 years); (b) Time-varying winning probability for uncensored samples.}
    \label{fig: HCC_ROC_WP}
\end{figure}

Given the relatively small sample size and few features, we specified two CFs with tuning parameters set to $\lambda_1 = 1$ and $\lambda_2 = 0.2$ in the competing model, with CF2 dominating over CF1 throughout the follow-up period, with a slight decline from around 90\% to 80\% (Figure \ref{fig: HCC_ROC_WP}(b)). The competing model shows around 8\% improvements in all performance metrics compared to the Cox PH and Weibull models, and 1--2\% increment to the disease mixture model. The estimated gene effects were largely consistent across the four models (Table \ref{tab: HCC_est} and Figure \ref{fig: HCC_ROC_WP}(a)).

\begin{table}[H]
\centering
\caption{Estimated coefficients with mean (SE), along with C-index and iAUC, from the competing Weibull model (survival time), Weibull model (survival time), Cox PH model (hazard) and disease mixture model (hazard) in the HCC study. The notation "-" means exclusion from the model.}
\resizebox{0.95\textwidth}{!}{
\begin{threeparttable}
\begin{tabular}{lrrrrrrr}
\toprule
\multirow{2}*{Features} &  \multicolumn{2}{c}{Competing Weibull} & \multicolumn{1}{c}{\multirow{2}*{
Weibull}} & \multicolumn{1}{c}{\multirow{2}*{Cox PH}} & \multicolumn{2}{c}{Disease Mixture}\\
\cmidrule{2-3} \cmidrule{6-7}
 &  \multicolumn{1}{c}{CF1} &  \multicolumn{1}{c}{CF2} &  &  &  \multicolumn{1}{c}{State 1 (CF2)} &  \multicolumn{1}{c}{State 0 (CF1)}\\
\midrule
Intercept   & 8.087 (1.323)  & 4.213 (1.654)  & 3.041 (0.919)  &- &-&- \\
\rowcolor[gray]{0.85} Age    & $-$2.216 (0.855) & -  & 0.007 (0.265)    & $-$0.005 (0.239) &-&1.154 (2.743)\\
GenderM    & 5.130 (0.778) & $-$1.791 (1.669)     & $-$0.535 (0.886) & 0.298 (0.784) &1.543 (1.809)&-\\
\rowcolor[gray]{0.85} CCDC107  & -  & 0.152 (0.262) & 0.251 (0.265) & $-$0.192 (0.239) &$-$0.129 (0.272)&-\\
CXCL12  & 0.231 (0.197)    & 0.219 (0.263)  & 0.265 (0.262)  &  $-$0.210 (0.231) &$-$0.260 (0.287)&$-$0.267 (1.071)\\
\rowcolor[gray]{0.85} GIGYF1   & - & $-$0.285 (0.216)   &$-$0.221 (0.231) & 0.231 (0.206)&0.349 (0.222)&-\\
GMNN  & - & $-$0.261 (0.061)  & 0.187 (0.310) & $-$0.169 (0.276) &$-$0.469 (0.347)&-\\
\rowcolor[gray]{0.85} IFFO1   & 5.707 (1.057) & 0.241 (0.311)  &  0.721 (0.297)     & $-$0.649 (0.267)  &$-$0.190 (0.315)&$-$6.115 (9.990)\\
 Scale ($\widehat{\sigma}$) & 0.229 (0.099)  & 0.984 (0.129)  & 1.132 (0.176)    & - &-&-\\
\hline
C-index   & \multicolumn{2}{c}{0.755}  &  0.673     & 0.682  &\multicolumn{2}{c}{0.747}\\
iAUC   & \multicolumn{2}{c}{0.766}  &  0.686     & 0.699 &\multicolumn{2}{c}{0.746}\\
\bottomrule
\end{tabular}
\end{threeparttable}
}\label{tab: HCC_est}
\end{table}

\subsection*{B.3 Breast Cancer Patients' Recurrence-free Time with GBCSG}\label{sec: GBCSG}
In this section, we analyse the recurrence-free survival time of breast cancer patients who received chemotherapy, with or without additional hormonal treatment, based on the trial conducted by the German Breast Cancer Study Group \citep[][GBCSG]{schumacher1994randomized}. The GBCSG enrolled 720 patients with primary node-positive breast cancer into a randomised trial, investigated the effectiveness of three versus six cycles of chemotherapy and the impact of additional hormonal therapy with tamoxifen. We use the subset of 686 patients (with 299 recurrence events) who had complete data at baseline, as provided by  \citet{sauerbrei1999building} with multivariable prognostic modelling. These variables include age, tumour size, number of positive lymph nodes, progesterone and estrogen receptor levels, menopausal status, and tumour grade, as described and summarised in Table \ref{tab: brcancer}.

\begin{table}[!htbp]
\centering
\caption{Description and summary of the features used in the breast cancer study.}
\resizebox{0.95\textwidth}{!}{
\begin{tabular}{*{3}{clc}}
\toprule
Features & Description &  Mean (range)/Counts (percentage) \\
\midrule

HormonT & Hormonal therapy indicator (1 for received therapy) & therapy: 246 (35.9\%) \\
\midrule
Menostat & \cellcolor[gray]{0.85} Menopausal status indicator (1 for postmenopausal)  & \cellcolor[gray]{0.85} post: 396 (57.7\%)  \\
\cmidrule{2-3}
Age45 & Age of the patients in years, proceed into an indicator (1 for $>45$) &$>45$: 533 (77.7\%)  \\
\midrule
Tsize & \cellcolor[gray]{0.85} Tumor size, measured in in mm & \cellcolor[gray]{0.85} 29.33 (3 to 120)  \\
\cmidrule{2-3}
\multirow{2}*{Tgrade} & Tumour grade (ordinal: I for well, II for moderate, III for poor),   & II: 444 (64.7\%) \\
 & \cellcolor[gray]{0.85} proceed into two indicators (tgrade2, tgrade3) & \cellcolor[gray]{0.85} III: 161 (23.5\%) \\
\cmidrule{2-3}
Pnodes & Number of positive nodes & 5.01 (1 to 51) \\
\midrule
Progrec & \cellcolor[gray]{0.85} Progesterone receptor concentration (fmol/mg protein) & \cellcolor[gray]{0.85} 110.00 (0 to 2,380) \\
\cmidrule{2-3}
Estrec & Estrogen receptor concentration (fmol/mg protein) & 96.25 (0 to 1144) \\
\bottomrule
\end{tabular} }
\label{tab: brcancer}
\end{table}

\begin{figure}[H]
    \centering
    \subfloat[]{
    \includegraphics[width=0.29\linewidth]{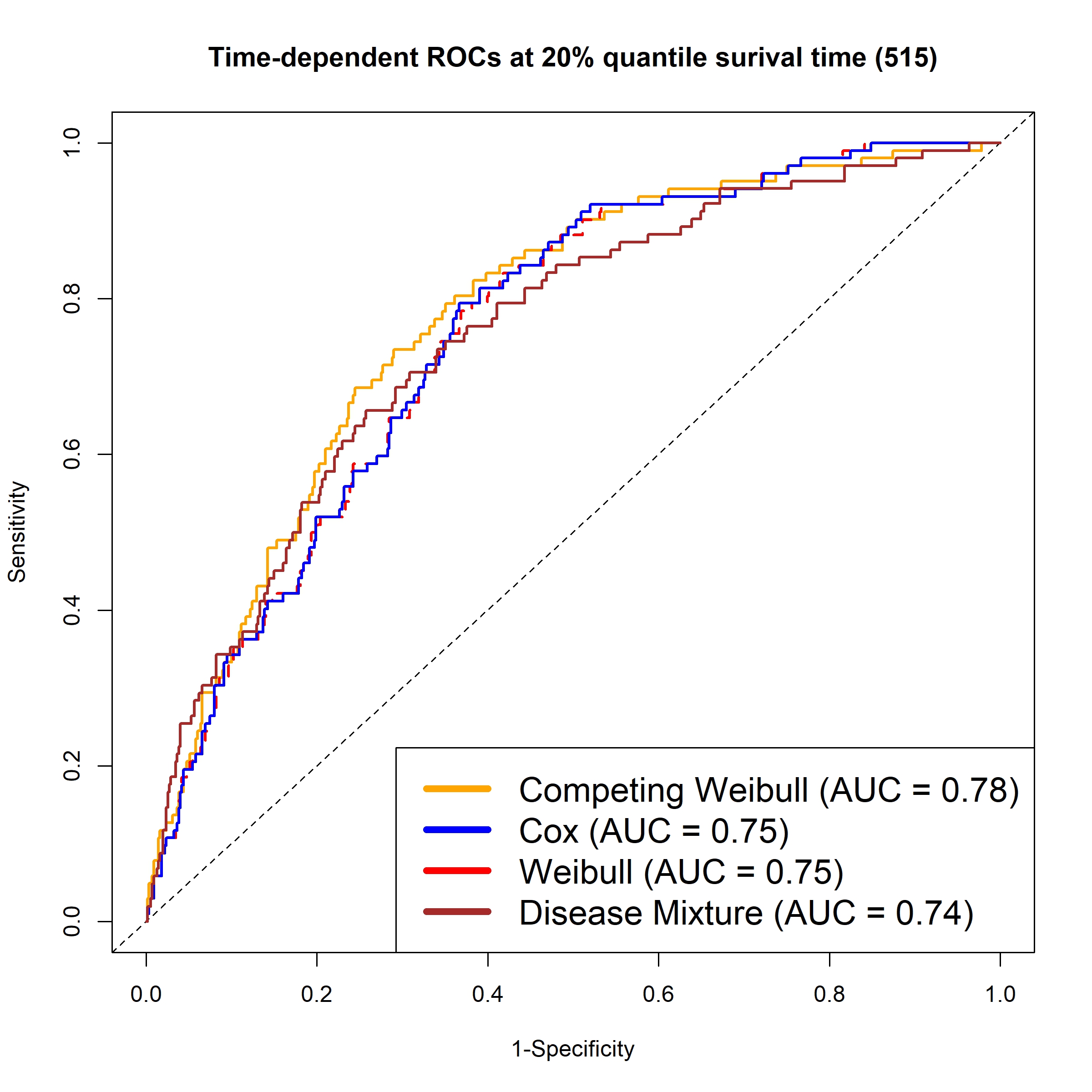}}
    \subfloat[]{
    \includegraphics[width=0.4\linewidth]{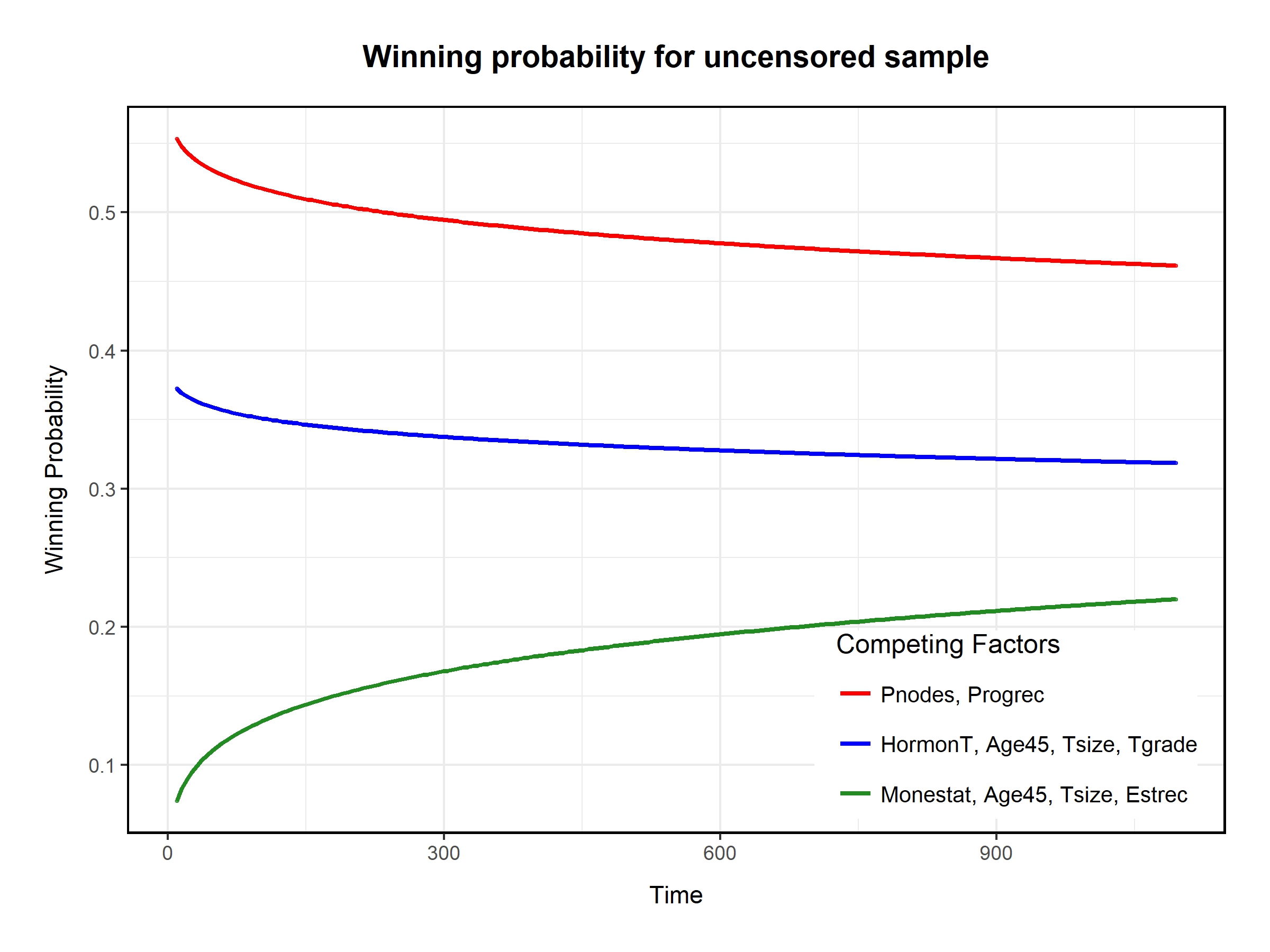}}
    \subfloat[]{
    \includegraphics[width=0.25\linewidth]{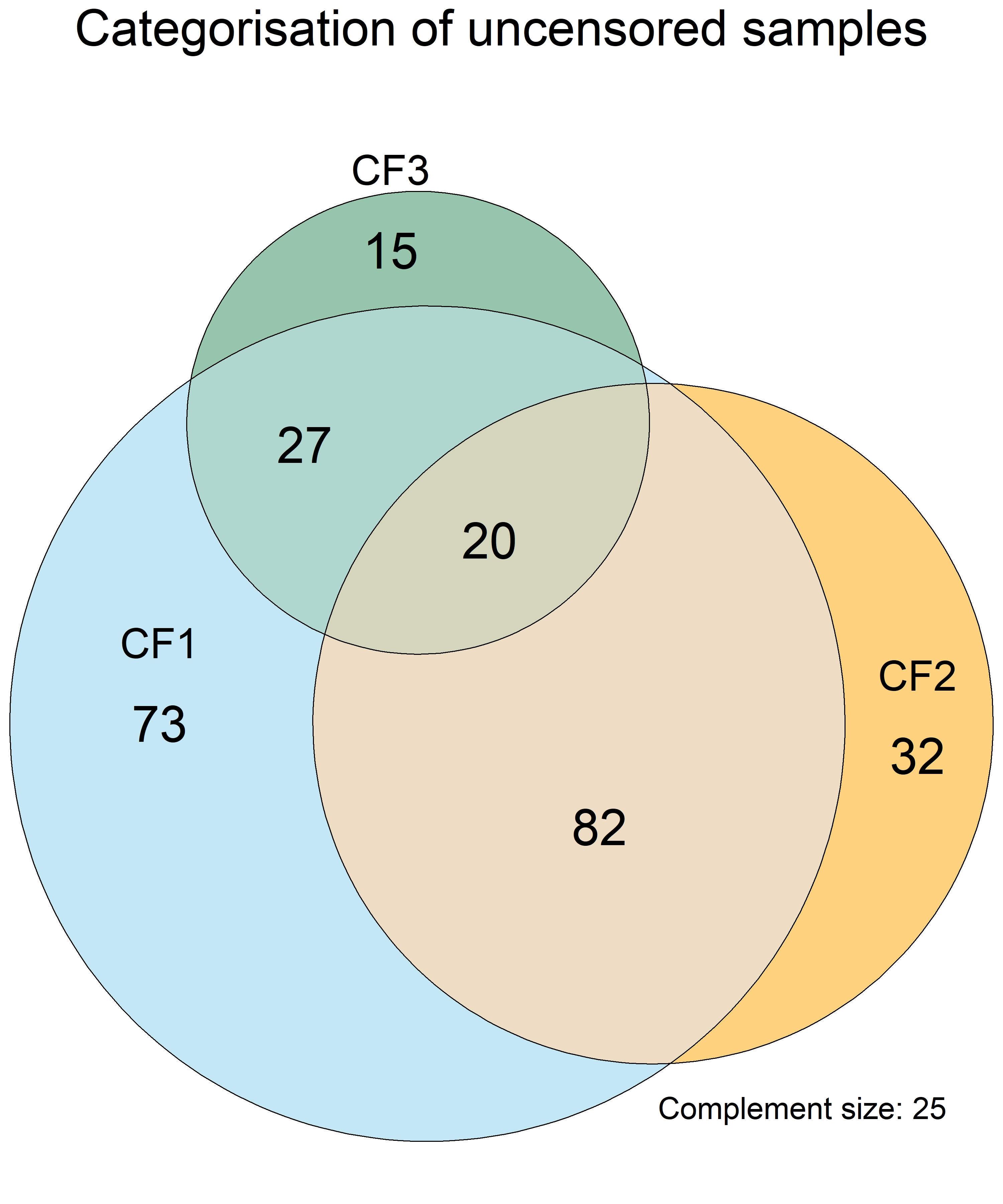}}
    \caption{Breast cancer study results: (a) Time-dependent ROCs at 20\% quantile (515 days) survival time; (b) Time-varying winning probability for uncensored samples; (c) Categorisation of uncensored samples based on IPW attributed survival times below the median.}
    \label{fig: brcaner_ROC_WP}
\end{figure}

We selected tuning parameters $\lambda_1 = 1$ and $\lambda_2 = 0.2$, and specified three CFs. The estimated parameters (Table \ref{tab: brcancer_est}) are generally consistent with those from the Weibull and Cox PH models. Our competing Weibull model achieves superior predictive performance, with a C-index of 0.705 and an iAUC of 0.760, compared to the Weibull model (C-index of 0.695, iAUC of 0.750), the Cox PH model (C-index of 0.696, iAUC of 0.750), and the disease mixture model (C-index of 0.687, iAUC of 0.737). It also demonstrates improved accuracy in estimating short survival times for severe cases, as illustrated by the time-dependent ROC curves at day 515, the 20\% quantile of recurrence-free time, in Figure \ref{fig: brcaner_ROC_WP}(a). Regarding cause attribution in Figure \ref{fig: brcaner_ROC_WP}(b), CF1 consistently contributes the most with approximately 50\% winning probability, followed by CF2 with about 35\%. CF3, while initially less dominant, exhibits a gradual increase over time, reaching approximately 20\% at later time points. For the 8 subtypes illustrated in Figure \ref{fig: brcaner_ROC_WP}(c), a larger proportion of cases are attributable to a single dominant CF1, as well as simultaneous abnormalities of CF1 \& CF2.

\begin{table}[H]
    \centering
    \caption{Estimated coefficients with mean (SE), along with C-index and iAUC, from the competing Weibull model (survival time), Weibull model (survival time), Cox PH model (hazard) and disease mixture model (hazard) in breast cancer study. The notation "-" means exclusion from the model.}
    \label{tab: brcancer_est}
    \resizebox{0.9\textwidth}{!}{
    \begin{threeparttable}
   \begin{tabular}{lrrrrrrr}
\toprule
\multirow{2}*{Features} &  \multicolumn{3}{c}{Competing Weibull} & \multicolumn{1}{c}{\multirow{2}*{
Weibull}} & \multicolumn{1}{c}{\multirow{2}*{Cox PH}} & \multicolumn{2}{c}{Disease Mixture}\\
\cmidrule{2-4} \cmidrule{7-8}
 &  \multicolumn{1}{c}{CF1} &  \multicolumn{1}{c}{CF2} &  \multicolumn{1}{c}{CF3} &  & &  \multicolumn{1}{c}{State 1 (CF1)} &  \multicolumn{1}{c}{State 0 (CF2 \& CF3)}\\
\midrule
    Intercept      & 9.438 (0.435)   & 8.504 (0.510)   & 10.276 (1.283)  & 7.983 (0.188)   & - & -& - \\
    \rowcolor[gray]{0.85} HormonT        & -               & 1.398 (0.530)   & -               & 0.269 (0.092)   & $-$0.348 (0.128) & -& 0.289 (0.223)\\
    Menostat       & -               & -               & $-$0.925 (0.635) & $-$0.255 (0.116) & 0.340 (0.162) & -& 1.115 (0.347)\\
    \rowcolor[gray]{0.85} Age45         & -               & 0.701 (0.208)   & $-$1.038 (1.299) & 0.305 (0.130)   & $-$0.415 (0.181) & -& 0.131 (0.488)\\
    Tsize          & -               & $-$0.169 (0.095) & $-$0.233 (0.090) & $-$0.084 (0.039) & 0.008 (0.003) & -& 0.184 (0.100)\\
    \rowcolor[gray]{0.85} Tgrade2       & -               & $-$0.892 (0.528) & -               & $-$0.479 (0.179) & 0.643 (0.249) & -& 0.092 (0.319)\\
    Tgrade3        & -               & $-$0.683 (0.572) & -               & $-$0.579 (0.193) & 0.774 (0.268)& -& 0.370 (0.386)\\
    \rowcolor[gray]{0.85} Pnodes        & $-$0.309 (0.038)& -               & -               & $-$0.204 (0.029) & 0.048 (0.007) & 0.274 (0.049)&-\\
    Progrec        & 3.371 (0.880)   & -               & -               & 0.329 (0.084)   & $-$0.002 (0.001) & $-$0.521 (0.165)& -\\
    \rowcolor[gray]{0.85} Estrec        & -               & -               & $-$0.133 (0.079) & $-$0.022 (0.049) & 0.0001 (0.0004) & -& 0.006 (0.109)\\
    Scale ($\widehat{\sigma}$) & 0.735 (0.046)   & 0.725 (0.053)   & 0.580 (0.055)   & 0.718 (0.048)   & - & -& -\\ \hline
     C-index        &  \multicolumn{3}{c}{0.705}              & 0.695   & 0.696 & \multicolumn{2}{c}{0.687}\\
      iAUC        &  \multicolumn{3}{c}{0.760}               & 0.750   & 0.750 &\multicolumn{2}{c}{0.737}\\
    \bottomrule
    \end{tabular}
    \end{threeparttable}
    }
\end{table}

\subsection*{B.4 Disease Mixture Model Specification and Diagnostic Plots for Real Applications}

{In this section, we describe the modifications applied to the disease mixture model \citep{moreno2017survival} to enable a fair comparison with our competing Weibull model in real applications. We also present the diagnostic plots for all applications in Figure \ref{fig: diag_app}.}

{The implementation generally follows the strategy outlined in simulation studies (see Appendix C.1), but with the "disease of interest" defined as the CF exhibiting the strongest dominance in the studied sample, as identified by our competing model stratification (e.g., CF1 in the sepsis study and CF3 in the LUAD study).}

{Specifically, a subject is treated as having the dominant CF as the underlying cause if it belongs to a subgroup containing that CF under our competing model-based stratification. For example, in the sepsis study, subjects in groups G1, G12, G13 or G123 are classified as having CF1 as the underlying cause; we assign $\pi=1$ for G1 and $\pi=0.75$ for G12, G13 and G123. For all other subjects, we assign $\pi \in \{0.25, 0.125, 0\}$, consistent with the simulation setting. The performance of the disease mixture model is generally robust to moderate variations in $\pi$.}

{For censored observations, where $\pi$ is undefined in the original framework, we assign $\pi$ using the average estimated winning probability of the dominant CF obtained from the competing model.}

{The same weighting strategy is applied to the HCC dataset with two CFs. However, to ensure numerical stability under the limited sample size in CF1, the gender variable is excluded from the second state (State 0).}

\begin{figure}[H]
    \centering
    \subfloat[]{
    \includegraphics[width=0.33\linewidth]{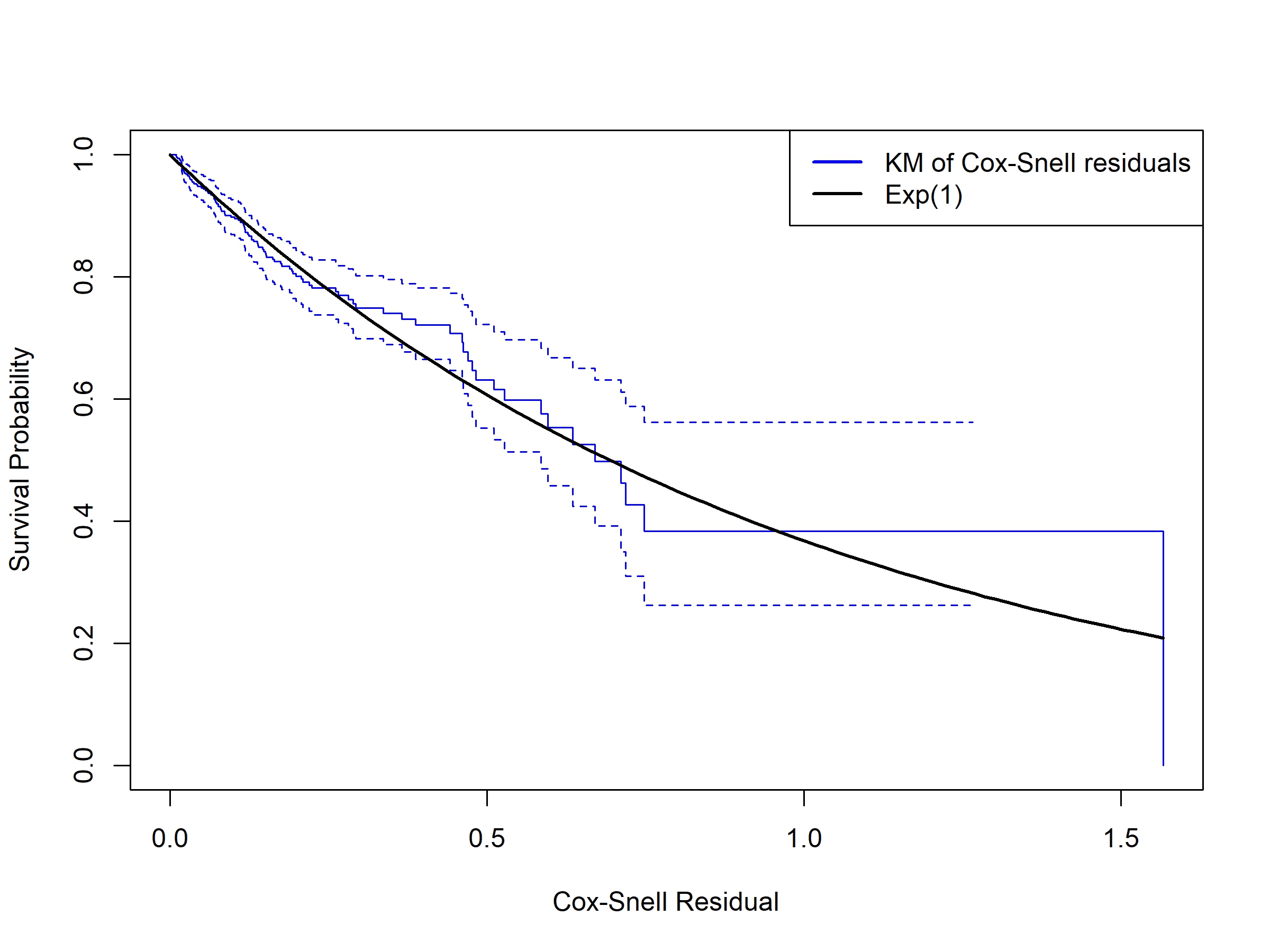}}
    \subfloat[]{
    \includegraphics[width=0.33\linewidth]{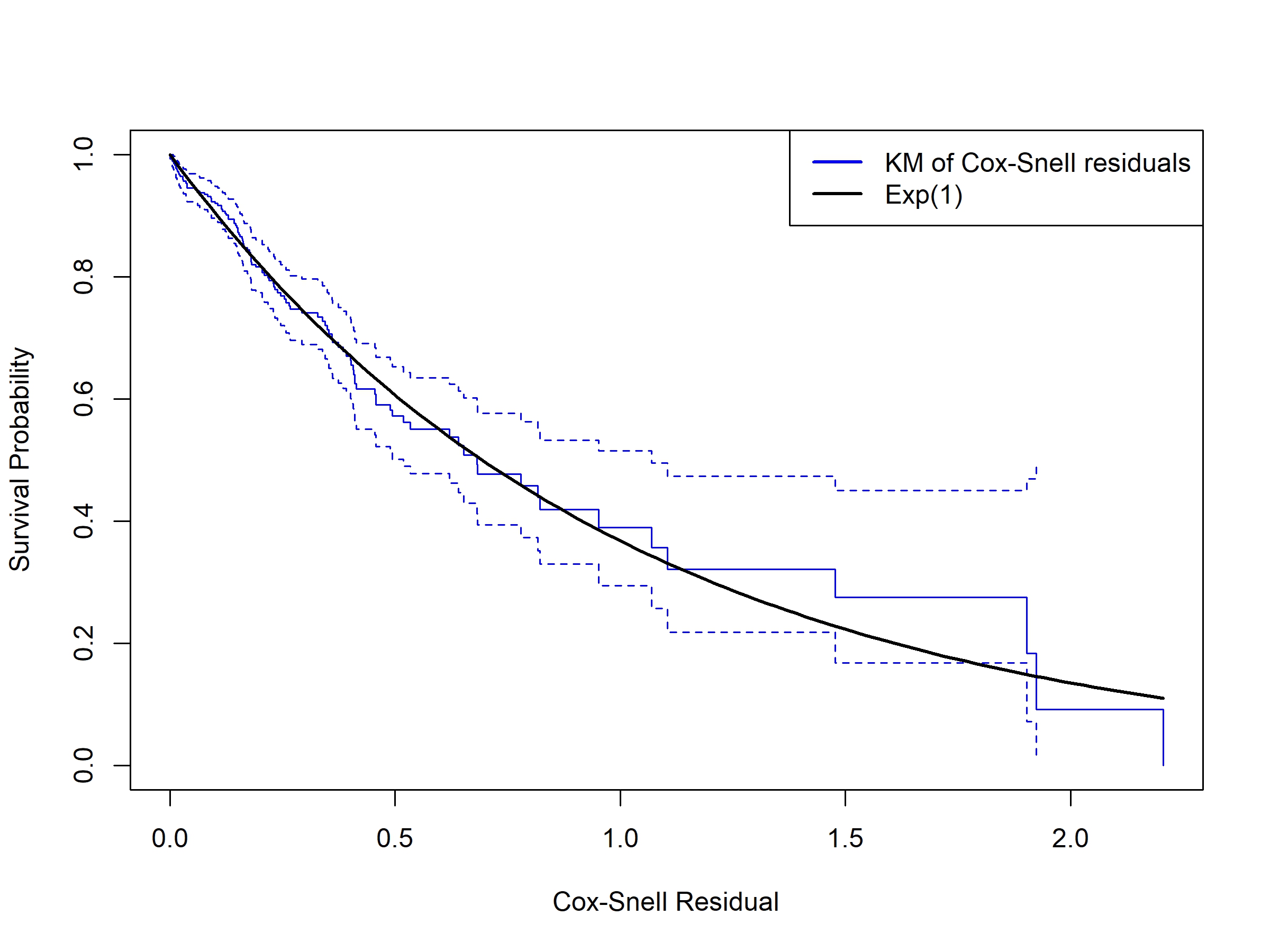}}
        \subfloat[]{
    \includegraphics[width=0.33\linewidth]{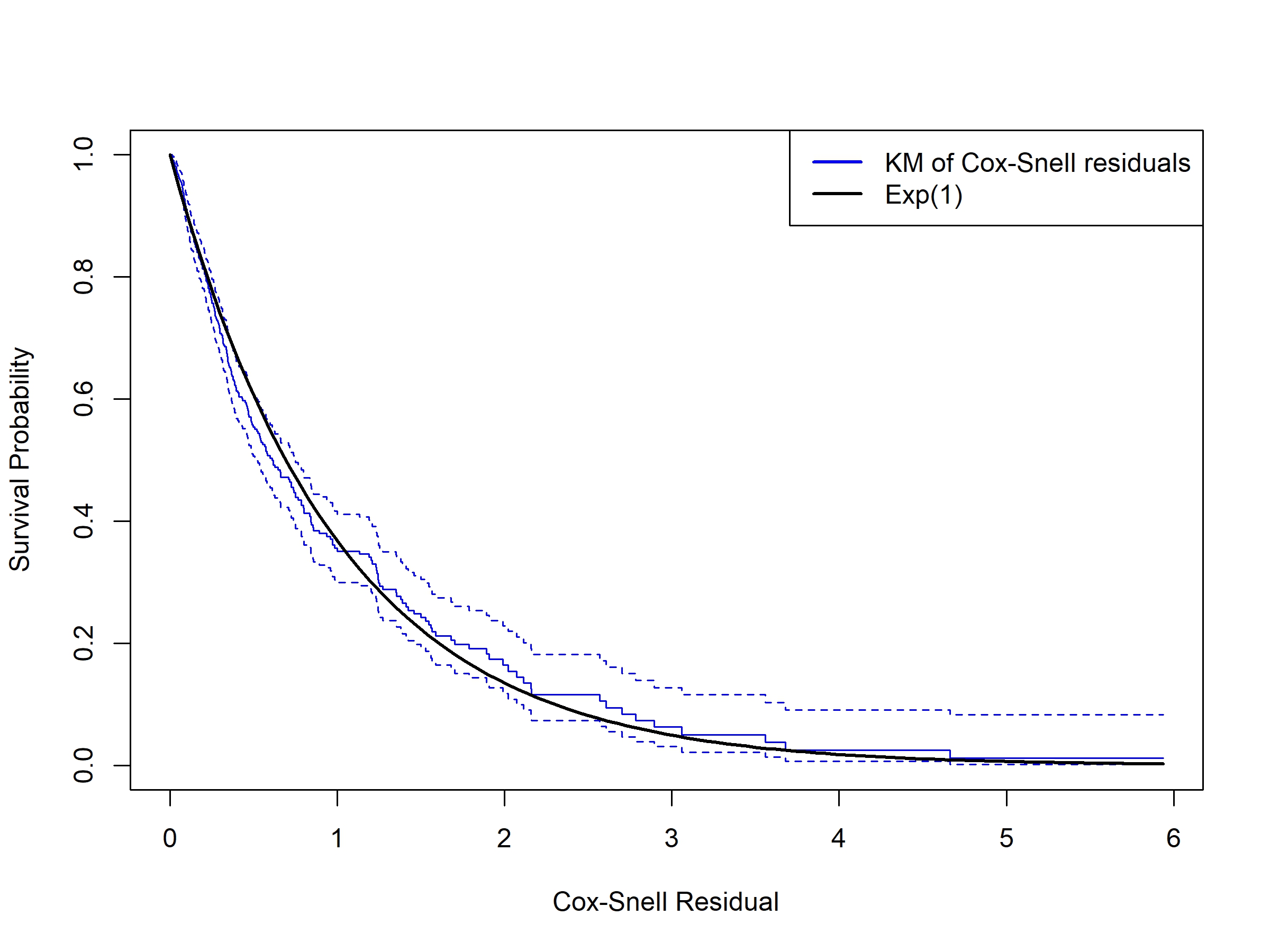}}\\
        \subfloat[]{
    \includegraphics[width=0.4\linewidth]{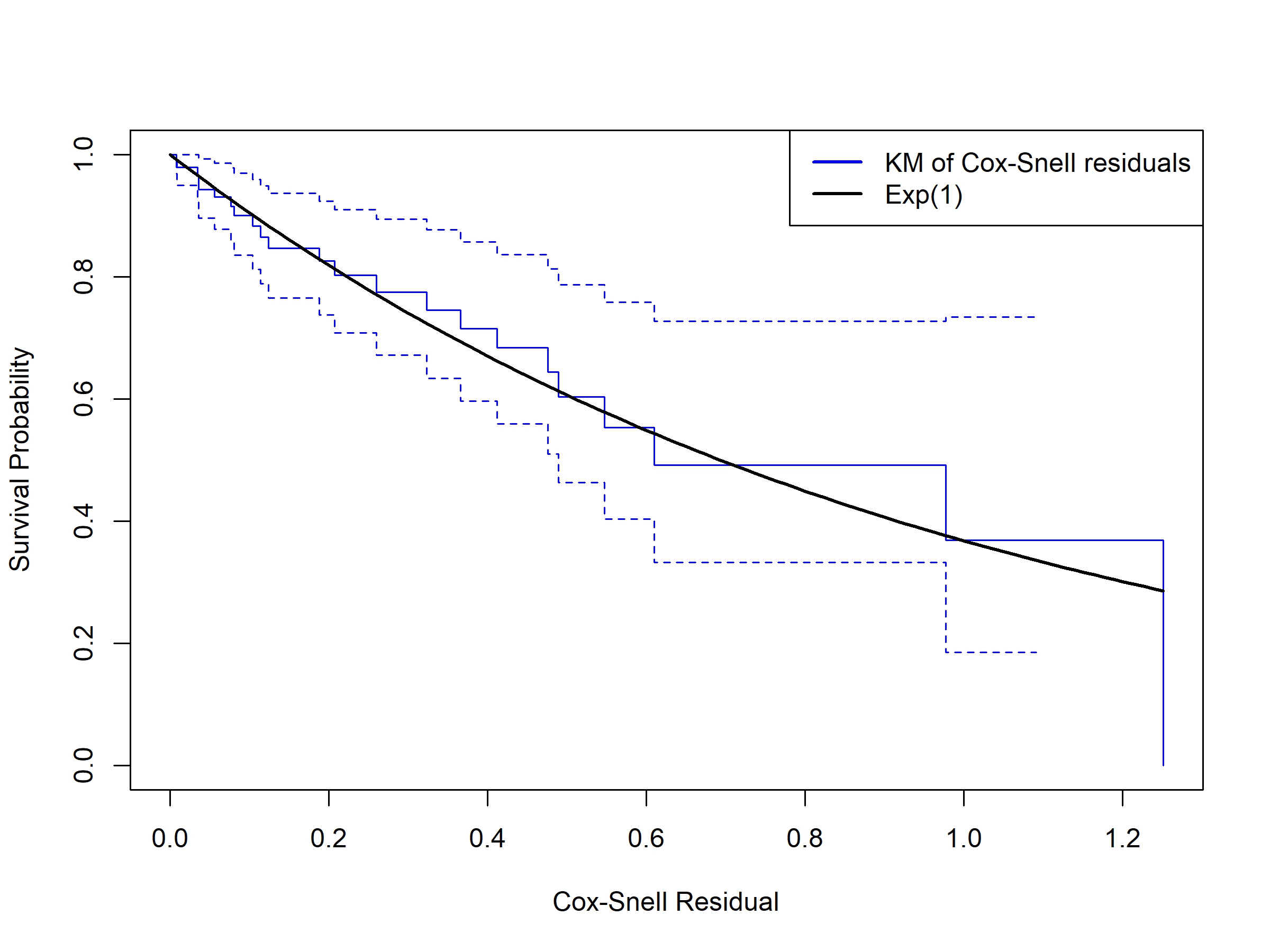}}
    \subfloat[]{
    \includegraphics[width=0.4\linewidth]{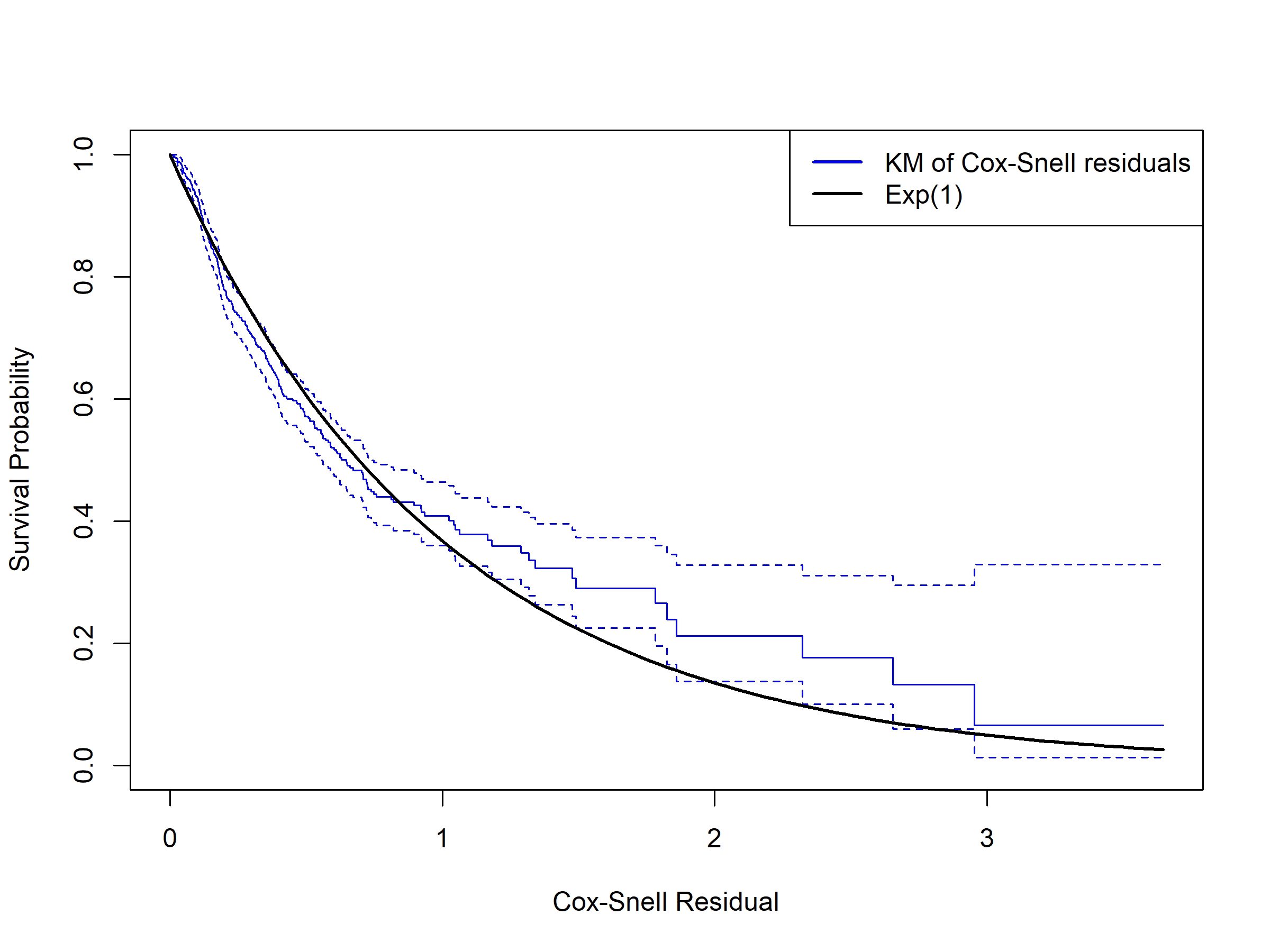}}
    \caption{Diagnostic plots of Cox-Snell residuals against the standard exponential distribution in (a) the sepsis study, (b) the LUAD study, (c) Alzheimer's disease study, (d) HCC study, and (e) breast cancer study.}
    \label{fig: diag_app}
\end{figure}

\section*{Appendix C}
In this section, we present additional simulation results for the competing Weibull model, including model comparison and diagnostic plots in C.1. Simulation for the competing log-normal model under dependent errors is described in C.2.

\subsection*{C.1 Simulation results of Competing Weibull Model \label{sec: simulation-Weibull}} 
\begin{figure}[!htbp]
    \centering
    \subfloat[]{
    \includegraphics[width=0.32\linewidth]{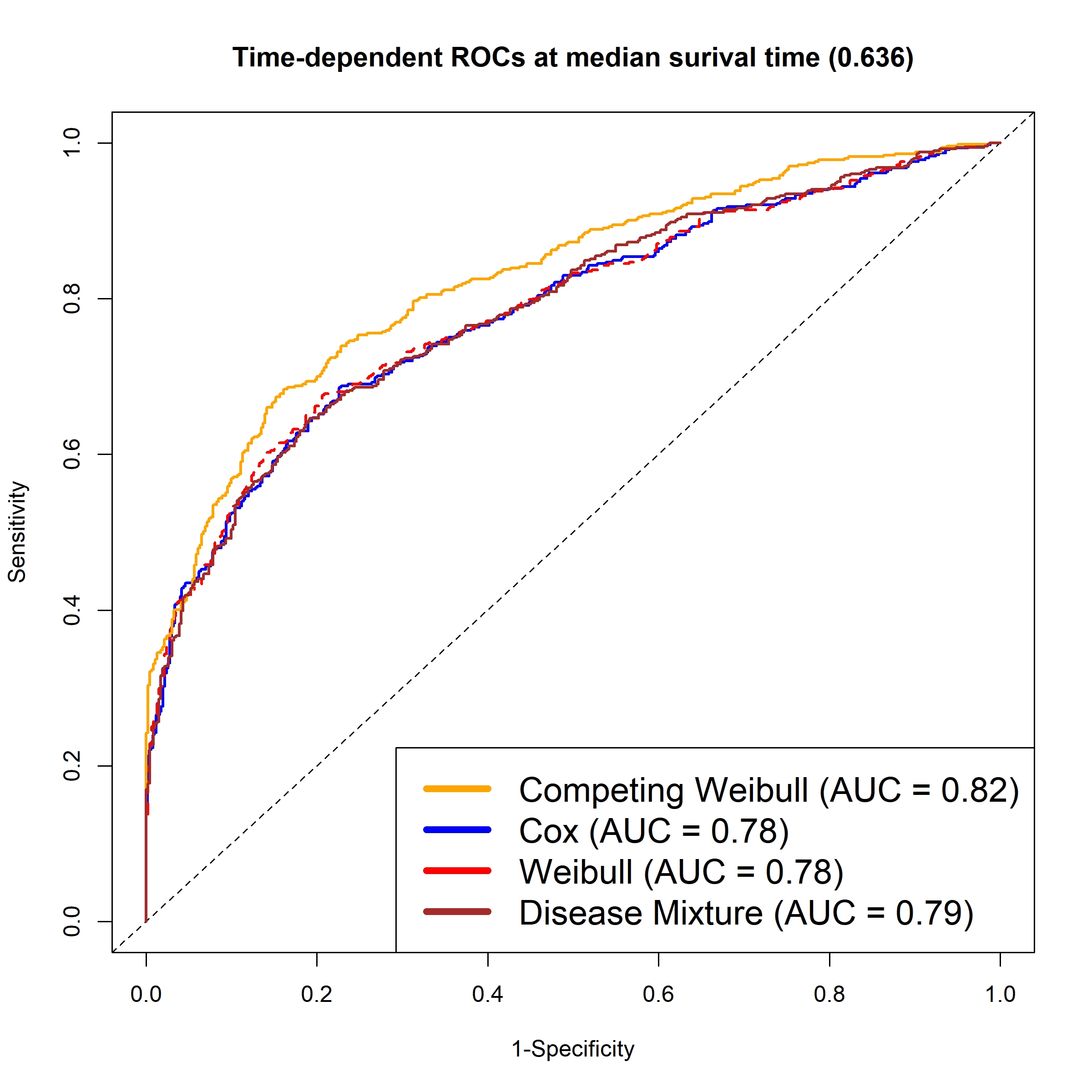}}
    \subfloat[]{
    \includegraphics[width=0.32\linewidth]{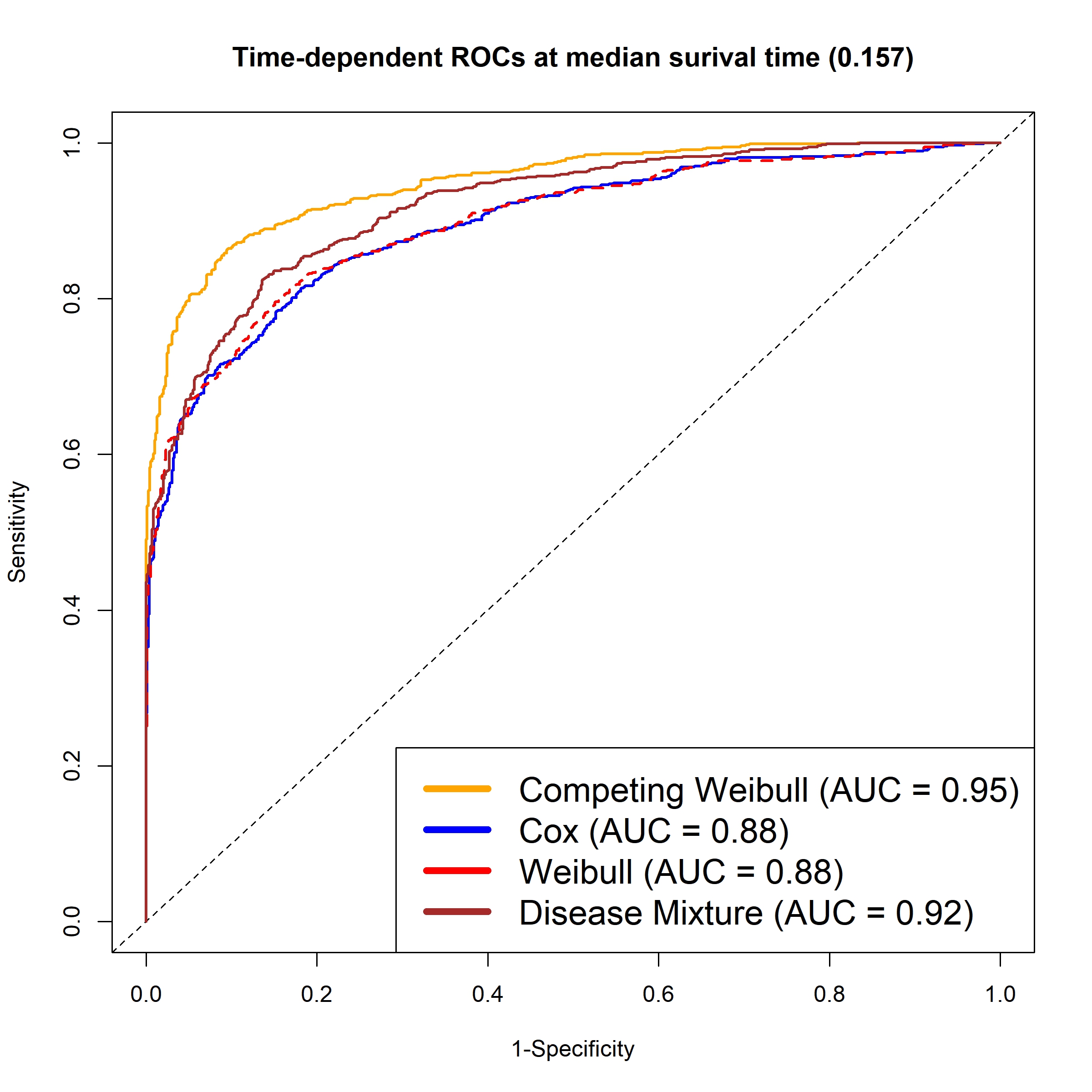}}
    \subfloat[]{
    \includegraphics[width=0.32\linewidth]{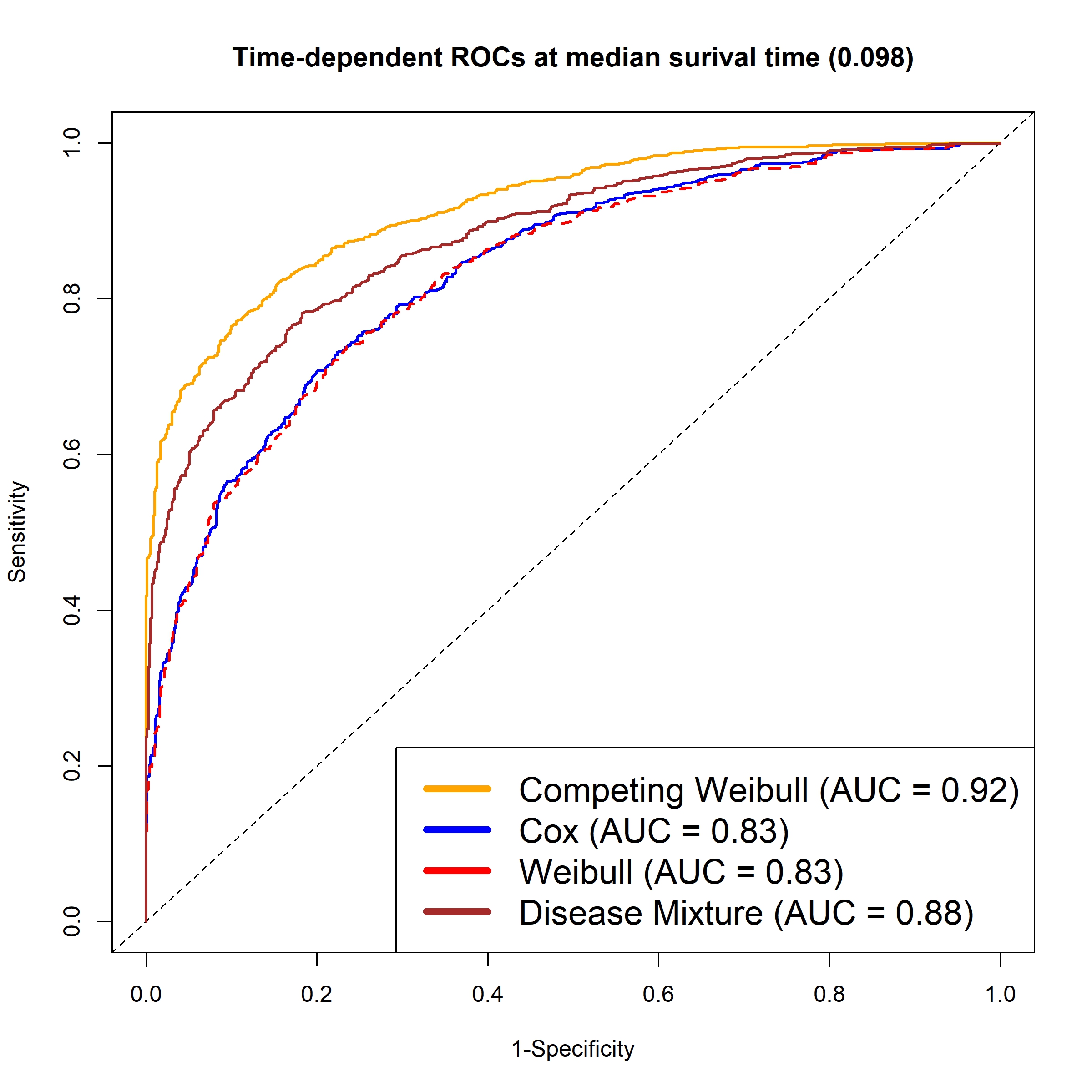}}\\
        \subfloat[]{
    \includegraphics[width=0.32\linewidth]{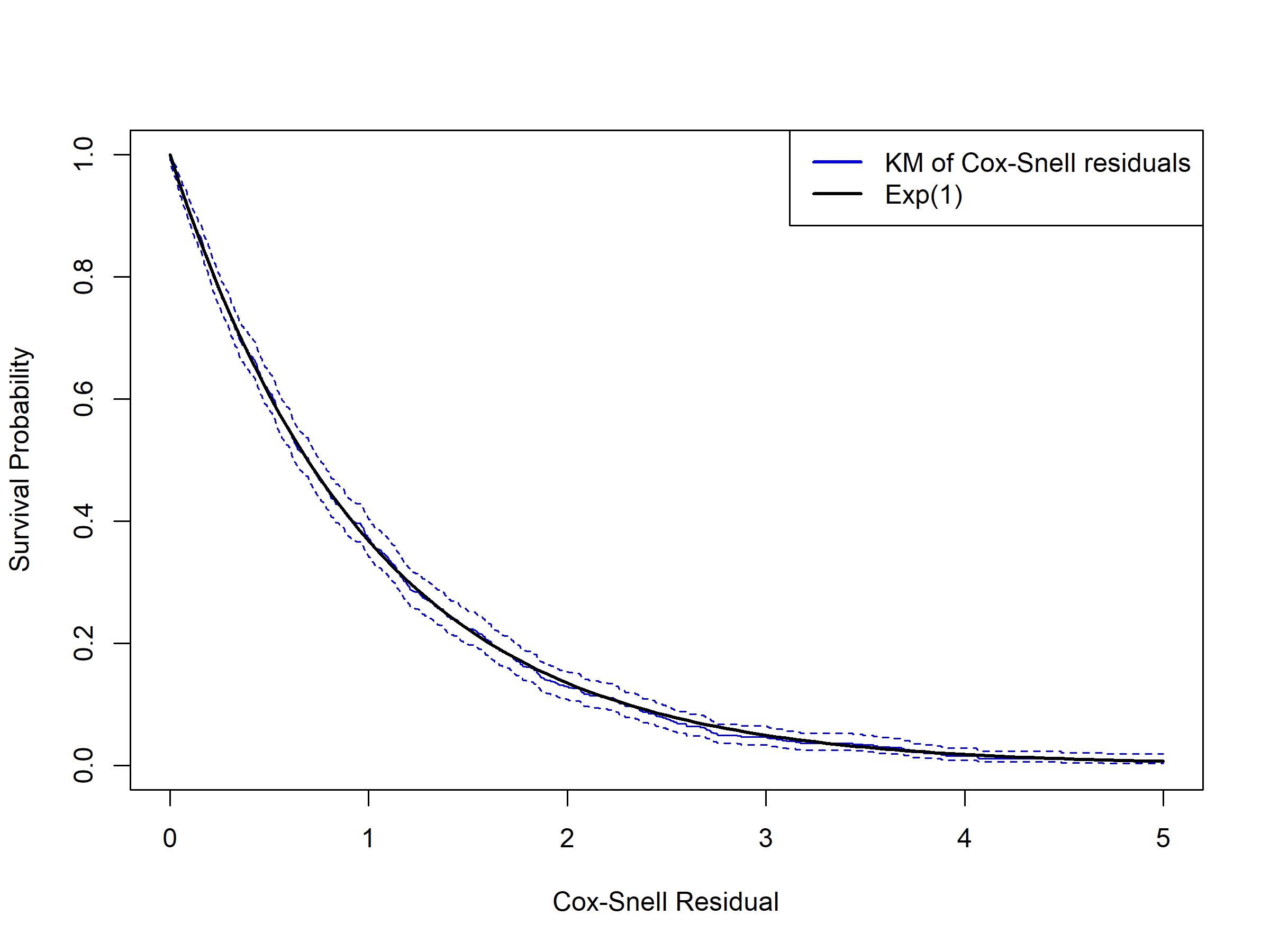}}
    \subfloat[]{
    \includegraphics[width=0.32\linewidth]{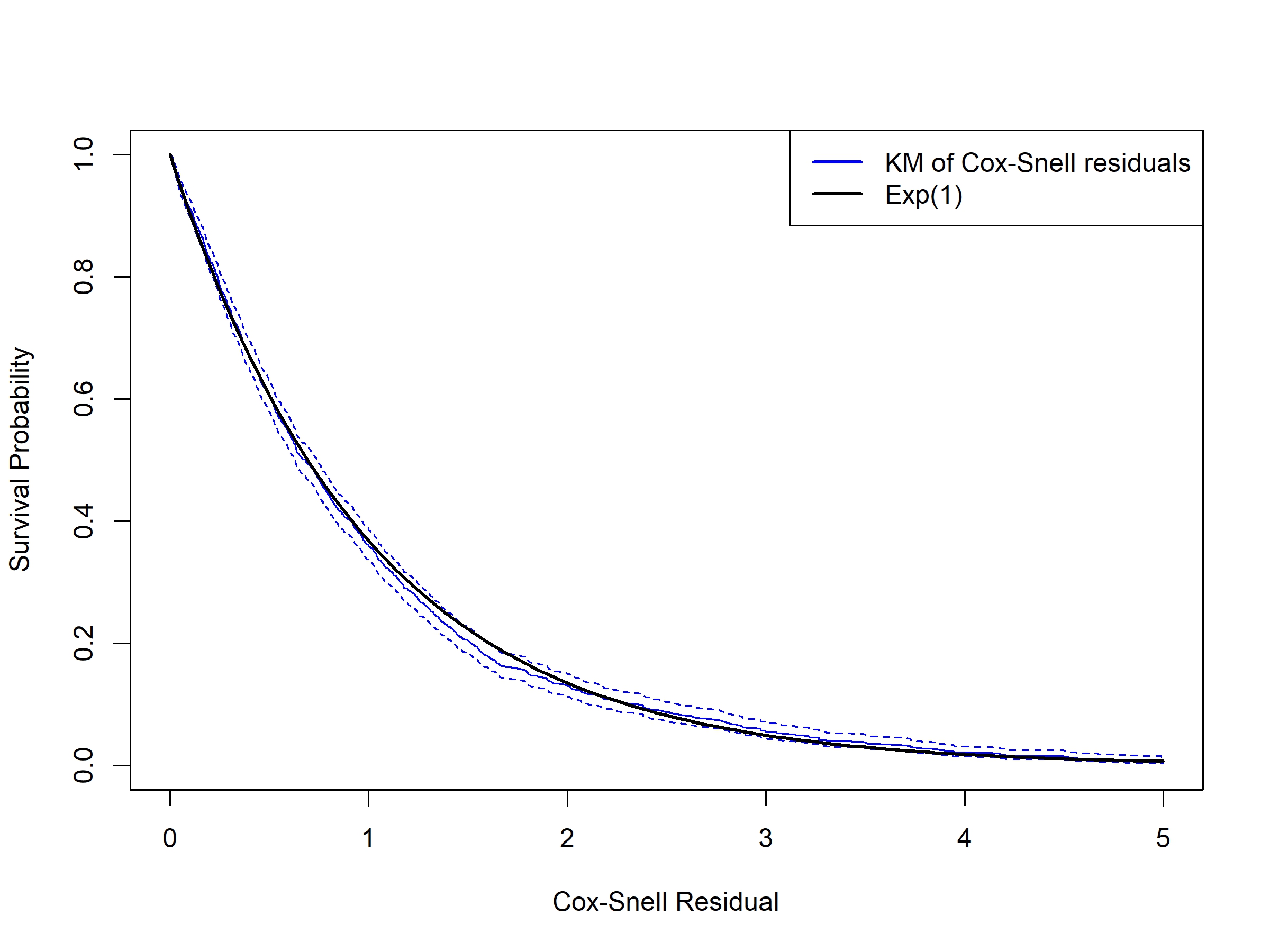}}
    \subfloat[]{
    \includegraphics[width=0.32\linewidth]{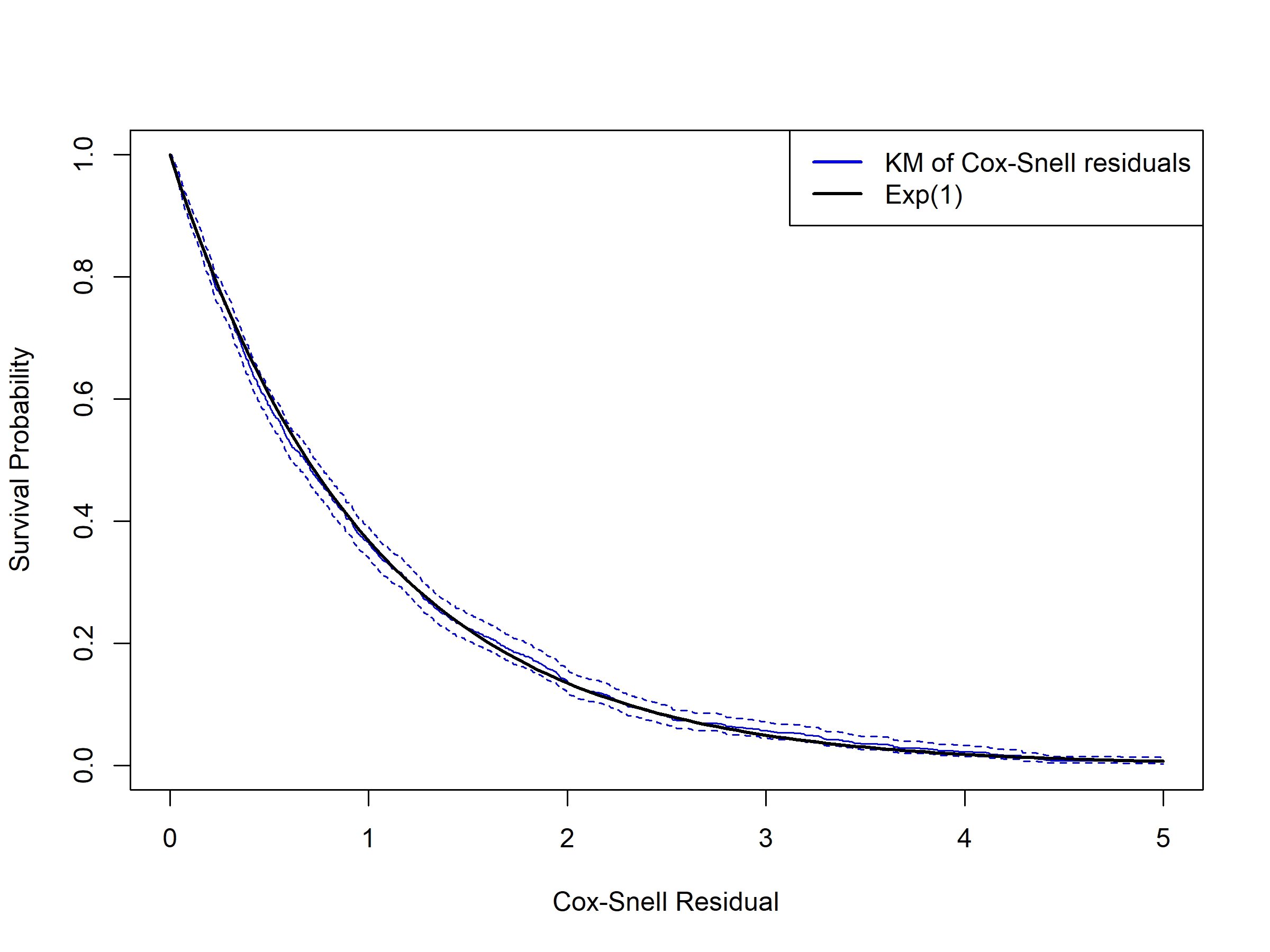}}
    \caption{Time-dependent ROCs (a, b, c) for competing Weibull, Cox PH, Weibull and disease mixture models at their median survival times. Diagnostic plots of Cox-Snell residuals (d, e, f) for the competing Weibull model. All cases refer to the 10\% random censorship for (a, d) Example 1 with sample size $n = 1000,$ and censored number $ c = 100$), (b, e) Example 2 ($n = 1500, c = 161$), and (c, f) Example 3 ($n = 1500, c = 150$).}
    \label{fig: ROC}
\end{figure}

{In this section, we describe the modifications applied to the disease mixture model \citep{moreno2017survival} to enable a fair comparison with our competing Weibull model. The corresponding results are reported in Table \ref{tab: comp} and Figure \ref{fig: ROC}, with the latter also including diagnostic plots.}

{The disease mixture model specifies the hazard as a two-component mixture,
$$
\lambda_\pi(t) = \pi\lambda_1(t) + (1-\pi)\lambda_0(t),
$$
where $\lambda_1(t)$ corresponds to a pre-specified disease of interest and $\lambda_0(t)$ aggregates all remaining causes. Its implementation requires externally assigned weights $\pi$, typically derived from death certificate information that distinguishes underlying and contributing causes. In particular, weights such as $\pi=1$ or $0.75$ are assigned when the disease is the underlying cause, while smaller values (e.g., $0.25$, $0.125$, or $0$) correspond to contributing or absent roles \citep{moreno2017survival}. Moreover, the implementation requires that both $\pi=1$ and $\pi=0$ cases are present in the data.}

{To adapt this framework to our setting with three CFs with relatively balanced dominance, we designate CF1 as the "disease of interest" and aggregate CF2 \& CF3 as the other source. In the simulated data, the latent failure times are observed, allowing us to identify whether CF1 is the dominant cause, i.e., $T_1 = \min(T_1,T_2,T_3)$. For such observations, we assign $\pi \in \{1, 0.75\}$ at uniformly random to mimic underlying cause scenarios. For all other failure cases, we uniformly randomly assign $\pi \in \{0.25, 0.125, 0\}$ to reflect weaker or absent contributions.}

{For censored observations, the original model leaves $\pi$ undefined \citep{moreno2017survival}. To enable risk prediction and comparison via concordance measures, we assign $\pi$ based on the average estimated winning probability of CF1 from our cAFT model. We also examined alternative specifications, including varying $\pi$ assignment of weights, e.g., fixed choices of $\pi=1$ for "underlying failures" and $\pi \in \{ 0.125, 0\}$ or $\pi=0$ for "contributing failures", as well as $\pi=0.5$ for censored observations, and found that the comparative performance remains stable across these choices.}

\begin{table}[H]
    \centering
    \caption{C-index and iAUC comparisons among competing Weibull, Cox PH, Weibull, and disease mixture models across different examples. The Cox PH and Weibull models include all explanatory variables in each example; the disease-mixture model treats CF1 as the pure state of disease and combines CF2 and CF3 into the pure state of other sources. Here, $n$ represents the total sample size and $c$ represents the number of censored data.
    }
\resizebox{\textwidth}{!}{
    \begin{threeparttable}
    \begin{tabular}{lclcclclcc}\toprule
      &  $c$  & Models & C-index &iAUC & & $c$ & Models & C-index &iAUC\\
        \midrule
       \multirow{4}{*}{\begin{tabular}{c}
      Example 1 \\
       ($n = 1000$)     \end{tabular}}& \multirow{4}{*}{$c=0$} & Competing Weibull & 0.744 &0.837& & \multirow{4}{*}{$c=100$}& Competing Weibull &  0.746 &0.836\\
   &  & Cox PH & 0.718&0.807& &  & Cox PH& 0.721&0.809 \\
   & & Weibull & 0.718&0.807 & &  & Weibull & 0.722&0.809\\
    & & Disease mixture & 0.714 &0.808 & &  & Disease mixture & 0.721 &0.809\\
    \midrule
       \multirow{8}{*}{\begin{tabular}{c}
      Example 2 \\
       ($n = 1500$)     \end{tabular}} &\multirow{4}{*}{$c=0$} & Competing Weibull & 0.865 &0.950& &\multirow{4}{*}{$c=161$}  & Competing Weibull &0.875 &0.955 \\
    & & Cox PH & 0.806 &0.902& &  & Cox PH& 0.816 &0.907 \\
    & & Weibull & 0.805 &0.902 & &  & Weibull & 0.816 &0.907\\
    & & Disease mixture & 0.766 &0.869 & &  & Disease mixture & 0.834 &0.927\\
   \cmidrule{2-5}\cmidrule{7-10}
     & \multirow{4}{*}{$c=289$}  & Competing Weibull & 0.879 &0.953  & &\multirow{4}{*}{$c=443$}  & Competing Weibull & 0.883 &0.952 \\
   & & Cox PH & 0.820 &0.904 & && Cox PH& 0.826 &0.903 \\
   &  & Weibull & 0.820 &0.904 &  & & Weibull & 0.826 &0.903\\
    & & Disease mixture & 0.842 &0.927 & &  & Disease mixture & 0.851 &0.927\\
      \midrule
    \multirow{4}{*}{\begin{tabular}{c}
      Example 3 \\
       ($n = 1500$)     \end{tabular}}&\multirow{4}{*}{$c=150$} & Competing Weibull & 0.845 &0.935& & \multirow{4}{*}{$c=446$}& Competing Weibull & 0.851 &0.928 \\
   &  & Cox PH & 0.755 &0.854 & & & Cox PH& 0.759 &0.842\\
   &  & Weibull & 0.754 &0.854 &  &  & Weibull & 0.759 &0.842\\
    & & Disease mixture & 0.798 & 0.895 & &  & Disease mixture & 0.817 &0.897\\
\bottomrule
    \end{tabular}
    \end{threeparttable}
}    \label{tab: comp}
\end{table}

\subsection*{C.2 Simulation with Competing log-normal Model} \label{sec: simulation-log-normal}
In this section, we evaluate the efficiency and accuracy of the proposed method under violations of the conditional independence assumption among latent survival times (independent errors). We simulate data with a sample size of 1500 and 10\% random censoring, incorporating overlapping covariates and varying correlations between latent failure times. For each sample, the covariate vector ($\vk X_i$) is independently drawn from $\mathcal{N}(0, \mathrm{I}_4)$, and the log-transformed latent times for three CFs, $(\log T_{i1}, \log T_{i2}, \log T_{i3})$, are generated from a multivariate normal distribution $\mathcal{MVN}(\boldsymbol{\mu}_i, \boldsymbol\Sigma)$, where the mean $\boldsymbol{\mu}_i = (\mu_{i1}, \mu_{i2},\mu_{i3})^\top$ is determined by the covariates and regression coefficients, i.e., $\mu_{il} = \alpha_l+ \vk X_{il}^\top\boldsymbol \beta_l,\ l=1,2,3$ (Table \ref{tab: estln}). The covariance matrix $\boldsymbol\Sigma = (\sigma_{jk})$ is specified as $\sigma_{jj} = \sigma_j^2, \sigma_{jk} = \rho\sigma_j\sigma_k, 1\le j\ne k\le 3$ with a constant $\rho\in [0,1)$.

\begin{table}[H]
    \centering
    \caption{Parameter setting and estimated value (bias) from the competing log-normal model with varying levels of dependence (correlation). Here, $c$ represents the number of censored data and $\rho$ represents the correlation.
    }
\resizebox{1\textwidth}{!}{
    \begin{threeparttable}
    \begin{tabular}{lccrrrrrr}\toprule
        & $c$ & & ${\sigma}$  & ${\alpha}$  & ${\beta}_1$  & ${\beta}_2$  & ${\beta}_3$  & ${\beta}_4$   \\
        \midrule
     \multirow{3}{*}{\begin{tabular}{c}
       Parameters\\
        Setting
    \end{tabular}} &  \multirow{3}{*}{10\%}& CF1& 1.0 & 1.0  &  $-$3.0& 2.0& - & 1.0\\
   & & CF2 & 1.0 &1.5 & 2.0& 2.0 & - & -\\
   & & CF3 & 1.5 &1.0 & $-$2.0& 3.0& 2.0 &-  \\
  \midrule
   \multirow{3}{*}{\begin{tabular}{c}
        Case 1 \\
        $(\rho = 0)$
    \end{tabular}} &  \multirow{3}{*}{$c=1342$}& CF1& 1.059 (0.059) & 0.981 (0.019)  &  $-$2.927 (0.073)& 2.043 (0.044)& - & 1.050 (0.050)\\
   & & CF2 & 1.014 (0.014) & 1.536 (0.036)& 2.038 (0.038)& 2.003 (0.003) & - & -\\
   & & CF3 & 1.445 (0.055)& 0.997 (0.003) & $-$1.977 (0.023)& 2.913 (0.086)& 2.056 (0.056) &-  \\
   \midrule
      \multirow{3}{*}{\begin{tabular}{c}
        Case 2 \\
        $(\rho = 0.2)$
    \end{tabular}} & \multirow{3}{*}{$c=1341$}& CF1& 0.982 (0.017) & 0.952 (0.047)  &  $-$2.998 (0.001)& 2.022 (0.022)& - & 0.902 (0.097)\\
   & & CF2 & 1.010 (0.010)& 1.746 (0.246)& 2.134 (0.134)& 2.016 (0.016) & - & -\\
  & & CF3 & 1.528 (0.028) & 1.338 (0.338) & $-$2.064 (0.064)& 3.020 (0.020)& 2.196 (0.196) &- \\
  \midrule
      \multirow{3}{*}{\begin{tabular}{c}
        Case 3 \\
        $(\rho = 0.5)$
    \end{tabular}} & \multirow{3}{*}{$c=1348$}& CF1& 0.998 (0.002) & 1.241 (0.241)  &  $-$3.159 (0.159)& 1.939 (0.060)& - & 1.127 (0.127)\\
   & & CF2 & 1.032 (0.032)& 1.753 (0.253)& 2.104 (0.104)& 1.932 (0.067) & - & -\\
  & & CF3 & 1.653 (0.153) & 1.050 (0.050) & $-$1.942 (0.057)& 2.880 (0.120)& 1.935 (0.064) &- \\
  \midrule
      \multirow{3}{*}{\begin{tabular}{c}
        Case 4 \\
        $(\rho = 0.8)$
    \end{tabular}} & \multirow{3}{*}{$c=1346$}& CF1& 1.050 (0.050) & 1.370 (0.370)  &  $-$3.187 (0.187)& 1.904 (0.095)& - & 1.147 (0.147)\\
   & & CF2 & 1.051 (0.051)& 1.866 (0.366)& 2.160 (0.159)& 1.953 (0.046) & - & -\\
  & & CF3 & 1.741 (0.241) & 1.280 (0.280) & $-$1.974 (0.026)& 2.927 (0.072)& 2.010 (0.010) &- \\
\bottomrule
    \end{tabular}
    \end{threeparttable}
}    \label{tab: estln}
\end{table}

With tuning parameters selected as $\lambda_1=2$ and $\lambda_2=1$ from a grid search over $\lambda_1 \in [0,2]$ and $\lambda_2 \in [0,1]$, the estimation results in Table \ref{tab: estln} show small biases in the covariate coefficients (most $<0.1$ and all $<0.2$), indicating relatively stable performance. In contrast, the intercept estimates exhibit larger biases as the correlation $\rho$ increases, reaching over 0.3 for $\rho=0.8$. These results suggest that, when the independence assumption is violated, the covariate coefficients remain relatively robust, whereas the intercept estimates are sensitive. Therefore, the relative importance of the CFs may be affected, and the attribution of failures to each CF should be interpreted with caution. 

\begin{figure}[!htbp]
    \centering
        \subfloat[]{
    \includegraphics[width=0.32\linewidth]{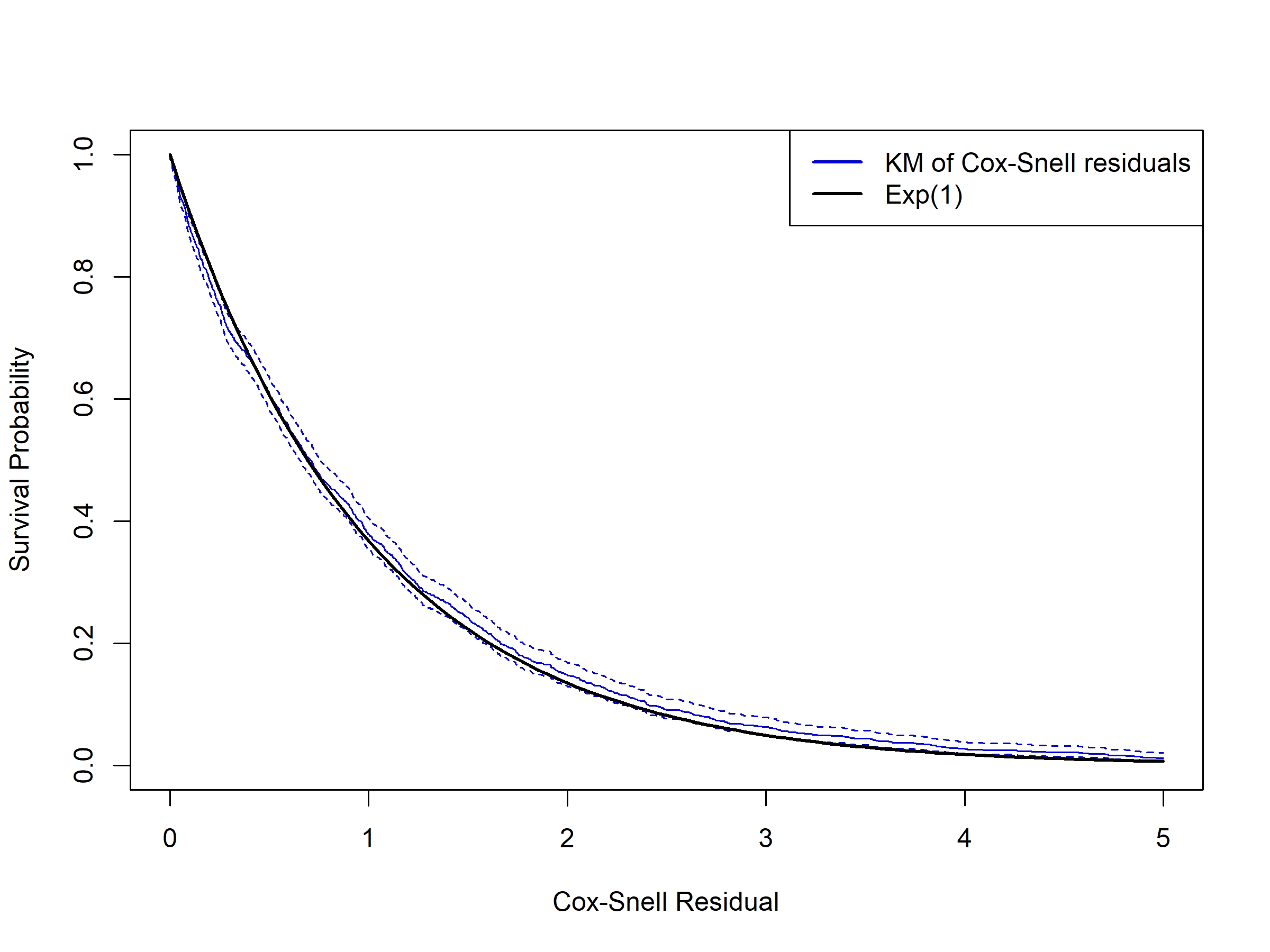}}
    \subfloat[]{
    \includegraphics[width=0.32\linewidth]{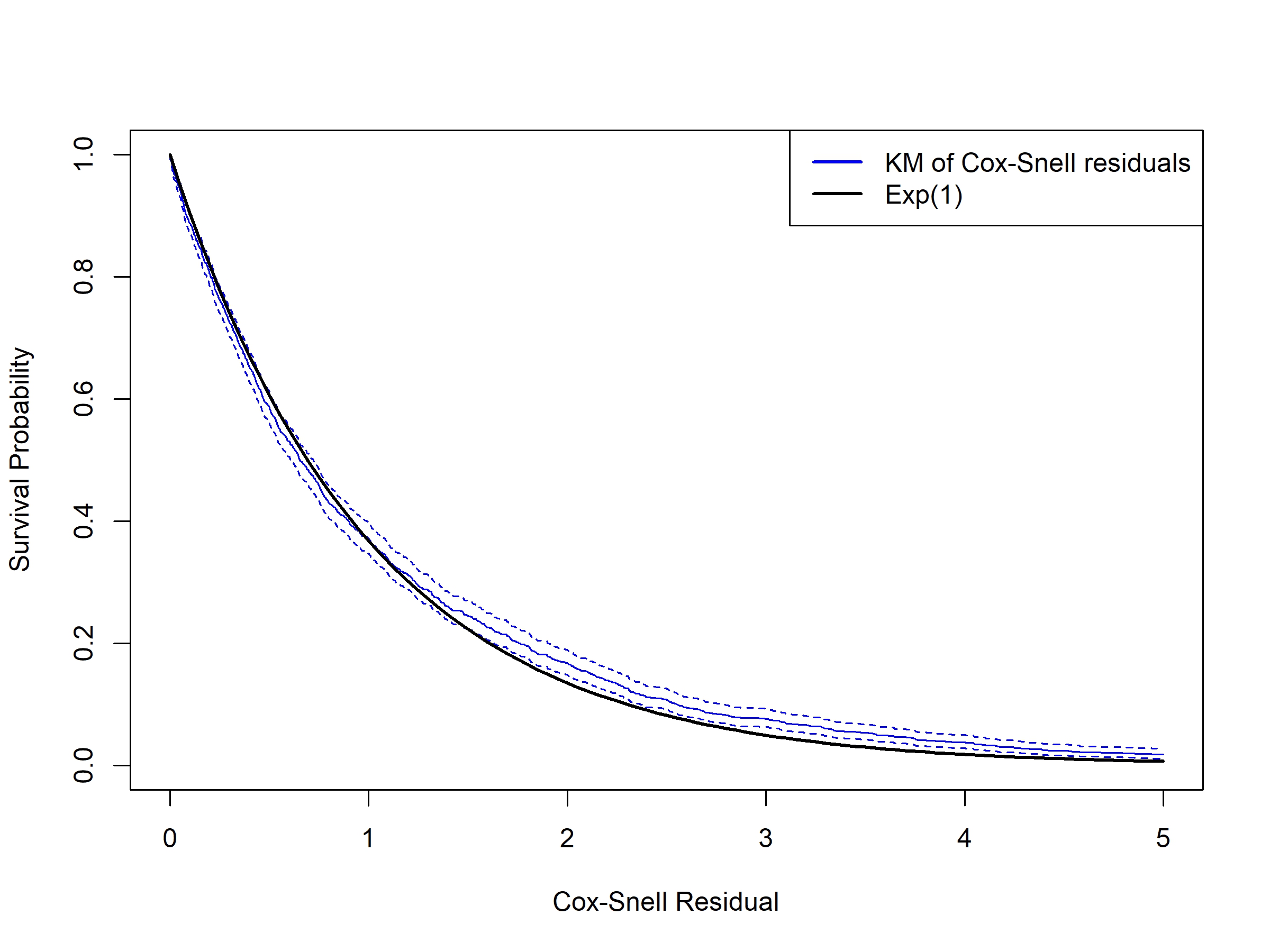}}
    \subfloat[]{
    \includegraphics[width=0.32\linewidth]{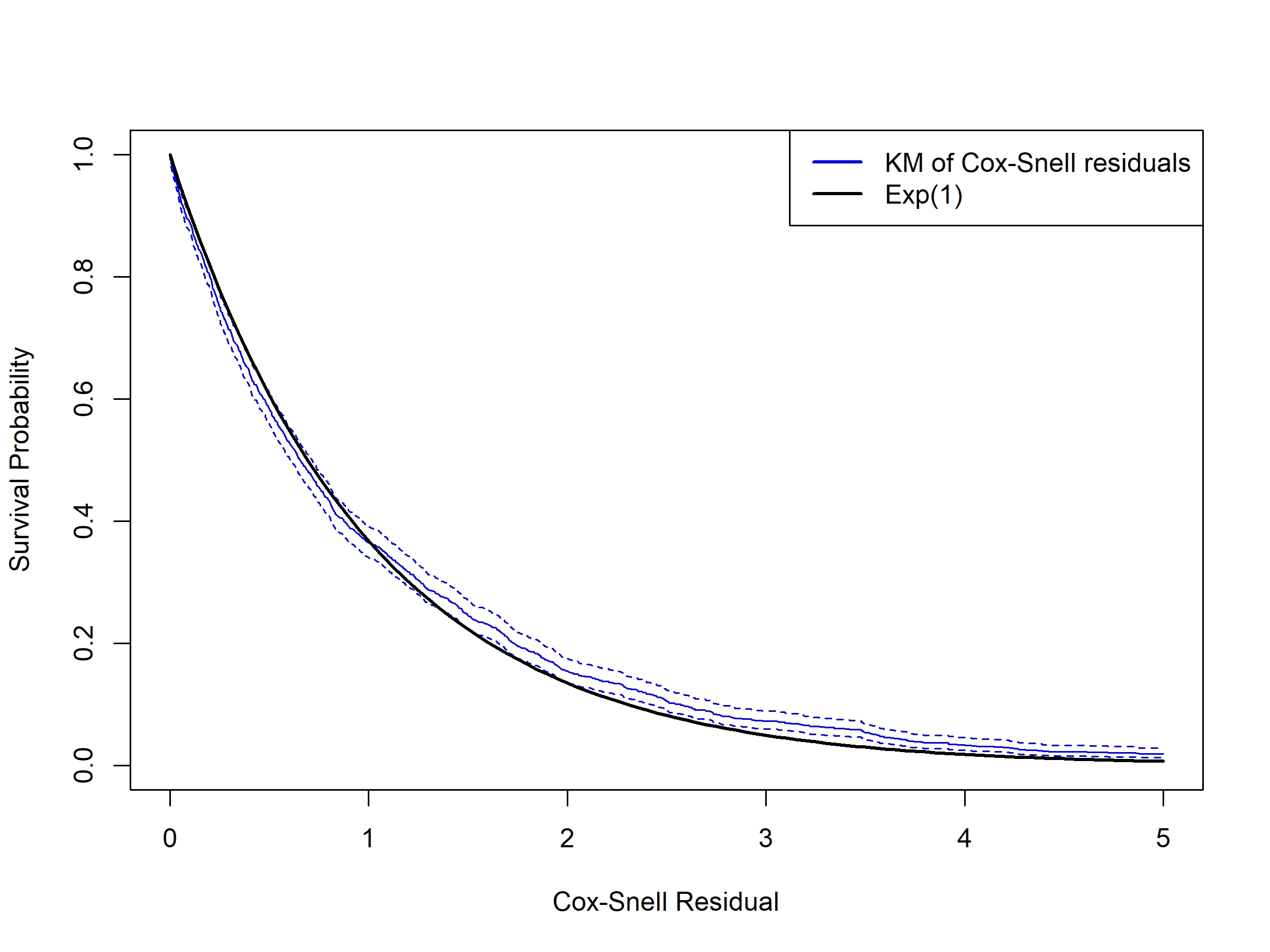}}
    \caption{Diagnostic plots of Cox-Snell residuals with dependent errors, with $\rho=0.2$ in (a), $\rho=0.5$ in (b), and $\rho=0.8$ in (c)}
    \label{fig: diagLN}
\end{figure}

{Additionally, the diagnostic plots (Figure \ref{fig: diagLN}) compare Cox-Snell residuals with the theoretical standard exponential distribution. While the weak dependence case ($\rho=0.2$) is a bit difficult to observe, moderate ($\rho=0.5$) and strong ($\rho=0.8$) dependence show clear deviations, demonstrating the effectiveness of the proposed diagnostic.}

\end{document}